\documentclass[12pt,a4paper]{article}
\usepackage{amsmath,amssymb,amsthm,epsfig}
\usepackage{epsfig}
\theoremstyle{plain}
\renewcommand{\theequation}{\thesection.\arabic{equation}}
\newlength{\extraspace}
\newlength{\extraspaces}
\newtheorem{theorem}{Theorem}[section]
\newtheorem{definition}{Definition}[section]
\newtheorem{proposition}{Proposition}[section]
\newtheorem{lemma}{Lemma}[section]
\newtheorem{remark}{Remark}[section]
\newtheorem{corollary}{Corollary}[section]
\newcommand{\be}{\begin{equation}
\addtolength{\abovedisplayskip}{\extraspaces}
\addtolength{\belowdisplayskip}{\extraspaces}
\addtolength{\abovedisplayshortskip}{\extraspace}
\addtolength{\belowdisplayshortskip}{\extraspace}}
\newcommand{\ee}{\end{equation}}
 
\newcommand{\ba}{\begin{eqnarray}
\addtolength{\abovedisplayskip}{\extraspaces}
\addtolength{\belowdisplayskip}{\extraspaces}
\addtolength{\abovedisplayshortskip}{\extraspace}
\addtolength{\belowdisplayshortskip}{\extraspace}}
\newcommand{\ea}{\end{eqnarray}}

\newcommand{\bas}{\begin{eqnarray*}
\addtolength{\abovedisplayskip}{\extraspaces}
\addtolength{\belowdisplayskip}{\extraspaces}
\addtolength{\abovedisplayshortskip}{\extraspace}
\addtolength{\belowdisplayshortskip}{\extraspace}}
\newcommand{\eas}{\end{eqnarray*}}
 
\newcounter{subequation}[equation]
\makeatletter

\expandafter\let\expandafter
\reset@font\csname reset@font\endcsname

\def\subeqnarray{\arraycolsep1pt
    \def\@eqnnum\stepcounter##1{\stepcounter{subequation}%
        {\reset@font\rm(\theequation\alph{subequation})}}
\jot5mm     \eqnarray}

\def\subarray{\arraycolsep1pt
    \def\@eqnnum\stepcounter##1{\stepcounter{subequation}%
        {\reset@font\rm(\alph{subequation})}}
\jot5mm     \eqnarray}

\makeatother
\newcommand{\newsection}[1]{
\vspace{15mm}
\pagebreak[3]
\addtocounter{section}{1}
\setcounter{equation}{0}
\setcounter{subsection}{0}

\addcontentsline{toc}{section}
{\protect\numberline{\thesection}{#1}}

\begin{flushleft}
{\large\bf \thesection. #1}
\end{flushleft}
\nopagebreak
\medskip
\nopagebreak}
 
\newcommand{\newsubsection}[1]{
\vspace{1cm}
\pagebreak[3]
 
\addtocounter{subsection}{1}
\addcontentsline{toc}{subsection}
{\protect\numberline{\thesection.\arabic{subsection}}{#1}}

\noindent{ \bf \thesection.\arabic{subsection} #1}
\nopagebreak
\vspace{2mm}
\nopagebreak}
\newcommand{\newappendix}[1]{
\vspace{15mm}
\pagebreak[3]
\addtocounter{section}{1}
\setcounter{equation}{0}
\setcounter{subsection}{0}

\addcontentsline{toc}{section}
{\protect\numberline{\thesection}{#1}}

\renewcommand{\theequation}{\Alph{section}.\arabic{equation}}
\begin{flushleft}
{\large\bf Appendix \Alph{section}: #1}
\end{flushleft}
\nopagebreak
\medskip
\nopagebreak}

\newcommand{\N}{\mathbb{N}}
\newcommand{\C}{\mathbb{C}}
\newcommand{\Z}{\mathbb{Z}}
\newcommand{\R}{\mathbb{R}}

\newcommand{\T}{\mathbb{T}}

\renewcommand{\H}{\mathbb{H}}
\newcommand{\up}{\uparrow}

\newcommand{\1}{\mbox{1\hspace{-.65ex}I}}
\newcommand{\bra}{\langle}
\newcommand{\ket}{\rangle}
\newcommand{\ra}{\rightarrow}

\newcommand{\rra}{\ \longrightarrow \ }

\newcommand{\is}{ &\! =\! & }
\newcommand{\nonum}{\nonumber \\[1.5mm]}
\newcommand{\sspace}{\makebox[1cm]{ }}

\newcommand{\nspace}{\!\!\!\!\!\!\!\!\!\!}

\newcommand{\Tr}{{\rm Tr}}

\newcommand{\inv}{^{-1}}

\renewcommand{\th}{{\theta}}
\newcommand{\eps}{\epsilon}
\newcommand{\lb}{\lambda}
\newcommand{\om}{\omega}
\newcommand{\sh}{{\rm sh}}
\newcommand{\ch}{{\rm ch}}
\newcommand{\dd}{{\partial}}

\newcommand{\cA}{{\cal A}}
\newcommand{\cB}{{\cal B}}
\newcommand{\cC}{{\cal C}}
\newcommand{\cD}{{\cal D}}
\newcommand{\cE}{{\cal E}}

\newcommand{\cG}{{\cal G}}
\newcommand{\cH}{{\cal H}}
\newcommand{\cJ}{{\cal J}}
\newcommand{\cL}{{\cal L}}
\newcommand{\cN}{{\cal N}}
\newcommand{\cM}{{\cal M}}

\newcommand{\cP}{{\cal P}}

\newcommand{\cS}{{\cal S}}
\newcommand{\cT}{{\cal T}}

\newcommand{\cZ}{{\cal Z}}

\newcommand{\bt}{{\bf T}} 
\newcommand{\bA}{{\bf A}}

\begin{document}

%%%%%%%%%%%%%%%%%%%%%%%%%%%%%%%%%%%%%%%%%%%%%%%%%%%%%%%%%%%%%%%%%%%%%%%
\begin{titlepage}

%footnotesymbols others than numbers
\renewcommand{\thefootnote}{\fnsymbol{footnote}}
\begin{flushright}
MPP-2006-5\\
to appear in Commun.~Math.~Phys.
\end{flushright}
\mbox{}
\vspace{2mm}

\begin{center}
\mbox{{\Large \bf Structure of the space of ground states}}
\\[4mm]
\mbox{{\Large \bf in systems with non-amenable symmetries}}
\vspace{2.5cm}

{{\sc M.~Niedermaier}%
\footnote{Membre du CNRS; e-mail: {\tt max@phys.univ-tours.fr} }}
and {{\sc E.~Seiler}}
\\[4mm]
{\small\sl Laboratoire de Mathematiques et Physique Theorique}\\
{\small\sl CNRS/UMR 6083, Universit\'{e} de Tours}\\
{\small\sl Parc de Grandmont, 37200 Tours, France}
\\[3mm]
{\small\sl Max-Planck-Institut f\"{u}r Physik}\\
{\small\sl F\"ohringer Ring 6}\\
{\small\sl 80805 M\"unchen, Germany}

\vspace{1.4cm}

{\bf Abstract}
\vspace{-3mm}

\end{center}

\begin{quote}
We investigate classical spin systems in $d\!\geq \!1$ dimensions whose 
transfer operator commutes with the action of a nonamenable unitary
representation of a symmetry group, 
here ${\rm SO}(1,N)$; these systems may alternatively be interpreted
as systems of interacting quantum mechanical particles moving on
hyperbolic spaces. In sharp contrast to the analogous situation with 
a compact symmetry group the 
following results are found and proven: 
(i) Spontaneous symmetry breaking already takes place for finite spatial 
volume/finitely many particles and even in dimensions $d=1,2$. The tuning 
of a coupling/temperature parameter cannot prevent the symmetry breaking.  
(ii) The systems have infinitely many non-invariant and non-normalizable 
generalized ground states. (iii) the linear space spanned by these ground 
states carries a distinguished unitary representation of ${\rm SO}(1,N)$, 
the limit of the spherical principal series. (iv) The properties 
(i)--(iii) hold universally, irrespective of the details of the interaction. 
\end{quote} 
\vfill

\setcounter{footnote}{0}
\end{titlepage}

%%%%%%%%%%%%%%%%%%%%%%%%%%%%%%%%%%%%%%%%%%%%%%%%%%%%%%%%%%%%%%%%%%%%%%%%%%%%%
\tableofcontents
\newpage
%%%%%%%%%%%%%%%%%%%%%%%%%%%%%%%%%%%%%%%%%%%%%%%%%%%%%%%%%%%%%%%%%%%%%%%%%%%%%

\newsection{Introduction}

Spontaneous symmetry breaking is typically discussed for compact internal
or for abelian translational symmetries, see e.g.~\cite{Ruelle,Sewell,NT}. 
Both share the property of being amenable \cite{Pat} and their spontaneous 
breakdown is a specific dynamical property of the interaction. Here we 
consider systems with a nonamenable symmetry, by which we mean that the 
dynamics is invariant under a nonamenable unitary representation 
\cite{bekka} of a locally compact group (which then by necessity is also 
nonamenable). One goal 
of this note is to show, roughly, that whenever the dynamics of a system of 
classical statistical mechanics is invariant under a nonamenable symmetry, 
this symmetry is {\it always} spontaneously broken, irrespective of the 
details of the interaction. Neither does the long or short ranged nature 
of the interaction matter, nor can the tuning of a (temperature) 
parameter prevent the symmetry breaking. Spontaneous symmetry breaking 
even occurs in one and two dimensions, where for compact symmetries this 
is ruled out by the Mermin-Wagner theorem. The phenomenon is not limited 
to a semiclassical regime and occurs already for systems with finitely 
many degrees of freedom.

The systems will be defined on a finite lattice $\Lambda$ of 
arbitrary dimension and connectivity. The dynamical variables are 
`spins' attached to the vertices of the 
lattice, taking values in some noncompact Riemannian symmetric space 
$Q=G/K$, where $G$ is the noncompact symmetry group and $K$ a maximally 
compact subgroup. The dynamics is specified by a transfer operator acting 
on the square integrable functions on the configuration manifold 
$Q^\Lambda$. Such a system can alternatively be interpreted as a quantum 
mechanical system of finitely many particles 
living on $Q$; we only have to interpret the transfer matrix as 
$\exp(-H)$ and thereby define the Hamiltonian $H$; the inevitable 
spontaneous symmetry breaking appears then as degeneracy of the 
generalized ground states of this system. Conversely, given a 
quantum mechanical system with a Hamiltonian $H$,
we can re-interpret the system as one of classical 
statistical mechanics with $\exp(- H)$ as the transfer matrix. 
To fix ideas one may take a Hamiltonian of the conventional form 
\be
H=-  \frac{1}{2} \sum_{i=1}^\nu\Delta_i+\sum_{i,k=1}^\nu V_{ik}\ ,
\label{i0}
\end{equation}
where $\Delta_i$ is the Laplace-Beltrami operator on $Q$ for the i-th 
particle and the $V_{ik}$ are some potentials describing the interaction of 
particle $i$ and $k$, depending only on the geodesic distances of the 
particles. Typically one would also require that the set $\Lambda=\{1,2 \ldots 
\nu\}$ has the structure of a lattice and that the interaction 
links neighboring sites only. The Hamiltonians (\ref{i0}) are however 
only one class of examples, many others are covered. Indeed  
apart from some technical conditions on the transfer matrix it is mostly 
the invariance that matters.  

To analyze these systems, it is necessary to perform something analogous 
to the well-known separation of the center of mass motion from the 
relative motion in Euclidean space. This turns out to be considerably more 
involved in our setting. It is convenient not to define an actual center 
of mass of the $\nu$ spins or particles, respectively, but rather take 
simply one of them as parameterizing the global position of the 
configuration. The universality of the resulting symmetry 
breaking is then related to the fact that the global motion of these 
systems never allows for a symmetric (proper or generalized) ground state.
 
Normally spontaneous symmetry breaking can only happen in the thermodynamic 
limit. We stress again that here the situation is different: spontaneous 
symmetry breaking takes place already on a finite spatial lattice.
The systems exhibit a 
remarkable universality in the structure of their generalized ground 
states. Namely, there are always {\it infinitely many} non-normalizable and 
non-invariant ground states which transform irreducibly under a 
preferred representation of the group -- the {\it same} for a large class 
of transfer operators! All generalized ground states can be 
generated by forming linear combinations of factorized wave functions, 
where one factor describes the global and the other one the relative 
motion; of course the second factor will be sensitive to the interaction 
as far as the relative motion is concerned; what is universal is the 
transformation law under global symmetry transformations of the first 
factor. 

In quantum one-particle systems described by an exactly soluble 
Schr{\"o}dinger equation an infinite degeneracy in the ground state energy 
has been found earlier: first of all in the well known problem of the 
Landau levels in the Euclidean plane; closer to our situation explicitly 
in  \cite{macf} for the supersymmetric {\rm SO(1,2)} invariant 
quantum mechanics and implicitly in \cite{lott} (p.172), \cite{zirnmig} 
and in \cite{comtet} for the lowest Landau level. The interplay between 
spontaneous symmetry breaking, nonamenability and properties of the 
transfer operator was understood in \cite{hchain}, initially for the 
hyperbolic spin chain.

The thermodynamic limit can usually only be taken on the level of 
correlation functions, so that the `fate' of the ground state orbit 
cannot directly be traced. By means of an Osterwalder-Schrader 
reconstruction one can in principle recover a Hilbert space description, 
however inevitably one with an exotic structure; cf.~\cite{hchain}. 

Our main example for $Q$ will be the hyperboloids $\H_N := 
{\rm SO}_0(1,N)/{\rm SO}(N)$, $N \geq2$, in part because of the importance of the 
Lorentz and de Sitter groups in physics, and in part because already for 
the groups ${\rm SO}_0(1,N)$ the harmonic analysis exhibits all of the 
characteristic complications. Most of the constructions however generalize 
to a large class of noncompact coset spaces and are actually easier to 
understand in a general setting. We thus specialize to $Q = \H_N$ only 
when needed. 

Let us now make things a little more explicit: the configuration space 
$\cM$ is the direct product of $\nu := |\Lambda|$ copies of the space $Q$. 
A hypercubical lattice $\Lambda \subset \Z^d$ of arbitrary dimension $d$ is 
a prime example, however neither the dimension nor the connectivity of the 
lattice is essential. The pure states of the system are described by 
elements of $L^2(\cM)$, 
i.e.~functions $\psi: \cM \ra \C$, square integrable with respect to the 
invariant measure $d\gamma$. The left diagonal $G$ action $d$ on $\cM$ 
induces a unitary representation $\ell_{\cM}$ of $G$ on $L^2(\cM)$ via 
$(\ell_{\cM}(g) \psi)(m) = \psi( g^{-1} m)$. Since $L^2(\cM) \simeq 
[L^2(Q)]^{\otimes \,\nu}$ it can be identified with the $\nu$-fold 
inner tensor product of the left quasiregular representation $\ell_1$.
On this $L^2$ space we consider bounded selfadjoint operators $\bA$ commuting with 
the group action, i.e.~$\ell_{\cM} \circ \bA = \bA \circ \ell_{\cM}$. 
Such operators $\bA$ can only have essential spectrum.

Specifically we consider so-called {\it transfer operators} defining 
the dynamics of the system. A precise 
definition is given in Definition \ref{transfer} below. Here 
it may suffice to say that a transfer operator is a bounded selfadjoint 
operator on $L^2$ which is positive as well as positivity improving; as 
usual the latter property is realized by taking for $\bt$ an integral 
operator with strictly positive kernel, $T(m,m') >0$ for all $m,m' \in 
\cM$. We are interested in an invariant dynamics, so we assume 
$\ell_{\cM} \circ \bt = \bt \circ \ell_{\cM}$. Important 
examples are $\bt = \exp(-H)$, with $H$ as in (\ref{i0}),
but Hamiltonians with more complicated `time derivative' 
terms and non-pair potentials would also be allowed. 
The latter is welcome because such complicated
Hamiltonians naturally arise as the result of blocking 
transformations. It is easy to see that an invariant transfer operator 
$\bt$ cannot be a compact operator, furthermore it cannot even  
have normalizable ground states, i.e.~solutions of $\bt \psi = 
\Vert \bt \Vert \psi$, with $\psi \in L^2$. (In fact $\Vert \bt \Vert$  
must either lie in the continuous spectrum of $\bt$ or be a limit of 
eigenvalues with infinite multiplicity). Instead our setting is such that
$\bt$ has a continuous extension to an operator from $L^p$ to $L^p$ for 
$1\le p\le\infty$ and we identify conditions under which $\bt$ has {\it 
generalized ground states}. These we take as almost everywhere defined 
{\it functions} (not just distributions) which are eigenfunctions 
with spectral value $\Vert \bt \Vert$. The set of these generalized 
ground states forms a linear space which we call the {\it ground state
sector} $\cG(\bt)$  of $\bt$. It is important that $\bt$ is defined as a 
selfadjoint operator on a Hilbert space, here $L^2(\cM)$, so that the 
spectral theorem can be applied to provide a resolution of the identity.
The notion of a generalized ground state is then unambiguous, although 
there is some freedom in the choice of the topological vector space in 
which the ground state wave functions live. 

The separation of global and relative configurations is achieved by 
writing  $\cM = Q \times \cN$, where $\cN$ collects the `relative degrees 
of freedom'; the construction is done in such a way that $\cM= 
(G\times \cN)/d(K)$, where $d(K)$ is the right diagonal action of $K$ on 
$G\times\cN$. The details of this construction will be given in Section 2. 
The global part of the configurations can now be Fourier transformed:
the $L^2$ functions on $G$ have a Plancherel-type decomposition
$\int^{\oplus} \! d\nu(\sigma) \, \cL_{\sigma} \otimes 
\check{\cL}_{\check\sigma}$, where the fiber spaces $\cL_{\sigma}$ 
carry the unitary (infinite dimensional) $\nu$-almost always 
irreducible representations $\pi_{\sigma}$, and $\nu$ is carried by the  
so-called restricted dual $\widehat{G}_r$ of $G$. We show in 
Section 3 that one can associate to an invariant selfadjoint operator 
$\bA$ on $L^2(\cM)$ a $\nu$-measurable 
field of bounded selfadjoint operators 
$\bA_{\sigma}$ on $L^2(\cN) \otimes \cL_{\sigma}$, $\sigma 
\in \widehat{G}_r$, via
\be 
\bA = \int \! d\nu(\sigma) (\1 \otimes \check\bA_{\check\sigma})\,, 
\sspace 
L^2(\cM) = \int \! d\nu(\sigma) \,\cL_{\sigma}^2(\cM)\,.
\label{i1}
\end{equation}
In the second formula we indicated that the state space $L^2(\cM)$ 
decomposes into fibers which carry the representation $\pi_{\sigma}$ and 
which are preserved under the action of $\1 \otimes 
\check\bA_{\check\sigma}$. Each of the fiber spaces is isometric to 
$\cL^2_{\sigma}(\cM) \cong  \cL_{\sigma} \otimes \check\cL_{\check\sigma} 
\otimes L^2(\cN)$; however its elements will be realized as functions on 
$\cM$. This is done by constructing for $v \in \cL_{\sigma}$ an 
(antilinear) map $\tau_{v \sigma} : 
L^2(\cN) \otimes \cL_{\sigma} \ra \cL^2_{\sigma}(\cM)$, 
$f \mapsto \tau_{v \sigma}(f)$. Roughly, the image function arises 
by reinterpreting the matrix element $(f(n), \pi_{\sigma}(g) 
v)_{\sigma}$ as a (generically not square integrable) function on 
$\cM$. The map is designed such that it is an isometry onto its image and 
has the following intertwining properties 
\begin{subeqnarray}
&& \tau_{v\sigma}(f)(g^{-1} m) = 
\tau_{\pi_{\sigma}(g)v,\sigma}(f)(m) \,,
\\
&& [\bA \tau_{v \sigma}(f)](m) = 
\tau_{v \sigma}(\bA_{\sigma} f)(m)\,.
\label{i2}
\end{subeqnarray}
According to the first equation the $G$-action on the argument of the function
just rotates the reference vector $v \in \cL_{\sigma}$ with the 
representation $\pi_{\sigma}$. In the second relation we anticipated 
that the action of $\bA$ can be extended to the (in general non-$L^2$)
functions $\tau_{v \sigma}(f)$. In view of (\ref{i2}) it 
is plausible that the spectral problems of $\bA$ and $\bA_{\sigma}$ 
are related as follows: suppose first that the eigenvalue equations 
$\bA_{\sigma} \chi = \lb \chi$ and $\bA \Omega = \lb \Omega$ are 
well-defined, with $\lb \in {\rm Spec}(\bA)$ and (generically non-$L^2$) 
eigenfunctions $\chi,\,\Omega$, and that second the map $\tau_{v \sigma}$ 
admits an extension to the 
generalized eigenfunctions $\chi$ of $\bA_{\sigma}$. Then by (\ref{i2}b) the 
image function $\tau_{v \sigma}(\chi)$ should be an eigenfunction of $\bA$ 
enjoying the equivariance property (\ref{i2}a).  

This construction principle can be implemented 
for a large class of invariant selfadjoint operators 
$\bA$ specified in Sections 3.3 and 3.4. In overview the result is 
that a set of generalized eigenfunctions $\Omega_{\lb \sigma}$ 
exists for almost all $\lb \in {\rm Spec}(\bA)$ and 
$\sigma \in \widehat{G}_r$ with the following properties: 
(i) they are almost everywhere defined functions (not 
distributions) $\Omega_{\lb,\sigma} : \cM \ra \C$.
(ii) they are $\sigma$-equivariant, i.e.~they lie in 
the image of the maps $\tau_{e_i\sigma}$, where $\{e_i\}$ is an 
orthonormal basis of $\cL_{\sigma}$ and where the domain is   
the linear hull of a complete set of eigenfunctions 
of $\bA_{\sigma}$ with spectral value $\lb \in {\rm Spec}\bA_{\sigma}$. 
(iii) $\Omega_{\lb\sigma}(g m ) \ra 0$, as $g$ leaves compact subsets 
of $G$. (iv) As $\lb$ runs through ${\rm Spec}(\bA)$ 
and $\sigma$ runs through $\widehat{G}_r$, the eigenfunctions 
$\Omega_{\lb \sigma}$ are complete, in the sense that 
any smooth function can be expanded in terms of the 
$\Omega_{\lb \sigma}$ and that a Parseval relation holds 
on $(L^1 \cap L^2)(\cM)$.  

The transfer operators $\bt$ considered are special cases 
of such invariant selfadjoint operators, which in addition 
are positive and positivity improving. The spectral value 
relevant for the ground states of a transfer operator $\bt$ is 
$\lb = \Vert \bt \Vert$. The important ``almost all $\lb 
\in {\rm Spec}(\bt)$'' clause in the above completeness result 
prevents one from getting all the generalized ground states simply 
by specialization. However, whenever for some $\sigma \in 
\widehat{G}_r$ a complete set of eigenfunctions of $\bt_{\sigma}$ with 
spectral value $\Vert \bt \Vert$ can be found, their images 
under $\tau_{e_i\sigma}$ will produce $\cG_{\sigma}(\bt)$, 
the space of $\sigma$-equivariant generalized ground states.

Off hand of course every $\sigma \in \widehat{G}_r$ could occur 
as a ``representation carried by the ground state sector'' 
in $\cG_{\sigma}(\bt)$. Remarkably this is not the case:
under fairly broad conditions only {\it one} representation 
occurs and always the same! For definiteness we formulate the 
following results for $G = {\rm SO}_0(1,N)$, $K = {\rm SO}(N)$, 
$M= {\rm SO}(N\!-\!1)$; many aspects however are valid for any noncompact 
linear reductive  Lie group.

\begin{theorem}\label{theorem1}   
Let $\bt$ be a transfer operator on $L^2(\cM, d\gamma)$ 
commuting with the unitary representation induced by a proper action $d$ 
of $G$. 
%and let $\cM/d(G)= \cN/d(K)$ be the space of orbits. 
In terms of the fiber decomposition (\ref{i1}) one has: 
\begin{itemize}
\item[(a)] $\cG_{\sigma}(\bt)$ is empty for all but the principal series 
representations whenever one of the following holds: (i) $\cG_{\sigma}(\bt)$ 
contains a strictly positive function. (ii) $\cG_{\sigma}(\bt)$ 
contains a $K$-singlet. (iii) $\bt_{\sigma}$ is compact.
\item[(b)] $\cG_{\sigma}(\bt)$ is non-empty for at most one of the
principal series representations -- the limit of the spherical
($M$-singlet) principal series.
\end{itemize}
\end{theorem}
Combined (a) and (b) imply that if there are generalized 
ground states which transform equivariantly according to some 
unitary irreducible representation $\pi_{\sigma}$, this 
representation must be -- under any of the conditions (i)--(iii) 
and possibly others -- the limit of the spherical principal series,
for which we write $\pi_{00}$.  

It remains to establish the existence of such generalized 
ground states. The known construction principles for generalized 
eigenfunctions (the classic ones \cite{gelfand,maurin}, as well as the 
one described above) are not sufficient to assure the existence of 
generalized ground states (neither as functions nor as distributions) -- 
so any of the fiber spaces in Theorem 1.1 could be empty, 
including $\cG_{00}(\bt)$.
In a follow-up paper \cite{NieSei3} we describe a construction principle 
which ensures the existence of generalized ground states in various    
situations; a preview is given in the conclusions. 
There is a simple case that is, however, important for applications,
in which the existence of generalized ground states is immediate: 
the case that all the fiber operators $\bt_{\sigma}$ are compact.  
The eigenspaces with eigenvalue $\Vert \bt_{\sigma} \Vert$ 
can in principle be constructed via the well-known 
projector s-$\lim_{t \ra \infty} (\bt_{\sigma}/\Vert \bt_{\sigma} \Vert)^t$ 
(where the limit exists in the strong operator sense). 
One then has the following concrete variant of Theorem 1.1:

\begin{theorem} Let $\bt$ be as in Theorem 1.1 and assume in addition 
that the fiber operators $\bt_{\sigma}$, $\sigma \in \widehat{G}_r$  are 
compact. Then all fiber spaces except $\cG_{00}(\bt)$ are empty. Further 
$\bt_{00}$ is itself a transfer operator which has a unique ground state 
in $\cG_{00}(\bt)\subset L^2(\cN)\otimes \cL_{\sigma =00}$. 
In the realization of $\cL_{\sigma =00}$ as $L^2(S^{N-1})$ this ground state can be 
represented by a unique a.e.~positive function  $\psi_0$. $\cG(\bt)$ is 
the linear span of functions of the form 
\be
\Omega(q,n) =     
\int_{S^{N-1}} \!\!dS(p) \,
\frac{\psi_0(n,\vec{p})}{(q_0 - \vec{q} \cdot \vec{p})^{\frac{N-1}{2}}}\,,
\label{i3}
\end{equation}
where $q = g q^{\uparrow} = (q_0, \vec{q})$. Here $\cM$ was identified
with $Q \times \cN$ in a way that replaces the original diagonal left 
$G$ action by $q \mapsto g q$. 
\end{theorem}

For comparison we mention here the corresponding results for amenable 
(compact or abelian) Lie groups. When $\bt$ is invariant under the 
action of a compact Lie group, the very same setting entails 
that the ground states have three concordant properties:
normalizability, uniqueness, and invariance. That is, there exists a 
normalizable, nondegenerate ground state, which is a group singlet. This 
ground state can be obtained by acting with a projector $P$ on an 
arbitrary $L^2$ function, where $P$ is obtained as the strong operator 
limit of the iterated (compact or trace class) transfer operator:  
$P = s-\lim_{t \ra \infty} (\bt/\Vert \bt \Vert)^t$. When 
$\bt$ does not have normalizable ground states this limit 
does not exist; the corresponding weak limit will be the null 
vector. A construction of generalized ground states based 
on a similar but more subtle fixed point principle will be discussed in 
\cite{NieSei3}; see the conclusions for a preview. 

When $\bt$ is invariant under a noncompact amenable Lie group, 
for instance in the Euclidean case $Q={\rm ISO}(N)/{\rm SO}(N)$, 
our construction resembles the well-known procedure of separating the 
center-of-mass motion (see for 
instance \cite{RS4}). It yields a ground state sector carrying the trivial 
representation of ${\rm ISO}(N)$ and for which the center-of-mass wave 
function is unique (the properties of the `internal' ground state sector 
replacing $\cG(\bt_{00})$ again depend on the details of the interaction). 
Of the three concordant properties above only the normalizability is lost. 
In appendix C we specialize our constructions to this degenerate 
situation, to contrast it with the non-amenable case. 

Both amenable cases have in common that the representation carried by the 
ground state sector is uniquely determined and always the same, namely the 
trivial one. The above results show which aspects of this picture 
generalize to the case of non-amenable  symmetries (namely, the 
uniqueness of the representation and its universality) and which do 
not (viz, the uniqueness of the ground state). In view of Theorems 1.1 
and 1.2 the limit of the spherical principal series appears to be the natural 
generalization of the singlet in the noncompact setting. 
 
The article is organized as follows.  In Section 2 we describe the 
setting in more detail and explain why the Gel'fand-Maurin theory 
is insufficient to account for the results we are aiming at. The fiber 
decomposition of $L^2(\cM)$ is introduced. Section 3 provides 
the adapted fiber decomposition of a wider class of selfadjoint operators 
${\bf A}$ and relates the spectral problem of ${\bf A}$ to that 
of the fiber operators ${\bf A}_{\sigma}$. These results are 
then applied in Section 4 to the ground state sector of transfer operators, 
giving a proof of Theorem 1.1 and ramifications of it. 
Appendices A and B provide the necessary background on the harmonic 
analysis of non-compact Lie groups. In appendix C we collect 
counterparts of some of the results in the trivial case of 
a flat symmetric space, for the sake of contradistinction. 
For further orientation we refer to the table of contents.  

%%%%%%%%%%%%%%%%%%%%%%%%%%%%%%%%%%%%%%%%%%%%%%%%%%%%%%%%%%%%%%%%%%%%%%%%%%%%%%
\newpage 
\newsection{Group decomposition of the state space}

Here we introduce the class of systems considered and prepare an 
orbit decomposition of the configuration manifold. The wave functions 
(square integrable functions on this manifold) are then subjected to a 
Plancherel decomposition with respect to their global position, which is 
parameterized by an element of $G$. Eventually this gives rise to a 
decomposition of the state space 
$L^2(\cM)$ into fibers labeled by irreducible unitary representations of 
the original group action.

\newsubsection{Generalized spin systems}

We consider generalized spin systems of the following type: 
the dynamical variables take values in an indecomposable 
Riemannian symmetric space $Q := G/K$, with $G$ a 
noncompact Lie group and $K$ a maximal compact subgroup.
The Lie groups will be taken to be linear reductive,
meaning that $G$ is a closed subgroup of ${\rm GL}(N,\R)$ 
or ${\rm GL}(N,\C)$ which is stable under conjugate transpose.
$K$ then is the isotropy group of the point $q^\uparrow =eK\in Q$. We 
further assume that there is an involution $\iota$ such that $K$ consists 
of its fixed points, and that $g \iota(g)^{-1}$ has unit determinant. 
Then $G$ and $K$ form a symmetric pair. Examples are 
${\rm SO}_0(1,N)/{\rm SO}(N)$, ${\rm SL}(N,\R)/{\rm SO}(N)$,
${\rm U}(p,q)/{\rm U}(p)\times {\rm U}(q)$. In Appendix C we 
will also consider the degenerate case $\R^N = {\rm ISO}(N)/{\rm SO}(N)$.
The configuration manifold $\cM$ is the direct product of $\nu := 
|\Lambda| \geq 2$ copies of 
this space. For much of the following $\Lambda$ only has to have the
structure of a point set;
it is assumed though that $\nu \ra \infty$ captures the physical 
intuition of a thermodynamic limit. A hypercubical lattice $\Lambda 
\subset \Z^d$ of arbitrary dimension $d$ is a prime example, however neither 
the dimension nor the structure of the lattice is essential.
Ordering the points in some way, we write $m = (q_1,\ldots , q_{\nu})$ 
for the points in $\cM$. Further we denote by $\gamma_Q$ and $d\gamma_Q$ 
the invariant metric and the measure on $Q$. Equipped with the product 
metric $\gamma(m) := \prod_i \gamma_Q(q_i)$ and the product measure 
$d\gamma(m) := \prod_i d\gamma_Q(q_i)$ the configuration space $\cM$ 
is a simply connected Riemannian manifold, and in fact 
a reducible Riemannian symmetric space. Further $\cM$ carries an action 
$d: G \times \cM \ra \cM$ of the group $G$,
via $d(g)(m) = (g q_1, \ldots g q_{\nu})$, where    
$q \ra g q$ is the left (transitive) action of $G$ on $Q$. Clearly 
$d(g)$ is an isometry and $d(g)(m) = m$ for all $m$ implies that 
$g$ is the identity in $G$, that is, the action of $G$ is effective. Since 
$d(G) := \{ d(g),\, g \in G\}$ is a subgroup of the full 
isometry group which is closed in the compact-open topology, 
$(\cM, \gamma)$ also is a proper Riemannian $G$-manifold in 
the sense of \cite{michor}, Section 5.  
In fact, the main reason for considering product manifolds of the 
above type is that they have a well defined orbit decompositon
$\cM = Q \times \cN$, $\cN/d(K) = \cM/d(G)$, to be described 
later. With certain refinements this generalizes to all proper 
Riemannian $G$-manifolds, see \cite{michor}.

The pure states of the system are described by elements 
of $L^2(\cM)$, i.e.~functions $\psi: \cM \ra \C$, square integrable 
with respect to $d\gamma$. The proper $G$ action $d$ on $\cM$ 
induces a unitary representation $\ell_{\cM}$ of $G$ on $L^2(\cM)$ via 
$(\ell_{\cM}(g) \psi)(m) = \psi(d(g\inv)(m))$. Since $L^2(\cM) \simeq 
[L^2(Q)]^{\otimes \,\nu}$ it can be identified with the $\nu$-fold 
inner tensor product of the left quasi-regular representation $\ell_1$
of $G$ on $L^2(Q)$; we write $\ell_{\cM} \simeq \ell_1^{\otimes \,\nu}$.

As outlined before, we call a bounded selfadjoint $\bt$ on $L^2(\cM)$ 
a transfer operator if it is positive as well as positivity improving. 
Positivity of $\bt$ means $(\psi, {\bf T} \psi) \geq 0$, which is 
equivalent to the spectrum being nonnegative. Typical transfer operators 
are also positivity improving because they arise as integral operators 
with a positive kernel; the formalization as a positivity improving map 
turned out to be useful, see \cite{RS4} p.201 ff. For convenience we 
recall the definitions: a function $\cM \ni m \mapsto \psi(m)
\in \C$ has some property almost everywhere (a.e.) if it holds for all 
$m \in \cM\backslash I$ with $\gamma(I) =0$. A nonzero function is called 
positive if $\psi \geq 0$ a.e.~and strictly positive if $\psi >0$ a.e. 
Then $\bt$ is called positivity preserving if $(\bt \psi)(m) \geq 0$ 
a.e.~and positivity improving if $(\bt \psi)(m) >0$ a.e.~for any positive 
$\psi$. Equivalently $\bt$ is positivity improving iff $(\phi, \bt \psi) 
>0$ for all positive $\phi,\psi \in L^2$; see \cite{RS4} p.202.     

Our notion of transfer operators requires an additional condition:

\begin{definition}\label{transfer}
A transfer operator $\bt$ is a positive integral operator, given by
\be
(\bt \psi)(m) = \int\! d\gamma(m')\, T(m,m') \psi(m')\ .
\end{equation}
where the kernel $T: \cM \times \cM \rightarrow \R_+$ is symmetric,
continuous and strictly positive, i.e.~$T(m,m')>0$ a.e. and satisfies
\begin{equation}
\sup_{m} \int \!d\gamma(m')\, T(m,m') < \infty\,.
\label{Tassumpt1}
\end{equation}
\end{definition}

The second condition is sufficient (but by no means necessary) to ensure
that ${\bf T}$ defines a bounded operator from $L^p$ to $L^p$ for
$1\leq p\leq\infty$; see \cite{Lax} p.173 ff. The operator norm $\Vert 
{\bf T} \Vert_{L^p \ra L^p} = {\rm sup}_{\Vert \phi \Vert_p =1} \Vert
{\bf T} \phi \Vert_p$ is bounded by the integral in (\ref{Tassumpt1}) and
coincides with it for $p=1,\infty$. Positivity of the kernel entails that 
${\bf T}$ is positivity improving. Positivity of the operator (that is, of 
its spectrum) does not follow from this. However if it is not satisfied we 
can switch to ${\bf T}^2$ and the associated integral kernel, where 
positivity is manifest. Without much loss of generality we assume
therefore the kernel to be such that ${\bf T}$ is positive. As a bounded 
symmetric operator on $L^2$ the integral operator defined by $T(m,m')$ has 
a unique selfadjoint extension which we denote by the same symbol ${\bf 
T}$. The kernel of ${\bf T}^t$ will be denoted by $T(m,m';t)$ for $t \in 
\N$. In this situation ${\bf T}$ and all its powers are transfer operators   
in the sense of the previous definition.

An invariant dynamics is specified by a $G$-invariant transfer operator,
i.e.~one which commutes with $\ell_{\cM}$ on $L^2$ 
\be 
\ell_{\cM}(g) \circ \bt = \bt \circ \ell_{\cM}(g) \,,\quad \forall \, g \in G\,.
\label{Trho}
\end{equation}
It is easy to see that $\bt$ then cannot have normalizable ground states 
(see Proposition \ref{spectrum}  below). In this situation one will 
naturally search for generalized eigenstates of $\bt$, which in our case
will be simply solutions of $\bt \Omega = \Vert \bt \Vert\Omega$ with 
$\Omega$ an almost everywhere defined function (not just a distribution) 
on $\cM$. The set of generalized ground states forms a linear space which 
we call the {\it ground state sector} $\cG(\bt)$ of $\bt$.

%%%%%%%%%%%%%%%%%%%%%%%%%%%%%%%%%%%%%%%%%%%%%%%%%%%%%%%%%%%%%%%%%%%%%%%%%%%%%%%
\newsubsection{The spectral problem}

Since $\bt$ commutes with the $G$ action $\ell_{\cM}$, one expects that the 
transfer operator and a set of operators $\cZ$ whose diagonal action 
characterizes an irreducible representation can be diagonalized 
simultaneously. In a sense this is correct and the generalized 
eigenfunctions with spectral value $\Vert \bt \Vert$ in fact belong to a 
special irreducible representation of $G$, see Theorem 1.1. The purpose of 
this interlude is to explain why the general results available in the
literature on such 
(simultaneous) spectral decompositions are insufficient to produce the 
generalized ground states sought for.    

Let $\bA$ be a bounded selfadjoint operator on the separable Hilbert space
$L^2(\cM)$. Let $\Phi \subset L^2(\cM) \subset \Phi'$ be a Gel'fand 
triple \cite{gelfand,bohm} (rigged Hilbert space) for $\bA$. An 
element $\Omega \in \Phi'$ 
is called an eigendistribution or generalized eigenstate of $\bA$ with 
spectral value $\lb \in {\rm Spec}(\bA)$ if $(\phi, (\bA - \lb) \Omega) =0$, 
for all $\phi \in \Phi$. The set of generalized eigenstates for some 
$\lb \in {\rm Spec}(\bA)$ forms a linear subspace of $\Phi'$ which is 
called the generalized eigenspace $\cE_{\lb}(\bA)$ for the 
spectral value $\lb \in {\rm Spec}(\bA)$. If $L^2(\cM)$ carries the 
unitary representation $\ell_{\cM}$ of a connected (noncompact) Lie group 
group $G$ there is a natural action of $G$ on the distributions 
$\Omega \in \Phi'$, 
viz $(\ell_{\cM}(g) \circ \Omega, \phi) := 
(\Omega, \ell_{\cM}(g)^{-1} \circ \phi)$ for all 
$\phi \in \Phi$ and $g \in G$. Naturally $\Omega$ is called invariant 
if $\ell_{\cM}(g) \circ \Omega = \Omega$ for all $g \in G$. If $\bA$ commutes 
with $\ell_{\cM}$, $\ell_{\cM}(g) \circ \bA = \bA \circ \ell_{\cM}(g)$ 
for all $g \in G$, 
one expects that the generalized eigenspaces can be decomposed into 
components irreducible with respect to $\ell_{\cM}$. Under mild extra 
conditions this is indeed the case. The nuclear spectral theorem 
(Gel'fand-Maurin theorem, \cite{gelfand,maurin}) guarantees the existence 
of direct integral decompositions of the form
\be 
L^2(\cM) = \int_{{\rm Spec}(\bA) \times \widehat{G}}\! d\mu(\lb,\sigma)\, 
\cE_{\lb,\sigma}(\bA)\,.
\label{nuclear1}
\end{equation}
Here $\widehat{G}$ is the dual of $G$ and $\mu(\lb,\sigma)$ 
is a measure on ${\rm Spec}(\bA) \times \widehat{G}$ defining the decomposition. 
To simplify the notation we identified elements $\pi_{\sigma}$ of $\widehat{G}$ 
with a set of parameters $\sigma$ uniquely specifying an equivalence 
class of unitary irreducible representations. The precise version of the 
nuclear spectral theorem can be found in \cite{gelfand,maurin,bohm}. 
The fiber spaces $\cE_{\lb,\sigma}(\bA)$ 
contain the generalized eigenfunctions of $\bA$ in $\Phi'$ transforming 
irreducibly under $G$. The nuclear spaces $\Phi$ are much smaller 
than $L^1$, the dual spaces $\Phi'$ therefore much larger than 
$L^{\infty}$, and the generalized eigenfunctions supplied by the 
Gel'fand type constructions may be genuine distributions.  
The fact that $\bA$ commutes with the elliptic Nelson operator 
of $G$ (built from the Casimirs of $G$ and $K$) entails 
\cite{maurin} that the (averaged) eigendistributions $\ell_{\cM}(g)
\circ \Omega$ are smooth functions in $g$, but little can be said
about their distributional type. A result by Berezanskii (described and proven
in \cite{maurin}) specifies sufficient conditions under which the dual space 
$\Phi'$ of a triple $\Phi \subset L^2 \subset \Phi'$ 
consists of almost everywhere defined functions (not 
distributions). We shall make use of this result for the 
`relative motion' alluded to in the introduction. Irrespective 
of the distributional type of the generalized eigenfunctions
the decomposition (\ref{nuclear1}) 
has however two important drawbacks:
\begin{itemize}
\itemindent -3mm
\item[{--}] generalized eigenfunctions are assured to exist only for 
$\mu$-almost all spectral values $\lb \in {\rm Spec}(\bA)$,  and the 
measure $\mu$ is usually not known explicitly. 
\item[{--}] whenever generalized eigenfunctions exist for a given 
$\lb \in {\rm Spec}(\bA)$, let $\widehat{G}_{\lb} \subset \widehat{G}$ 
denote a set that carries the restricted measure in (\ref{nuclear1}). 
Then the spaces $\cE_{\lb,\sigma}(\bA)$ are assured to be irreducible only 
for $\mu$-almost all $\sigma \in \widehat{G}_{\lb}$. 
\end{itemize}
The first of these `almost all' caveats presents a major obstruction 
if one wants to apply the general framework to a specific spectral value, 
like $\Vert \bt \Vert$, the ground state value, which is our main concern 
here.

The following example illustrates the problem. Let $\bt$ be the 
integral operator on $L^2(\R_+)$ defined by the kernel $\cT(x,y) = 
e^{-|x-y|}$. It can be seen to be a transfer operator in the above sense
with spectrum ${\rm Spec}(\bt) = [0,2]$. For all spectral values 
different from $\lb =2$ there exist generalized $L^{\infty}$ eigenstates, 
yet the operator does {\it not} have a generalized ground state. The 
point to observe is that for all $\psi \in L^2(\R_+)$ the image function 
$(\bt\psi)(x)$ is twice differentiable with  
\be 
(\bt \psi)^{''} = (\bt \psi) - 2 \psi\,.
\label{example1}
\end{equation}
All solutions of $\bt \psi = \lb \psi$ therefore must be 
linear combinations of $e^{\pm i \omega x}$ with $\om= 
\sqrt{ 2/\lb(\om) -1}$, i.e. $\lb(\om)=2/(1+\om^2)$. One finds 
\be 
\psi_{\om}(x) = \frac{1}{\sqrt{1 + \om^2}}[ \sin \om x 
+ \om \cos \om x ] = \cos(\om x-b) \,, \quad \om \geq 0\,, \quad 
b = {\rm arccot}\ \om\,,
\label{example2}
\end{equation} 
where the normalization has been chosen such that $\Vert \psi_{\om}
\Vert_{\infty} =1$. The explicit construction shows that $\psi_{\om} 
\in L^{\infty}$, although the space of test functions is slightly smaller 
than $L^1$ in that twice differentiable $L^1$ functions $\phi$ have to 
satisfy $\phi(0) = \dd_x \phi(0)$ (which can be seen by averaging 
Eq.~(\ref{example1}) with a test function). The fact that the 
generalized eigenfunctions (\ref{example2}) also satisfy 
$\psi_{\om}(0) = \dd_x \psi_{\om}(0)$ ensures their completeness; it is 
easy to verify that
\ba 
&& \int_0^{\infty} \!d\om \,\psi_{\om}(x) \psi_{\om}(y) = 
\frac{\pi}{2}\ \delta(x-y)\,,\quad x,y\ge 0\,,
\nonum
&& \int_0^{\infty} \!dx \,\psi_{\om_1}(x) \psi_{\om_2}(x) = 0\,, \quad 
\om_1\neq \om_2\,. 
\label{example3}
\end{eqnarray}
Using Eq. (\ref{example3}) we find the following spectral resolution of 
the integral kernel $\cT$:
\be
\cT(x,y;n)=\frac{2}{\pi}\int_0^\infty d\om\ 
\left(\frac{2}{1+\om^2}\right)^n 
\psi_\om(x)  \psi_\om(y)\ .
\end{equation}
%this implies
%\be
%\sup_x \cT(x,x;n) = \frac{2}{\pi}\int_0^\infty 
%d\om\ \left(\frac{2}{1+\om^2}\right)^n\ . 
%\end{equation}
For example 
\ba
\cT(x,y;2) \is e^{-|x-y|}(1+ |x-y|) - \frac{1}{2}e^{-(x+y)}\,.
%\nonum
%\cT(x,y;4) \is xxxx \,.
\label{example4}
\end{eqnarray}

Despite these nice properties no generalized ground state exists.
This is because candidates for it must be contained in the set 
(\ref{example2}); however the relevant limit ($\lb \ra 2$ i.e.~$\om \ra 0$)
vanishes pointwise, while the closest maximum of $\psi_{\om}$, lying at 
$b(\om)/\om\sim\pi/(2\om)$, moves out to $\infty$. 

The upshot is that the `almost all' caveat in the general 
theorems is crucial for their validity and renders them at the same time
useless for the construction e.g.~of the ground state sector. 
Even transfer operators with a complete system of regular 
(here: $L^{\infty}$) generalized eigenfunctions may fail to 
have a ground state. This explains why in the constructive 
Theorem 5.1 certain subsidiary conditions must be present; we do 
expect however that the ones given can still be weakened.

In Section 3 we will analyze the spectral problem for invariant 
selfadjoint operators $\bA$ as defined in Definition 3.1.
Under mild subsidiary conditions ((C) in Section 3.3
and (C1), (C2) in Section 3.4) a complete set of eigenfunctions 
$\Omega_{\lb \sigma}$ in $\cE_{\lb \sigma}(\bA)$ can be found. 
Some of the properties of the $\Omega_{\lb \sigma}$ have been 
anticipated in the introduction. In this context it is 
worth emphazising two points. First, in contrast to the familar situation with 
normalizable eigenfunctions  the existence of a $\Omega \in \Phi'$ such 
that $\bA \Omega = \lb \Omega$  does in itself {\it not} imply 
$\lb \in {\rm Spec}(\bA)$. A simple counterexample is the    
hyperbolic spin chain discussed in detail in \cite{hchain}: there 
the constant functions are eigenfunctions 
of the transfer operator $\bt$, but the corresponding eigenvalue lies 
above the spectrum of $\bt$. When solving the spectral problem 
$\bA \Omega = \lb \Omega$ with a nonnormalizable $\Omega \in \Phi'$,
the information that $\lb$ is a point in the spectrum therefore
has to be supplied independently. A second point worth repeating is 
that all known construction principles for generalized eigenfunctions
(including the one presented in Section 3) are guaranteed to 
work only for {\it almost all} points in the spectrum. 
For a prescribed $\lb \in {\rm Spec}(\bA)$ additional 
considerations are necessary to show that sufficiently 
many eigenfunctions exist. This applies in particular to the 
$\sigma$-equivariant eigenspaces $\cE_{\lb \sigma}(\bA)$ of
an invariant selfadjoint operator $\bA$ and to the ground state fibers 
$\cG_{\sigma}(\bt):= \cE_{\Vert \bt \Vert, \sigma}(\bt)$ of a transfer 
operator.  

Whenever the fiber spaces $\cE_{\lb,\sigma}(\bA)$ in (\ref{nuclear1}) 
are nonempty for a {\it fixed} $\lb \in {\rm Spec}(\bA)$ one can match the 
decomposition in (\ref{nuclear1}) with the purely group theoretical 
one. Since $\ell_\cM$ is a unitary representation on general grounds it 
can be decomposed into irreducible components \cite{Dixmier1}.   
That is, there exists a measure $\mu_{\ell_{\cM}}$ on $\widehat{G}$ 
such that \be 
L^2(\cM) = \int_{\widehat{G}}^\oplus \!d\mu_{\ell_\cM}(\sigma) 
\, \cL^2_{\sigma}(\cM)\,,
\label{nuclear2}
\end{equation}
where the fibers $\cL^2_{\sigma}(\cM)$ are irreducible 
for $\mu_{\ell_{\cM}}$-almost all $\sigma \in \widehat{G}$ in the support 
of the measure. On the other hand from (\ref{nuclear1}) one can define 
generalized eigenspaces $\cE_{\lb}(\bA)$ by 
\be 
\cE_{\lb}(\bA) = \int_{\widehat{G}_{\lb}} \!d\mu(\lb,\sigma) \,
\cE_{\lb,\sigma}(\bA)\,,
\label{nuclear3}
\end{equation}     
where $\widehat{G}_{\lb}$ denotes the part of $\widehat G$ for which there 
is a non-empty eigenspace $\cE_{\lb,\sigma}(\bA)$.

Thus $\int_{{\rm Spec}(\bA)} d\mu(\lb,\sigma) = d\mu_{\ell_{\cM}}(\sigma)$.
There are no obvious strategies to determine the representation content 
$\widehat{G}_{\lb} \subset \widehat{G}$ 
of a given spectral value $\lb$ (for which generalized eigenfunctions exist). 
For a transfer operator $\bt$ we identify its ground state 
sector $\cG(\bt)$ with $\cE_{\Vert \bt \Vert}(\bt)$, and similarly 
for the equivariant fibers $\cG_{\sigma}(\bt) = \cE_{\Vert T \Vert, 
\sigma}(\bt)$. We shall thus apply the decomposition (\ref{nuclear3}) 
also for the ground state sector and write $\widehat{G}_{\Vert T\Vert}$ for 
its representation content. One of the main goals later on will 
be to show that under moderate extra assumptions $\widehat{G}_{\Vert T\Vert}$
consists of a {\it single} point (a single representation) only, which is
always the {\it same} for all transfer operators considered.

%%%%%%%%%%%%%%%%%%%%%%%%%%%%%%%%%%%%%%%%%%%%%%%%%%%%%%%%%%%%%%%%%%%%%%%%%%%
\newsubsection{Orbit decomposition}

A simple but crucial fact about the configuration manifolds $\cM = Q \times
\ldots \times Q$ is that they have a well defined orbit decomposition 
which eventually carries over to the states and the operators acting on them. 
The idea of the decomposition is to single out one of the variables 
in $m = (q_1,\ldots, q_{\nu})$, say $q_1$, to parameterize the location on 
the orbits and to define coordinate functions $n_i$ transversal to it 
to describe the relative location of the points such that they change 
only by elements of $K$ as one moves along an orbit. To this end
we fix some $q^\up \in Q=eK$ with isotropy group $K$, i.e.~$k q^\up = 
q^\up$ for all $k \in K$. Based on it we wish to define a section $g_s: Q 
\ra G$ such that $q = g_s(q) q^{\uparrow}$ for all $q \in Q$. Clearly 
this condition defines $g_s$ only up to right multiplication by some 
$k_s= k_s(g,q) \in K$, 
\be 
g_s(g q) = g g_s(q) k_s(g,q)\,.
\label{section}
\end{equation} 
For $g_s$ to be well-defined the element $k$ has to be uniquely 
determined for given $g$ and $q$. It is easy to see that this 
is the case whenever $G$ admits an Iwasawa decomposition, which is the 
case for all connected simple noncompact Lie groups, in particular 
the ones considered. Consistency requires the 
cocycle condition $k_s(g_1 g_2, q) = k_s(g_2, q) k_s(g_1, g_2 q)$, 
in particular  $k_s(e,q)=e$ for all $q \in Q$ and $e \in G$ the identity. 
If we normalize $g_s$ such that $g_s(q^\up) =e$ it follows that 
$k_s(g_s(q),q^\up) =e = k_s(g_s(q)^{-1},q)$ and $k_s(g,q) = k_s(g g_s(q), 
q^\up)$. 
For elements $k \in K$ of the subgroup one has $k_s(k,q^\up) = k^{-1}$.  
Generally the Iwasawa decomposition entails that the cocycle 
$k_s(\,\cdot\,,q): G \ra K$ is surjective for all $q \in Q$.

Our main example for the symmetric space $Q$ will be 
$\H_N = {\rm SO}_0(1,N)/{\rm SO}(N)$, the $N$-dimensional 
hyperboloid with the Riemannian metric $\gamma_{\H_N}$ induced by the 
indefinite metric $q\cdot q = (q^0)^2 - (q^1)^2 - \ldots -(q^N)^2 $
in the imbedding linear space. Explicitly 
$\H_N = \{ q \in \R^{1,N} \,|\,q\cdot q = (q^0)^2 - (q^1)^2
- \ldots - (q^N)^2 =1,\, q^0 > 0\}$. The invariant measure on $\H_N$ is 
$d\gamma_{\H_N}(q) = d^{N+1} q \delta(q^2 -1) \theta(q^0)$ and will be 
denoted by $dq$ for short. The $\nu$-fold product $\cM = \H_N^{\nu}$ 
equipped with the product metric then is a simply connected  
Riemannian manifold (and, in fact, a reducible symmetric space). 
We denote the product measure $\prod_{i=1}^\nu dq_i$ on $\cM$
by $d\gamma_\cM$.
For $G = {\rm SO}_0(1,N)$ and $K = {\rm SO}(N)$ the section 
$g_s(q)$ is just the familar expression for the pure boost mapping 
$q^{\uparrow} =(1,0,\ldots,0)$ to $q=(q_0,q_1,\ldots, q_{N-1})
=: (q_0, \vec{q})$. Explicitly 
\ba 
&& g_s(q) = \left( 
\begin{array}{c|c} q_0 & \vec{q}^{\;T} \\[1mm]
\hline
               \mbox{} & \mbox{ }\\[-1mm]
               \vec{q} & \1 + \frac{1}{q_0 +1} \;\vec{q}\; \vec{q}^{\;T}
\end{array} \right)\,.
\label{sectionHN}
\end{eqnarray}
It satisfies $g_s(q) =e$ iff $q = q^{\uparrow}$. The cocycle 
$k_s(g,q) = g_s(q)^{-1} g^{-1} g_s(g q)$ based on this section 
is known as `Wigner rotation' \cite{weinberg} and satisfies 
\be 
k_s(k,q) = k^{-1} \quad \mbox{for all}\;\; q \in Q,\; k \in K\,,
\label{Wigner}
\end{equation} 
that is, not only for $q = q^{\uparrow}$. 

Using the section $g_s$ we now define the following diffeomorphism 
on $\cM = Q^{\nu}$:  
\ba
&& \vartheta(q_1,\ldots, q_{\nu}) = 
(q_1, g_s(q_1)^{-1} q_2, \ldots, g_s(q_1)^{-1} q_{\nu}) =: 
(q_1, n_2, \ldots , n_{\nu})\,,  
\nonum
&& \vartheta^{-1}(q_1, n_2, \ldots, n_{\nu})
= (q_1, g_s(q_1) n_2, \ldots, g_s(q_1) n_{\nu})\,.
\label{iso1Ms}
\end{eqnarray}
This diffeomorhism is measure preserving due to the invariance of the 
measures $dq_i$: let $f\in L^1(d\gamma_\cM)$ and $dn:= d\gamma_{\cN}(n) := 
\prod_{i \neq 1} d\gamma_Q(n_i)$. Then
\ba  
&& \int (f\circ\vartheta)(q_1,n) d\gamma_\cM(m) = \int dq_1\int\prod_{i\neq 1} 
dq_i\ f(q_1,g_s(q_1)^{-1} q_2, \ldots, g_s(q_1)^{-1} q_\nu)\cr 
\nonum 
&& \quad = \int\! dq_1\int\! \prod_{i\neq 1} dq_i \, f(q_1,\ldots,q_{\nu})= 
\int \!d\gamma_\cM(m)\, f(m)\, .     
\label{invar1}
\end{eqnarray}
So the measure $d\gamma_\cM$ can also be factorized as
\be 
d\gamma_\cM(m) = dq_1\, dn\,,\quad 
\label{mufac}
\end{equation}
The product $Q \times \cN$ equipped with the product metric $\gamma_Q \times
\gamma_{\cN}$ and the measure $dq\ dn$ is a Riemannian manifold $\cM_s$ 
which by construction is isometric to $\cM$, and with the 
isometry given by the above $\vartheta$:  
\be 
\vartheta: (\cM, \gamma) \rra (\cM_s, \gamma_s) := (Q \times \cN, 
\gamma_Q \times \gamma_{\cN})\,.
\label{iso2Ms}
\end{equation}
The manifold $\cM_s$ also has the structure of a $G$ space which it 
inherits from $\cM$. Recall that the transversal coordinate functions 
$n_i = n_i(q_1,q_i) \in Q$ are defined by $n_i := g_s(q_1)^{-1} q_i$, 
for $i=2, \ldots, \nu$. On them $\ell_{\cM}$ acts 
by $\ell_{\cM}(g)(n_i) = g_s(g\inv q_1)^{-1} g\inv q_i = k_s(g\inv, q_1)
\inv n_i$. 
The original diagonal action $d(g)m = (gq_1, \ldots, gq_{\nu})$ 
becomes a twisted action $d_s$ (depending on a choice of section) 
in the coordinates $(q,n)$, i.e.~$\vartheta \circ d = d_s \circ 
\vartheta$, with
\be 
d_s(g\inv)(q,n) = (g\inv q, k_s(g\inv,q)\inv n)\,.
\label{dtwist}
\end{equation}
As $d_s(g_s(q))(q_0,n) = (q,n)$ and $d_s(g_s(q)\inv)(q,n) = (q_0,n)$ 
it acts transitively on the first variable. However as one moves along 
$Q$ the action $d_s$ co-rotates the $n$ variables in a $q$-dependent way.
For a generic symmetric space this happens even on the subgroup $K$;
for $\H_N$ and the Wigner rotation one has $d_s(k\inv)(q,n) = (k^{-1}q, 
k^{-1} n)$, though. In the terminology of \cite{Zimmer}, Section 4, 
$\cM_s := Q\times_{k_s} \cN$ 
is the skew product $G$-space induced from the $K$-space $\cN$. 
Indeed, $\cN$ equipped with the diagonal action of $K$: 
$d_{\cN}(k)n = (k^{-1}n_2,\ldots, k^{-1} n_{\nu})$ is a $K$-space,
and by construction $(\cM, d)$ and $(\cM_s, d_s)$ are isometric 
as $G$ spaces. Since the cocycle is surjective for fixed $q$ 
we gained a less redundant description of the space of orbits: 
\be 
\mbox{space of orbits:}\quad \cM/d(G) = \cM_s/d_s(G) = \cN/d_{\cN}(K)\,,
\label{dorbits}
\end{equation}
where, importantly, $K$ is compact. On the other hand, the twisted action 
(\ref{dtwist}) is cumbersome when one tries to decompose the unitary 
representation based on $d_s$ into irreducible components. 
But the left twisted action $d_s$ can be traded for an untwisted 
right action $r$ by the following construction (based on a remark in 
\cite{Zimmer}, p.75). Consider 
\be 
\cM_r := (G \times \cN)/d(K),
\label{Mprime}
\end{equation}
that is, the space of equivalence classes $(g,n) \sim (k^{-1} g, k^{-1}n)$,
$k \in K$ in $G \times \cN$. In order not to clutter the notation we 
also write $(g,n)$ for the equivalence class generated by a point in 
$G \times \cN$. Maps and functions on $G \times \cN$ that are constant on 
$d(K)$ orbits then lift unambiguously to maps and functions on $\cM_r$. On 
$G \times \cN$ and $\cM_r$ we define a right $G$-action $r$ by 
\be 
r(g') (g,n) = (g g', n)\,,
\label{rho1def}
\end{equation}
which is just the standard right action of $G$ on itself leaving 
the $n$ variables untouched. The action (\ref{rho1def}) is constant
on the equivalence classes because the right $r(G)$ action 
and the left $d(K)$ action on $G \times \cN$ commute, $r(g) d(k) = d(k) 
r(g)$. In fact $\cM_r$ equipped with the right $G$ action is isomorphic 
to the original manifold $\cM=Q^\nu$ with the diagonal left action 
$\ell_{\cM}(G)$ (and thereby also to the skew product $G$ space $\cM_s = Q 
\times_{k_s} \cN$). The isomorphism is given by first considering the 
following map $\tilde \phi: \cM\to G\times \cN$:
\ba   
&& \tilde \phi \!: \,\cM \rra G \times \cN \,,
\nonum
&& \tilde\phi(q_1, \ldots , q_{\nu}) =
(g_s(q_1)^{-1},g_s(q_1)^{-1}q_2\ldots, g_s(q_1)^{-1}q_{\nu}) \,.
\label{chitheta}
\end{eqnarray}
This map is injective, but not surjective.  Because $g_s$ is a 
global section of $G/K$ its range intersects each $d(K)$ orbit exactly 
once, so that it determines uniquely a diffeomorphism 
\be
\phi: \cM\!\rra\!\cM_r\,.
\end{equation}
We define an inverse of $\tilde\phi^{-1}:G\times\cN\rra\cM$ (initially only 
defined on the range of $\tilde\phi$) by
\be
\tilde\phi^{-1}(g,n) = d(g\inv)(q^\up, n)\,.
\end{equation}
We can immediately interpret this as a map from all of $G\times\cN$ to 
$\cM$ which is constant on the equivalence classes under $d(K)$
\be
\tilde\phi^{-1}(kg,kn)=d(kg)(q^\up,kn)=(g^{-1}q^\up,g^{-1}n)=
\tilde\phi^{-1}(g,n)\,, 
\end{equation}
and hence lifts to a map $\phi^{-1}: \cM_r \rra \cM$. By direct computation 
one verifies 
\be
\tilde\phi\circ\tilde\phi^{-1}={\rm id}\ , 
\end{equation}
whereas $\tilde\phi^{-1}\circ \tilde\phi$ only maps orbits of $d(K)$ into 
themselves. But this is enough to see that $\phi$ and $\phi^{-1}$ are 
really inverse to each other.

According to (\ref{rho1def}) the map $\phi$ intertwines the left action 
$d(G)$ with the right action $r(G)$:
\be
\phi\circ d = r\circ \phi\ .
\label{intertw}
\end{equation}
In addition the map $\phi$ is measure preserving; this can be seen similarly as
the measure preserving property of $\vartheta$:
consider a $L^1$ function $f$ on $\cM$ and let $f_r=f\circ \phi^{-1}$. 
Then, using the $G$-invariance of the measure $dn$ and the fact that 
the invariant measure $dq$ is the push-forward of Haar measure $dg$ 
under the canonical projection $G\!\rra\!G/K$, one sees that 
\be
\int\!\prod_{i=1}^\nu d\gamma_Q(q_i) f(q_1,\ldots,q_{\nu}) 
=\int\! dg\ dn\,f_r(g,n)\,. 
\label{meas}
\end{equation}

For completeness we also note explicitly the isometry 
$\chi=\phi\circ \vartheta^{-1}$ between 
the skew product $Q$ space $\cM_s=Q\times_{k_s}\cN$ and the 
$G$-manifold $\cM_r$ with the diagonal right action:
\ba 
\chi \!&:&\! Q \times_{k_s} \cN \ra \cM_r\;,\quad \quad 
\chi(q,n) = (g_s(q)^{-1},n) \,,
\nonum
\chi^{-1}\!&:&\! \cM_r \ra  Q \times_{k_s} \cN\,,\quad\, 
\chi^{-1}(g,n) = (g^{-1} q^\up, k_s(g^{-1}, q^\up)^{-1} n)\,.
\label{MMprime}
\end{eqnarray}

In summary we have two equivalent descriptions of the 
original $G$-manifold $(\cM, d)$, namely $(\cM_s, d_s)$ and 
$(\cM_r, r)$. The second one is more convenient for 
the reduction problem because the space of orbits $\cM/d(G) = \cM_r/r(G)$ 
is now described by equivalence classes with respect to the usual right
action of $G$ on itself. 

This structure of course carries over to the function spaces and 
the unitary representations on them induced by the G-actions.    
We have
\begin{subeqnarray}
L^2(\cM) \ni \psi &\;\mapsto\;&
(\ell_\cM(g) \psi)(m) =\psi(d(g\inv)(m)) \,,
\\
L^2(\cM_s) \ni \psi_s &\;\mapsto\;& 
(\ell_s(g) \psi_s)(q,n) =\psi_s(d_s(g\inv)(q,n) \,,
\\
L^2(\cM_r) \ni \psi_r &\;\mapsto\;& 
(\rho(g_0) \psi_r)(g,n) =\psi_r(r(g_0)(g,n)),\quad \ell_r(K)\psi_r=\psi_r\,, 
\label{Lreps}
\end{subeqnarray}
where $\ell_r(k) \psi_r(g,n) = \psi_r(k^{-1} g, k^{-1} n)$. 
As Hilbert spaces of course all three $L^2$ spaces are isometric 
to $L^2(Q \times \cN)$ and the three representations
$\ell_\cM, \ell_s$, and $\rho$ are likewise unitarily equivalent.
Explicitly
\ba 
&& \Phi : L^2(\cM_r) \rra L^2(\cM)\,,
\quad (\Phi \psi_r)(m) := \psi_r(\phi(m))\,, 
\nonum
&& \rho = \Phi^{-1} \circ \ell_{\cM} \circ \Phi\, 
\label{Phidef}
\end{eqnarray}
and similarly for $L^2(\cM_s)$. By (\ref{Lreps}c) $L^2(\cM_r)$ can also 
be identified with the subspace invariant under $\ell_r(K)$ of 
$L^2(G\times\cN)$. 
By (\ref{meas}) the map $\Phi$ is indeed an isometry. 
As $(\Phi^{-1} \circ\ell_{\cM}(g_0)\circ \Phi \psi_r)(g,n) = \psi_r(\phi  
\circ d(g_0\inv)\circ \phi^{-1}(g,n))$ the unitary equivalence of the 
representations follows from (\ref{intertw}).  

We summarize our results in 
\begin{proposition}\label{orbit}
There is a diffeomorphism $\phi$ from the configuration manifold 
$\cM=Q^\nu$ to $\cM_r = (G\times Q^{\nu-1})/d(K)$ such that the diagonal 
left action $d(G)$ on $\cM$ gets mapped into the right action 
$r(G)$ on the first factor of $\cM_r$.
$\phi$ is measure preserving for the natural measures on $\cM$ and 
$\cM_r$ and therefore induces a natural isomorphism of the spaces
$L^2(\cM)$ and $L^2(\cM_r)$.
\end{proposition}
\addtocounter{proposition}{1}

%%%%%%%%%%%%%%%%%%%%%%%%%%%%%%%%%%%%%%%%%%%%%%%%%%%%%%%%%%%%%%%%%%%%%%%
\newsubsection{The reduction of $\rho(G)$ on $L^2(\cM_r)$}

With these preparations at hand we can now address the 
reduction problem of $\ell_{\cM}(G)$ in the variant where it 
acts as $\rho(G)$ on $L^2(\cM_r)$. As noted above, the latter is the 
$\ell_r(K)$ invariant subspace of $L^2(G \times \cN)$, which carries the 
{\it commuting} unitary representations $\rho(G)$ and 
$\ell_r(K)$. Moreover $\rho(G)$ for fixed $n \in \cN$ is just the 
right regular representation of $G$ mapping $\psi_r(g,n)$ into 
$\rho(g_0)\psi_r(g,n)= \psi_r(gg_0,n)$. Its decomposition into 
unitary irreducible representations $\pi_{\sigma}$, $\sigma \in 
\widehat{G}_r$, is thus 
given by the Plancherel decomposition (\ref{Planchrho}). The precise 
form used and the notations are summarized in appendix A. 
In particular $d\nu$ is the Plancherel measure on $\widehat{G}_r$, the 
restricted dual of $G$, and $g \mapsto \pi_{\sigma}(g)$ denotes the 
irreducible representation associated with some $\sigma \in 
\widehat{G}_r$. It acts on a separable Hilbert space $\cL_{\sigma}$ 
with inner product $(\cdot,\cdot)_{\sigma}$.

Applying the expansion (\ref{Planch1}) to the $G$-part 
of a function in $\psi: G \times \cN \ra \C$ gives 
\ba 
\psi(g,n) \is \int_{\widehat{G}_r}\!d\nu(\sigma) \, 
{\rm Tr}[\pi_{\sigma}(g)\, \widehat{\psi}(\sigma,n)] \,,
\nonum
\widehat{\psi}(\sigma,n) \is \int_G \! dg \, \pi_{\sigma}(g^{-1})
\psi(g,n) \,.
\label{psi1}
\end{eqnarray}
For functions $\psi$ that are $\ell_r(K)$ invariant this will lead to 
the desired decomposition of $L^2(\cM_r)$ into irreducible 
components. We consider here first the decomposition of the larger space
$L^2(G \times \cN)$. Provided suitable conditions are imposed on the 
function $\psi$ (which we describe shortly) the transforms 
$\widehat{\psi}(\sigma,n)$ for fixed $n$ are trace class or 
Hilbert-Schmidt operators on $\cL_\sigma$. Further they satisfy
\be
\label{psi2}
[\rho(g_0)\ell(g_1) \psi]^{\widehat{\phantom{-}}}(\sigma,n) 
= \pi_{\sigma}(g_0) \widehat{\psi}(\sigma,n) \pi_{\sigma}(g_1^{-1})
= (\pi_{\sigma} \times \pi_{\check{\sigma}})(g_0,g_1) 
\widehat{\psi}(\sigma,n)\,,
\end{equation} 
using (\ref{piHS}) and the notation $\check{\pi}_{\sigma} = 
\pi_{\check{\sigma}}$ in the last equation.  
This states that the map $\psi \ra \widehat{\psi}$ 
intertwines the outer tensor product $\rho \times \ell$ of the left 
and the right regular representation of $G$ with $\pi_{\sigma} \times 
\pi_{\check{\sigma}}$.

Throughout we shall adopt the following conventions for compact  
operators $A,B$ on some separable Hilbert space $\cH$ with orthonormal basis 
$e_i,\,i \in \N$, and its dual space $\check\cH$ with dual basis 
$\check{e}_i,\,i\in \N$:
\be
A = \sum_{ij} e_i A_{ij} \check{e}_j\,,\sspace 
[A^{\dagger}]_{ij} = A^*_{ji}\,.
\label{conventions}
\end{equation}
Compact operators that are even trace class arise for example as 
Fourier transforms of functions $\psi \in \cD$, where $\cD$ is the space 
of functions $\psi(g,n)$ that are smooth with compact support in $g \in 
G$ and square integrable in $n\in \cN$: for such $\psi$ the 
Fourier transform $\widehat{\psi}(\sigma,n)$ is a trace class operator on 
a separable Hilbert space $\cL_{\sigma}$, for all $\psi \in \cD$, 
and almost all $\sigma \in \widehat{G}_r$, $n \in \cN$. Moreover the 
Fourier expansion (\ref{psi1}) then is valid pointwise in $g$. 
If $\psi$ is in $L^2(G) \cap L^1(G)$ as a function of $g$ and square 
integrable in $n$ the Fourier coefficients $\widehat{\psi}(\sigma)$ are 
still Hilbert-Schmidt operators for almost all $\sigma \in 
\widehat{G}_r$, $n \in \cN$. We identify the trace class operators with a 
subspace of $\cL_{\sigma} \otimes \check\cL_{\check{\sigma}}$, 
which in turn can be identified with the space of Hilbert-Schmidt 
operators on $\cL_{\sigma}$; see Appendix A3. This 
means the coefficients are functions $\widehat{\psi}(\sigma, 
\, \cdot \,): \cN \ra \cL_{\sigma} \otimes  
\check\cL_{\check{\sigma}}$. The Parseval identity 
\be 
\int_G \!dg \, \phi(g,n)^* \psi(g,n') = 
\int_{\widehat{G}_r}\! d\nu(\sigma) \, 
{\rm Tr}[ \widehat{\phi}(\sigma,n)^{\dagger} \psi(\sigma,n')]\,,
\label{Planchpsi}
\end{equation}
is valid for all functions in $L^2(\cM_r)$ which for fixed $n$ lie in 
$L^1(G) \cap L^2(G)$. It implies that the trace ${\rm Tr}
[\widehat{\psi}(\sigma, n)^{\dagger} \widehat{\psi}(\sigma,n)]$
is integrable with respect to $d\nu(\sigma)dn$; hence it is an 
integrable function on $\cN$ for almost all $\sigma\in \widehat G_r$. 

This suggests to equip the fibers at fixed $\sigma \in \widehat{G}_r$ 
with the structure of a Hilbert space which we denote by 
$L^2_{\sigma\check\sigma}(\cN)$. We shall also need various 
subspaces of $L^2_{\sigma\check\sigma}(\cN)$ and for convenient 
reference we collect them in the following definition.

\begin{definition} The Hilbert spaces 
\begin{subeqnarray}
\mbox{}\nspace L^2_{\sigma \check{\sigma}}(\cN) &:=& 
\Big\{ F: \cN \ra \cL_{\sigma} 
\otimes \check\cL_{\check{\sigma}}\;\Big|  
\;(F,F)_{\sigma \check{\sigma}} < \infty \Big\}\cong 
\cL_{\sigma}\otimes \check\cL_{\check{\sigma}}\otimes L^2(\cN)\,,
\\ \mbox{}\nspace L^2_{\sigma\check\sigma}(\cN)_0 &:=& \Big\{ F: \cN \ra \cL_{\sigma}
\otimes \check\cL_{\check{\sigma}}\;\Big|\;(F,F)_{\sigma \check{\sigma}} 
< \infty \,, F(k^{-1}n)\pi_\sigma(k)^\dagger= F(n)\Big\}\,,
\label{LNdef1}
\end{subeqnarray}
where $(F_1, F_2)_{\sigma \check{\sigma}} := 
\int \! dn \,{\rm Tr}[F_1^{\dagger}(n) F_2(n)]$ are called the 
fiber spaces of $L^2(G \times \cN)$ and $L^2(\cM_r)$, respectively. 
Let $\widehat{K}$ be the unitary dual of $K$, $\kappa \in \widehat{K}$, 
and $V_{\check \kappa} \subset \cL_{\check\sigma}$ the subspaces
in Eq.~(\ref{LMred3}) below. Then
\begin{subeqnarray}
L^2_{\sigma \check{\kappa}}(\cN) & := & \{ f: \cN \ra \cL_{\sigma}
\otimes V_{\check{\kappa}}\;| \;(f,f)_{\sigma \check{\kappa}} < \infty \}
\cong\cL_\sigma\otimes V_{\check\kappa}\otimes L^2(\cN)\,,
\\
L^2_{\sigma \check{\kappa}}(\cN)_0 &:=& \{ f: \cN \ra \cL_{\sigma} 
\otimes V_{\check{\kappa}}\;| \; f(kn) = f(n) r_\kappa(k)^\dagger\,,\;
(f,f)_{\sigma \check{\kappa}} < \infty \}\,,
\label{LMred5}
\end{subeqnarray}
where $(f_1, f_2)_{\sigma \check{\kappa}} :=
\int \! dn {\rm Tr}_{V_{\check\kappa}}[f_1(n)^{\dagger} f_2(n)]$ 
are called the $\kappa$-channels of $L^2_{\sigma\check\sigma}(\cN)$
and $L^2_{\sigma\check\sigma}(\cN)_0$, respectively.  
The adjoints of the singlet channels $\kappa =0$ lead to spaces 
\begin{subeqnarray}
L^2_{\sigma}(\cN)& :=& \Big\{ f: \cN \ra \cL_{\sigma}\;\Big| \; 
\int \! dn (f(n),f(n))_{\sigma} < \infty \Big\}
\cong \cL_{\sigma}\otimes L^2(\cN)\,,
\\
L^2_{\sigma}(\cN)_0 & :=& \{ f \in L^2_{\sigma}(\cN)\;| f(kn) = 
\pi_{\sigma}(k) f(n)\}\,. 
\label{LNdef}
\end{subeqnarray}
\end{definition}

With the definition (\ref{LNdef1}a) the Fourier transformation 
(\ref{psi1}) becomes an isometry 
\ba 
\cD \ni \psi &\longmapsto&  \widehat{\psi} \in \int^{\oplus} 
\!d\nu(\sigma)\, 
L^2_{\sigma \check{\sigma}}(\cN)\,,
\nonum 
\int \! dg dn \, \psi(g,n)^* \psi(g,n)\nspace\; &=& \nspace\; 
\int_{\widehat{G}_r} \!d\nu(\sigma) \, 
(\widehat{\psi}, \widehat{\psi})_{\sigma \check{\sigma}}\,,
\label{Planchpsi2}
\end{eqnarray}
which extends uniquely to an isometry between Hilbert spaces. 
Since the trace class operators form an ideal 
in the algebra of all bounded linear operators on $\cL_{\sigma}$
for all  $\psi \in \cD$ the trace 
${\rm  Tr}[\pi_{\sigma}(g)\,\widehat{\psi}(\sigma,n)]$ 
is defined pointwise for all $(g,n) \in G \times \cN$ 
and it is a continuous function in $g$. For the same reason 
$[\rho(g_0) \ell(g_1)\psi]^{\widehat{\phantom{-}}}(\sigma,n) =
\pi_{\sigma}(g_0)\widehat{\psi}(\sigma,n) \pi_{\sigma}(g_1)^\dagger$ 
is a trace class operator for all $g_0,g_1$ if $\widehat{\psi}(\sigma,n)$
is. As a consequence $L^2_{\sigma\check{\sigma}}(\cN)$ carries 
a unitary representation $\pi_{\sigma\check{\sigma}}$ of $G \times G$
\ba 
&& \pi_{\sigma\check{\sigma}}(g_0,g_1) F(n) 
:= \pi_{\sigma}(g_0) F(n) \pi_{\sigma}(g_1)^\dagger\,,
\nonum
&& \big(\pi_{\sigma\check{\sigma}}(g_0,g_1) F_1, 
\pi_{\sigma\check{\sigma}}(g_0,g_1) F_2\big)_{\sigma \check{\sigma}} 
= \big(F_1, F_2 \big)_{\sigma\check{\sigma}}\,.
\label{LNdef2} 
\end{eqnarray} 
It coincides with $\pi_{\sigma} \times \pi_{\check{\sigma}}$, the outer tensor
product of the two representations (which is irreducible \cite{folland}, 
Thm 7.20]) whenever both of the factors are.   
The isometry (\ref{Planchpsi2}) therefore also provides the decomposition 
of $\rho \times \ell$, the outer tensor product of the right and the 
left regular representation of $G$ into a direct integral of 
($\nu$-almost everwhere) irreducible representations, 
\be 
\rho \times \ell = \int \! d\nu(\sigma) \, 
\pi_{\sigma} \times \pi_{\check{\sigma}}\,.
\label{Planchrl}
\end{equation}

With these preparations at hand we can turn to the decomposition 
of $L^2(\cM_r)$, which we naturally identified with the subspace 
of $L^2(G \times \cN)$ consisting of left $K$ invariant functions. 
Clearly the left $K$ invariance of the function $\psi$ translates into 
the following condition on the Fourier coefficients
\be 
{[\ell_r(k)\psi]}^{\widehat{\phantom{-}}}(\sigma,n) 
= \widehat{\psi}(\sigma,k^{-1}n)\pi_\sigma(k)^\dagger
\stackrel{\displaystyle{!}}{=}\widehat{\psi}(\sigma, n)\,.
\label{LMred1}
\end{equation}
We also introduce the corresponding $K$-singlet subspace of 
$L^2_{\sigma\check\sigma}(\cN)$ as in (\ref{LNdef1}b). 
Since $\pi_{\sigma}^\dagger$ `acting from the right' is unitarily 
equivalent to $\pi_{\check\sigma}$, in representation theoretical terms 
(\ref{LMred1}) means
\be 
\ell_\cN \times \pi_{\check\sigma}|_K \stackrel{\displaystyle{!}}{=} 
{\rm id}
\label{LMred2}
\end{equation} 
where $\ell_{\cN}(k) F(n):= F(k^{-1} n)$. 
The condition $(!)$ can be understood as the projection onto the subspace 
of left $K$ singlets in a decomposition of $L^2(G) \otimes L^2(\cN)$ 
which we prepare now.

First recall that the restriction of 
$\pi_{\check\sigma}$ to the subgroup $K$ decomposes as follows 
\be 
\pi_{\check\sigma}|_K = \bigoplus_{\widehat{K}_{\check\sigma}} 
m_{\check\kappa} r_{\check\kappa} \,,\sspace 
\cL_{\check\sigma} = 
\bigoplus_{\check\kappa \in \widehat{K}_{\check\sigma}} 
m_{\check\kappa} V_{\check\kappa} \,. 
\label{LMred3}
\end{equation}
Here the subset $\widehat{K}_{\check\sigma} \subset \widehat{K}$ 
for which the irreducible representations $r_{\check\kappa}$ on 
the finite dimensional vector space $V_{\check\kappa}$ occurs 
with nonzero multiplicity $m_{\check\kappa}$ is called the 
$K$ content of $\pi_{\check\sigma}$; see appendix A. 
Often it is convenient to use a basis of $\cL_{\check\sigma}$ obtained by 
concatenation of the bases $e_{\check\kappa s},\,s =0, \ldots, m_{\kappa} 
\dim V_{\check\kappa} -1$, of $m_{\check\kappa} V_{\check\kappa}$. (Here 
of course for fixed $\check\kappa$ the basis vectors $e_{\check\kappa s}, 
\, s= 0, \ldots, \dim 
V_{\check\kappa}\! -\!1$, $e_{\check\kappa s}, \, s=\dim V_{\check\kappa}, 
\ldots, \dim 2 
V_{\check\kappa}\! -\!1$, etc are likewise orthogonal.) We shall call 
\be 
e_{\check\kappa s},\quad s =0, \ldots, m_{\check\kappa} \dim 
V_{\check\kappa} -1, 
\quad \check\kappa \in \widehat{K}_{\check\sigma}\,,
\label{adaptb}
\end{equation}
the $K$-adapted basis of $\cL_{\check\sigma}$. 
For operators $F \in L^2_{\sigma \check\sigma}(\cN)$ the components with 
respect to an orthonormal basis $\{e_i,\,i \in \N\}$, 
and its dual $\{ \check{e}_i = (e_i,\cdot\,)_{\sigma}, \,\,i \in \N\}$ 
are $F_{ij} := \check{e}_i(F e_j)$,
so that $F = \sum_{ij} e_i F_{ij} \check{e}_j$. 
In the $K$-adapted basis these become $F_{\kappa s, \kappa's'} = 
e_{\check{\kappa}s}(F e_{\kappa' s'})$.

In view of (\ref{LMred2}) one has to decompose 
the representation $\rho \times \ell|_K \times \ell_{\cN}$ of 
$G\times K$ in order to decompose $L^2(G \times \cN)\cong L^2(G) 
\otimes L^2(\cN)$. Since the group that is represented is really only 
$G\times K$, the second tensor product should be read as an inner one; 
however for our analysis it is convenient first to regard it 
as an outer one too: combining (\ref{Planchrl}) with 
(\ref{LMred3}) gives for the first factor
\be 
\rho \times \ell|_K = \int^{\oplus} \!d\nu(\sigma) \, 
\bigoplus_{\kappa \in \widehat{K}_{\check\sigma}} \pi_{\sigma} \times 
m_{\check{\kappa}} r_{\check{\kappa}}\,.
\label{LMred4a}
\end{equation}
For the second factor we write 
\be 
\ell_{\cN} = \bigoplus_{\kappa} m_{\cN \!\kappa} \,r_{\kappa} \;,
\quad L^2(\cN) = \bigoplus_{\kappa} m_{\cN\!\kappa} L^2_{\kappa}(\cN)\,,
\label{LMred4b}
\end{equation}
where $L^2_{\kappa}(\cN)$ is the subspace of functions transforming 
irreducibly according to $f(k^{-1}n) = r_{\kappa}(k)f(n)$ and 
$m_{\cN \! \kappa}$ are some multiplicities. Combining (\ref{LMred4a}) 
and (\ref{LMred4b}) results in 
\ba 
L^2(G) \otimes L^2(\cN) = \int \! d\nu(\sigma) \, 
\bigoplus_{\kappa, \kappa'}  \cL_{\sigma} \otimes 
m_{\check{\kappa}} V_{\check{\kappa}} \otimes 
m_{\cN\! \kappa'} L_{\kappa'}^2(\cN)\,.
\label{LGNred1}
\end{eqnarray}
The Fourier coefficients transform according to 
\be 
(\rho \times \ell|_K \times \ell_{\cN})(g,k,k') 
\widehat{\psi}(\sigma, n) =  
\pi_{\sigma}(g) \widehat{\psi}(\sigma,n) 
(r_{\check{\kappa}} \times r_{\kappa'})(k,k')\,.  
\label{LGNred2}
\end{equation}
It remains to implement the condition (!) in (\ref{LMred1}), (\ref{LMred2}).
Since $\psi(k^{-1}g, k^{-1} n) = (\ell|_K \times \ell_{\cN})(k,k) \psi(g,n)$ 
this amounts to considering now the inner tensor product 
$\ell|_K \otimes \ell_{\cN}$ and projecting onto the singlet 
sector. The reduction of $\ell|_K \otimes \ell_{\cN}$ produces 
in a first step a direct double sum over $\kappa,\,\kappa'$ of terms 
of the form $r_{\check{\kappa}} \otimes r_{\kappa'}$ with 
multiplicities $m_{\kappa} m_{\cN \!\kappa'}$. In the next step 
we use that $r_{\check{\kappa}} \otimes r_{\kappa'}$ contains the 
singlet if and only if $\kappa = \kappa'$. The latter readily follows 
from the general result of \cite{bekka,BHV} (described in Appendix A4)
on the singlet content of a tensor product of two unitary 
representations. As a consequence the direct double sum in 
(\ref{LGNred1}) reduces to a single sum and one can check that the 
condition (!) comes out correctly: $\widehat{\psi}(\sigma,n) 
(r_{\check{\kappa}} \times r_{\kappa})(k,k) = 
\widehat{\psi}(\sigma, k^{-1} n) r_{\kappa}(k^{-1}) = 
\widehat{\psi}(\sigma,n)$. Viewed as functions of $n$ alone 
the Fourier coefficients obey $f(k^{-1} n) = f(n) r_{\kappa}(k)$ 
or $f(kn) = f(n) r_\kappa(k)^\dagger$.

Therefore, in order to go from $L^2(G\times \cN)$ to $L^2(\cM_r)$ 
it is useful to consider the subspaces  $L^2_{\sigma\check{\kappa}}(\cN)$ 
of $L^2_{\sigma\check{\sigma}}(\cN)$ and $L^2_{\sigma\check{\sigma}}(\cN)_0$ 
defined in (\ref{LMred5}). 
Here we interpret the elements of $\cL_\sigma\otimes 
V_{\check\kappa}$ as linear maps from $V_\kappa\subset \cL_\sigma$ to 
$\cL_\sigma$; the trace in the inner product (\ref{LNdef1}) for generic 
Hilbert-Schmidt operators reduces to a trace on $V_{\kappa}$ for 
operators with values in $\cL_{\sigma} \otimes V_{\check{\kappa}}$. 
In components 
\be
{\rm Tr}_{V_{\kappa}}[f_1(n)^{\dagger} f_2(n)] =
\sum_{i, s} f_1(n)^*_{i \kappa s}\, f_2(n)_{i \kappa s}\,,\sspace 
f(n) = \sum_{i, s} e_i \,f(n)_{i \kappa s} \check{e}_{\kappa s}\,,
\label{LMred6}
\end{equation}
where $e_i,\,i\in \N$, is a basis on $\cL_{\sigma}$ and 
$\check{e}_{\kappa s}$, $s =0, \ldots, m_{\kappa} \dim V_{\kappa} \!-\!1$, 
is a basis of linear forms on $m_{\kappa} \,V_{\kappa}$. The 
space $L^2_{\sigma \check{\kappa}}(\cN)$ carries a unitary representation 
$\pi_{\sigma \check{\kappa}}$ of $G \times K$, 
\ba 
&& \pi_{\sigma\check{\kappa}}(g,k) f(n) 
:= \pi_{\sigma}(g) f(n) r_\kappa(k)^\dagger\,,
\nonum
&& {\rm Tr}_{V_{\kappa}}[ [\pi_{\sigma\check{\kappa}}(g,k) f_1(n)]^{\dagger} 
\pi_{\sigma\check{\kappa}}(g,k) f_2(n)]
= {\rm Tr}_{V_{\kappa}}[f_1(n)^{\dagger} f_2(n)]\,.
\label{LMred7} 
\end{eqnarray} 
Note that the product $\pi_{\sigma}(g) f(n)$ still transforms 
nontrivially under the left diagonal action of $K$ according to 
$\pi_{\sigma}(g) f(n) \mapsto \pi_{\sigma}(kg) f(kn) = 
\pi_{\sigma}(k)[\pi_{\sigma}(g) f(n)] 
r_{\kappa}(k^{-1})$. However, introducing $P_\kappa$ as the orthogonal 
projection from $L^2(G)$ onto the subspace transforming according to 
$\kappa$, traces of the form  
${\rm Tr}_{V_{\kappa}}[P_{\kappa} \pi_{\sigma}(g) f(n)]$ are invariant 
under the left diagonal action of $K$.

Finally we arrive at the desired decomposition of $L^2(\cM_r)$ 
into a direct integral of irreducible spaces
\be 
L^2(\cM_r) = \int_{\widehat{G}_r} \!d\nu(\sigma) \,
\bigoplus_{\kappa \in \widehat{K}_{\sigma}} m_{\cM \check{\kappa}} 
L^2_{\sigma\check{\kappa}}(\cN)_0\,,
\label{LMred9}
\end{equation}
for some multiplicities $m_{\cM \check{\kappa}}$. The representation 
$\rho$ itself has been decomposed into a direct integral of 
primary representations (that is, \cite{folland}, p.206, ones which are 
direct sums of identical copies of some irreducible representation). 
Here $\rho$ acts as the right regular representation of $G$ on the 
first argument of the functions $L^2(\cM_r)$, i.e.~$\rho(g_0) \psi(g,n) 
= \psi(gg_0, n)$. The left $K$ invariance of the functions is 
broken up into the different $\kappa$ `channels' on the Fourier 
coefficients; generic left $K$ invariant functions are built from 
simple ones. Consistency with the initial decomposition 
(\ref{Planchpsi2}) fixes the multiplicities 
\be 
m_{\cM \kappa} = m_{\kappa}\,,
\label{LMmulti1}
\end{equation}   
where $m_{\kappa}$ are the multiplicities occuring in the 
decomposition of $\pi_{\sigma}|_K$, see (\ref{LMred3}). 
Indeed, disregarding the breakup into the irreducible 
$\kappa$ channels the fiber spaces have the following 
isometric descriptions
\be
\cL_{\sigma}^2(\cM_r) = \bigoplus_{\kappa \in 
\widehat{K}_{\sigma}} m_{\cM \check\kappa} L^2_{\sigma\check\kappa}(\cN)_0
\cong L^2_{\sigma\check\sigma}(\cN)_0 \subset 
L^2_{\sigma\check\sigma}(\cN) \cong
\cL_{\sigma} \otimes \bigoplus_{\kappa \in 
\widehat{K}_{\sigma}}  m_{\cM \check\kappa} V_{\check\kappa} 
\otimes L^2(\cN)\,,
\label{LMmulti2}
\end{equation}
where we used (\ref{LMred5}) in the first isometry and inferred 
(\ref{LMmulti1}) from the required match with  $L^2_{\sigma\check\sigma}(\cN)
\cong \cL_{\sigma} \otimes \check\cL_{\check\sigma} \otimes L^2(\cN)$.

The significance of the breakup into the irreducible $\kappa$ 
channels can be seen more clearly by identifying the pre-images of the  
functions in $\int \!d\nu(\sigma)\,m_{\cM \check{\kappa}} 
L^2_{\sigma\check{\kappa}}(\cN)_0$ 
with respect to the isometry $\psi \mapsto \widehat{\psi}$ on 
$L^2(G \times \cN)$. Here we take $L^2_{\sigma\check{\kappa}}(\cN)$   
to consist of the zero vector only, if $\kappa \notin \widehat{K}_{\sigma}$.
Roughly, the pre-image consists of functions invariant under 
left convolution with the character $\chi_{\check{\kappa}}$ of 
$\check{\kappa} 
\in \widehat{K}_{\sigma}$. Specifically we set 
\ba
&& 
(E_{\kappa'}* \psi)(g,n) := d_{\kappa'} \int \! dk \,\chi_{\kappa'}(k^{-1}) 
\psi(k g,n) \,,
\nonum
&& L^2_{\kappa}(\cM_r) :=\Big\{ \psi_{\kappa} \in
L^2(\cM_r) \;\Big|\; E_{\kappa'} * \psi_{\kappa}=
\delta_{\kappa \kappa'} \psi_{\kappa} \Big\}\,.  
\label{LMred10}
\end{eqnarray}
The properties of the functions in this subspace are summarized 
in the following Lemma.

\begin{lemma}\label{l2kappa}
(a) For $\psi_{\check{\kappa}} \in L^2_{\check{\kappa}}(\cM_r)$, 
the Fourier coefficients obey 
\ba
&& \widehat{\psi_{\check{\kappa}}} P_{\check{\kappa}'} =
\left\{ \begin{array}{cl} 0 \quad & \mbox{for}\;\; \kappa' \neq \kappa\,,
\\[1mm]
\widehat{\psi_{\check{\kappa}}} \in L^2_{\sigma \check{\kappa}}(\cN)_0 
\quad& \mbox{for} \;\; \kappa' = \kappa \,.
\end{array} \right. 
\label{LMred11}
\end{eqnarray}
In terms of components with respect to the $K$-adapted basis (\ref{adaptb})
this is equivalent to 
$\widehat{\psi_{\check{\kappa}}}(\sigma,n)_{\kappa_1 s_1, 
\kappa_2 s_2} = f_{\check{\kappa}}(n)_{\kappa_1 s_1,s_2} 
\delta_{\kappa \kappa_2}$, with $f_{\check{\kappa}} \in 
L^2_{\sigma \check{\kappa}}(\cN)_0$. 

(b) For $\psi_{\check{\kappa}} \in \cD \cap L^2_{\check{\kappa}}(\cM_r)$  
(i.e.~$g \mapsto \psi_{\check{\kappa}}(g,n)$ is smooth in $g$ with compact 
support) we have 
\be  
\psi_{\check{\kappa}}(g,n) = \int_{\widehat{G}_r} \! d\nu(\sigma)\, 
{\rm Tr}_{V_{\check{\kappa}}}[P_{\check{\kappa}} \pi_{\sigma}(g)\,  
\widehat{\psi_{\check{\kappa}}}(\sigma, n)]\,.
\label{LMred13}
\end{equation}
(c) For   $\phi_{\check{\kappa}}, \,\psi_{\check{\kappa}} \in 
L^2_{\check{\kappa}}(\cM_r)$,  
a Parseval identity holds 
\be
\int \! dg dn \phi_{\check{\kappa}}(g,n)^* \psi_{\check{\kappa}}(g,n) 
= \int_{\widehat{G}_r} \!d\nu(\sigma)  \int \!dn 
{\rm Tr}[\widehat{\phi}_{\check{\kappa}}(\sigma, n)^{\dagger} 
\widehat{\psi}_{\check{\kappa}}(\sigma, n)]\,.
\label{LMred14}
\end{equation}
\end{lemma} 

\begin{proof} (a) We compute $(E_{\kappa}*\psi)(g,n)$ by inserting 
the Fourier decomposition (\ref{psi1}) for $\psi \in L^2(\cM_r)$.  
This gives 
\ba 
(E_{\kappa}*\psi)(g,n) &=& \int_{\widehat{G}_r} \!
d\nu(\sigma) \,{\rm Tr}\Big[\widehat{\psi}(\sigma,n) 
\;d_{\kappa}\! \int_K \!dk\, \chi_{\kappa}(k^{-1}) 
\pi_{\sigma}(k)\; \pi_{\sigma}(g)\Big]\,.
\nonum
&=& \int_{\widehat{G}_r} \!d\nu(\sigma) \,{\rm Tr}_{V_{\kappa}}
[\pi_{\sigma}(g) \widehat{\psi}(\sigma,n)P_{\kappa}]\,. 
\label{LMred11a}
\end{eqnarray} 
In the second step we identified the $K$ integral as 
the projector (\ref{Pkappa}) onto the $m_{\kappa} V_{\kappa}$
subspace of $\cL_{\sigma}$. This shows $\widehat{\psi}_{\check{\kappa}}
P_{\check{\kappa}'} = \delta_{\kappa \kappa'}\widehat{\psi}_{\check{\kappa}}$,
or in components with respect to the $K$-adapted basis 
$\widehat{\psi_{\check{\kappa}}}(\sigma,n)_{\kappa_1 s_1, 
\kappa_2 s_2} = f_{\check{\kappa}}(n)_{\kappa_1 s_1, s_2} 
\delta_{\kappa \kappa_2}$, with $f_{\check{\kappa}} \in 
L^2_{\sigma \check{\kappa}}(\cN)_0$.  
The equivariance property in the definition of 
$L^2_{\sigma \check{\kappa}}(\cN)_0$ follows from (\ref{LMred1})
and $\widehat{\psi}(\sigma,kn) P_{\check{\kappa}} = 
\widehat{\psi}(\sigma,n) \pi_{\sigma}(k)^\dagger  P_{\check{\kappa}} = 
\widehat{\psi}(\sigma,n) P_{\check{\kappa}} 
r_{\check{\kappa}}(k)^\dagger$. 
The fact that $f_{\check{\kappa}}$ is square integrable in the norm
$(\;,\;)_{\sigma\check{\kappa}}$ follows from the Plancherel 
identity (\ref{LMred13}).

(b) Eq.~(\ref{LMred13}) follows from (\ref{LMred11a}) and
the definition of $L^2_{\check\kappa}(\cM_r)_0$. Note that the trace is 
constant 
on the equivalence classes $(g,n) \sim (k g, kn)$, although the product 
$\pi_{\sigma}(g) \widehat{\psi_{\check{\kappa}}}(\sigma,n)$ 
itself transforms nontrivially under $(g,n) \mapsto (kg,kn)$. 

(c) This follows from (a) and (\ref{Planchpsi}). 
\end{proof}
We add some comments on Lemma 2.1. First, in view of (\ref{LMred9}) 
the result can be summarized by stating that the map $\psi_{\check{\kappa}}
\mapsto \widehat{\psi_{\check{\kappa}}}$ in (\ref{LMred13}) provides 
a partial isometry 
\be 
L^2_{\check{\kappa}}(\cM_r) \rra \int_{\widehat{G}_r}^\oplus \! 
d\nu(\sigma) 
m_{\cM \check{\kappa}} \,L^2_{\sigma\check{\kappa}}(\cN)_0\,.
\label{LMred19}
\end{equation}
Next we discuss two special cases, where in the decomposition (\ref{LMred13}) 
only ``class 1'' representations occur (i.e.~representations which contain 
a vector invariant under the action of the subgroup $K$). The first case is 
that of $\rho(K)$ singlets in $L^2(\cM_r)$. In the decomposition (\ref{psi1}) 
functions obeying $\psi(gk,n) = \psi(g,n)$ are characterized by 
Fourier coefficients with $P_0 \widehat{\psi}(\sigma,n) =  \widehat{\psi}(\sigma,n)$,
where $P_0$ is the projector onto the singlet sector (`$\kappa =0$'). 
In this case Eq.~(\ref{psi1}) reduces to 
\ba 
\label{Ksinglets1}
\psi(gK,n) \is \int_{\widehat{G/K}}\!d\nu(\sigma) \, 
P_0[\widehat{\psi}(\sigma,n) \,\pi_{\sigma}(gK)]\,
\\
P_0 \widehat{\psi}(\sigma,n) \is   
\int_{G/K} \! d\gamma_{G/K}(gK) \,\psi(gK,n) \, 
P_0 \pi_{\sigma}(K^{-1}g^{-1})\,.
\nonumber
\end{eqnarray}
Since the the functions $g \mapsto P_0 \pi_{\sigma}(g)$ are left $K$-invariant
by definition only `class 1' representations (with respect to $K$) 
occur in (\ref{Ksinglets1}). Specifically, consistency with the harmonic 
analysis on $Q=G/K$ requires that only those (class 1) irreducible 
representions occur which enter the harmonic analysis on $G/K$. We 
have written $\widehat{G/K} \subset \widehat{G}_r$ for this subset;
as described in Appendix A.7 it contains the spherical principal series 
representations only.

For Fourier coefficients obeying in addition (\ref{LMred1}) the projection 
$P_0[\widehat{\psi}(\sigma,n) \,\pi_{\sigma}(gK)]$ is invariant 
under $(g,n) \mapsto (kg,kn)$. In this case the 
functions in (\ref{Ksinglets1}) define a subspace of $L^2(\cM_r)$. 
Note however that it is not $\rho(G)$ irreducible; the latter 
requires Fourier coefficients satisfying $P_0 \widehat{\psi}(\sigma,n) P_{\kappa}
= P_0 \widehat{\psi}(\sigma,n)$ for some $\kappa \in \widehat{K}$. 
To proceed recall that the section $g_s: Q \ra G$ provides an injective 
imbedding of $Q$ in $G$. We set 
\ba 
&& E_{\sigma,\check{\kappa}}(q) := P_{\check{\kappa}} 
\pi_{\sigma}(g_s(q)) P_0\,,
\sspace 
E_{\sigma,\check{\kappa}}(q)^{\dagger}  = 
P_0 \pi_{\sigma}(g_s(q)^{-1}) P_{\check{\kappa}}\,,
\nonum
&& \psi_{\check{\kappa}}(q,n) := \psi_{\check{\kappa}}(g_s(q)^{-1},n)\,,
\label{Ksinglets2}
\end{eqnarray}
in terms of which the expansion (\ref{LMred13}) takes the form 
\ba
\psi_{\check{\kappa}}(q,n) \is \int_{\widehat{G/K}} \! d\nu(\sigma)\, 
{\rm Tr}[ E_{\sigma,\check{\kappa}}(q) 
\widehat{\psi_{\check{\kappa}}}(\sigma,n)]\,,
\nonum
\widehat{\psi_{\check{\kappa}}}(\sigma,n) \is 
\int_Q \! d\gamma_Q(q) \,\psi_{\check{\kappa}}(q,n) 
E_{\sigma,\check{\kappa}}(q)^{\dagger}\,.
\label{Ksinglets3}
\end{eqnarray}

Another subspace of $L^2(\cM_r)$ in whose decomposition only class 1 
irreducible representations appear is the singlet ($\kappa=0$) sector
of (\ref{LMred9}). On account of Lemma 2.1 this sector arises from functions
$\psi_0 \in L_0(\cM_r)$, that is, functions obeying $\psi_0(g, k^{-1} n) = 
\psi_0(g,n) = \psi_0(k g, n)$. The Fourier decomposition takes the form   
\ba 
\label{Ksinglets4}
\psi_0(g,n) \is \int_{\widehat{G/K}}\! d\nu(\sigma)\,
P_0[\pi_{\sigma}(g) \widehat{\psi_0}(\sigma,n)]\,,
\nonum 
\widehat{\psi_0}(\sigma,n) \is  \widehat{\psi_0}(\sigma,n) P_0 = 
\int \!dg\, \psi_0(g,n) \pi_{\sigma}(g^{-1}) P_0\,.
\end{eqnarray}

We can now summarize the results on the group theoretical 
decomposition. Since the action of $\rho(G)$ on $L^2(\cM_r)$ corresponds
to the original $\ell_{\cM}$ action of $G$ on 
$L^2(\cM)$ (not to be confused with the left regular representation of $G$ 
on itself denoted by $\ell$) we have found the desired 
group theoretical decomposition (\ref{nuclear2}): only a subset 
of $\widehat{G}$, the restricted dual $\widehat{G}_r$ appears, 
and each $\sigma \in \widehat{G}_r$ occurs with infinite 
multiplicity ${\rm dim}\ (\cL_\sigma)$. The measure 
$d\mu_{\ell_\cM}$ is the Plancherel 
measure on $\widehat{G}_r$ and the copies are counted with the 
counting measure on $\widehat{K}$. We continue to use the realization 
$L^2(\cM_r)$, where the group acts via the right regular representation 
on the first argument of the functions $\psi  =\psi_r$. 
For convenient reference we collect the results in:  

\setcounter{proposition}{1}
\begin{proposition}
(a) The Hilbert space $L^2(\cM_r)$ decomposes
under the action of the unitary representation $\rho(G)$ according to 
\be 
L^2(\cM_r) = \int_{\widehat{G}_r} \!d\nu(\sigma) \,
\bigoplus_{\kappa \in \widehat{K}_{\sigma}} m_{\check{\kappa}}\,
L^2_{\sigma\check{\kappa}}(\cN)_0 =: 
\int_{\widehat{G}_r} \!d\nu(\sigma) \,\cL^2_{\sigma}(\cM_r)\,,
\label{LMdecomp}
\end{equation}
with $L^2_{\sigma,\check{\kappa}}(\cN)_0$ defined in (\ref{LMred5}),
and $m_{\kappa}$ the multiplicities in (\ref{LMred3}).\\ 
(b) Disregarding the breakup into irreducible $\kappa$ channels 
the fiber spaces have the following descriptions
\be
\cL_{\sigma}^2(\cM_r) = \bigoplus_{\kappa \in 
\widehat{K}_{\sigma}} m_{\check\kappa} L^2_{\sigma\check\kappa}(\cN)_0
\cong  L^2_{\sigma\check\sigma}(\cN)_0 
\subset  L^2_{\sigma\check\sigma}(\cN)\cong \cL_{\sigma} \otimes 
\check\cL_{\check\sigma} \otimes L^2(\cN) \,,
\label{LMdecomp2}
\end{equation}
with $ L^2_{\sigma\check\sigma}(\cN)$ defined in (\ref{LNdef1}). \\
(c) On the subspace of $L^2(\cM_r)$ containing a $\rho(K)$ singlet 
the decomposition has support only on the spherical principal 
series representations. 
\end{proposition}

The content of Proposition 2.2 can be illustrated by matching 
(\ref{LMdecomp}) against a direct decomposition  of 
$L^2(\cM) \simeq L^2(Q)^{\otimes \nu}$ into 
irreducibles. From (\ref{GKharmonic}) one has 
\be 
L^2(Q)^{\otimes \nu} \simeq \int \!\frac{d\om_1}{|c(\om_1)|^2} 
\ldots \frac{d\om_1}{|c(\om_1)|^2}\, \cL_{\om_1} \otimes \ldots 
\otimes \cL_{\om_{\nu}}\,,
\label{LMdecomp3} 
\end{equation}
where $\cL_{\om}\,,\om \in \widehat{Q}\subset \R^{\dim A}$ carries 
the spherical principal series represenation $\pi_{\om,0}$. 
The decomposition problem essentially amounts to decomposing 
arbitary tensor products $\cL_{\om_1} \otimes \ldots 
\otimes \cL_{\om_{\nu}}$ of spherical principal series representations.  
This can be done inductively by first decomposing $\cL_{\om_1} 
\otimes \cL_{\om_2} = \int\! d\mu(\om_1,\om_2|\sigma) \cL_{\sigma}$,
then $\cL_{\sigma} \otimes \cL_{\om_3}$ for any $\sigma \in 
\widehat{G}$ in the support of the measure $d\mu(\om_1,\om_2|\sigma)$, 
and so on. Depending on what is known about the support of  
$d\mu(\om_1,\om_2|\sigma)$ in $\widehat{G}$ this strategy requires 
knowledge of large portions of the complete branching rules.
Since the complete branching rules are known only for a few 
noncompact groups (like ${\rm SL}(2,\R)$, see \cite{Mukunda}) 
this would be tedious, to say the least. 

Information about the support of the measure $d\mu(\om_1,\om_2|\sigma)$, 
seems to be available only in a few cases, like ${\rm SO}_0(1,N)$ from 
\cite{Dobrevetal}. According to Theorem 10.5 given there, a tensor product 
of two spherical principal series unitary irreducible representations (UIR) 
of ${\rm SO}_0(1,N)$ decomposes into principal series UIR only, 
for $N \geq 4$ even.~(For $N$ odd this holds trivially and for $N\!=\!2$ it 
is manifestly not true, see \cite{Mukunda}). For more than two 
tensor 
copies (or for $N\!=\!2$), however, the result is insufficient to exclude 
the occurrence of UIR other than principal series representations. For $\nu
=3$ one would need to know how the tensor product of a spherical and a generic 
(non-spherical) principal series decomposes. This might include discrete series
representations and possibly others. So, for the direct decomposition
of $\nu \geq 4$ fold tensor products of the spherical principal series, one  
needs to know essentially the complete branching rules.  

As a spin-off of Proposition 2.2 we have  

{\bf Corollary 2.3.}{\it
(a) Let $\cL_{\om_1} \otimes \ldots \otimes \cL_{\om_{\nu}}$ 
be an arbitrary  tensor products of spherical principal series 
representations. Then for almost all $\om_1,\ldots,\om_{\nu}$
with respect to the measure in (\ref{LMdecomp3}) its 
decomposition does not contain the singlet (as the only finite 
dimensional UIR). 
(b) Let $Q(\om_1,\ldots,\om_{\nu}|\sigma)(q_1,\ldots ,q_{\nu})$ be 
an intertwiner from $\cL_{\om_1} \otimes \ldots \otimes \cL_{\om_{\nu}}$ 
to $\cL_{\sigma}$, $\sigma \in \widehat{G}$. Then, for almost 
all $\om_1,\ldots,\om_{\nu}$
\be 
\int_K \! dk\, Q(\om_1,\ldots,\om_{\nu}|\sigma)(kq_1,\ldots ,kq_{\nu}) 
=0\,,
\end{equation}
unless $\sigma$ is again a spherical principal series representation.
}

Part (a) generalizes a result by Fulling \cite{Fulling} for tensor 
products of 
spherical principal representations to all reductive 
linear Lie groups. Part (b) provides a nontrivial `sum-rule' whenever 
explicit expressions for the interwiners are available. 
The `almost all' caveat is needed because the Harish-Chandra c-function 
defining the measure in (\ref{LMdecomp3}) is a meromorphic function 
over the complexification of $\widehat{Q}$ (viz $a^*_\C$, where 
$a$ is the Lie algebra of the subgroup A in the Iwasawa decomposition.) 
See e.g.~\cite{Helgason2}, Chapter, II.3. At the position 
of a pole the decomposition (\ref{LMdecomp3}) yields no information about 
the fiber spaces, but otherwise all spherical principal series 
representations occur \cite{Helgason2}, Chapter VI.3. 
The location of the poles of $c$ can be analyzed from the 
explicit expressions \cite{Helgason2}, which could be used to specify the 
exceptional sets.

%%%%%%%%%%%%%%%%%%%%%%%%%%%%%%%%%%%%%%%%%%%%%%%%%%%%%%%%%%%%%%%%%%%%%%%%
\newsection{The spectral problem on the invariant fibers} 

\setcounter{proposition}{0}
\setcounter{lemma}{0}
\setcounter{definition}{0}

We proceed to study the consequences of the orbit decomposition 
for selfadjoint operators commuting with the group action. 
It is natural to consider the same class of integral operators as in 
Definition 2.1, but drop the positivity requirements. 

\begin{definition}
A standard invariant selfadjoint operator $\bA$ on $L^2(\cM)$ is an 
integral operator with 
a symmetric continuous kernel $A: \cM \times \cM \ra \C$, 
which obeys (\ref{Abound}) below and satisfies $A(d(g)m, d(g)m') = 
A(m,m')$ for all $g \in G$, $m,m'\in \cM$. Its image $\Phi^{-1} \circ \bA \circ
\Phi$ under the isometry in (\ref{Phidef}) is also denoted by $\bA$ and is an 
integral operator on $L^2(\cM_r)$ with kernel $\cA: G \times \cN \times 
\cN \ra \C$, subject to (\ref{Tr2}) below and (\ref{Abound}). Here  
\be
\sup_m \int \!dm' |A(m,m'| = \sup_n\int\! dn'dg\, 
|\cA(g,n,n')| =:\ K_{\cal A} <  \infty\ .
\label{Abound}
\end{equation}
\end{definition}
As in Definition 2.1 the condition (\ref{Abound}) entails that $\bA$ is 
well-defined as a bounded selfadjoint operator from $L^p$ to $L^p$ for 
all $1 \leq p \leq \infty$. The structure of the kernels $A$ and $\cA$ 
of course implies that the operators $\bA$ commute with $\ell_{\cM}(G)$ 
and $\rho(G)$, respectively.

%%%%%%%%%%%%%%%%%%%%%%%%%%%%%%%%%%%%%%%%%%%%%%%%%%%%%%%%%%%%%%
\newsubsection{Basic consequences of the $\rho(G)$ invariance}  
 
As a first result we have:  

\begin{proposition}\label{spectrum} 

(a) Every standard invariant selfadjoint operator ${\bf A}$ on $L^2(\cM)$ 
has purely essential spectrum, Spec(${\bf A}$) =     
\mbox{ess-Spec}(${\bf A}$).\\
(b) A transfer operator commuting with
$\ell_{\cM}(G)$ cannot have normalizable ground states; hence $\Vert \bt 
\Vert \in$ c-Spec$(\bt)$, or $\Vert \bt \Vert$ is a limit point of 
eigenvalues of infinite multiplicity.
\end{proposition}

\begin{proof} 
(a) Recall from Eqs.~(\ref{nuclear2}) and
(\ref{nuclear3}) that the measure $\mu_{\ell_\cM}$ on $\widehat{G}$ 
defining the
decomposition of the unitary representation $\ell_{\cM}$ is related
to that defining the generalized eigenspaces of ${\bf A}$ by
$\int_{Spec({\bf A})} d\mu(\lb,\sigma) = d\mu_{\ell_\cM}(\sigma)$.
By Proposition 2.2 the measure $\mu_{\ell_\cM}(\sigma)$ can be identified 
with the Plancherel measure. For the noncompact
Lie groups considered the support of the Plancherel
measure contains only infinite dimensional representations
(see Appendix A). Therefore, whenever a generalized eigenspace
$\cE_{\lb, \sigma}({\bf A})$ occurs in (\ref{nuclear1}) its spectral value
$\lb$ has infinite multiplicity. This excludes that $\lb$ lies in the
discrete spectrum of the operator; hence $\lb \in$ ess-Spec$({\bf A})$.
Note that it is not excluded that ${\bf A}$ has point spectrum of infinite
multiplicity, i.e.~infinite multiplets of normalizable eigenfunctions for
some spectral value $\lb$.

(b) Let now $\bt$ be a transfer operator on $L^2(\cM)$ 
commuting with $\ell_{\cM}(G)$. Assume it has a normalizable ground state, 
i.e.~a solution of $\bt \psi = \Vert \bt \Vert \, \psi$, with 
$\psi \in L^2(\cM)$. By a well known result, based on the 
Perron-Frobenius theorem (see for instance \cite{GlimmJaffe,RS4}), this ground state 
would be unique and therefore invariant under the action of $\ell_\cM(G)$, 
i.e. an $\ell_\cM$ singlet. This entails that $(\psi,\psi)$ diverges as the 
infinite volume of the group is `overcounted'. Thus $\Vert \bt \Vert$ 
cannot be a eigenvalue. By part (a) it can also not be a limit point of 
the discrete spectrum (since there is none), leaving only the possibilities: 
$\Vert \bt \Vert \in$ c-Spec$(\bt)$ or  $\Vert \bt \Vert$ a limit point 
of eigenvalues of infinite multiplicity.
\end{proof}
To proceed, let us momentarily denote the selfadjoint operator $\bA$ in the 
realization acting on $L^2(\cM_r)$ by $\bA_r := \Phi^{-1}\circ \bA \circ \Phi$, with 
$\Phi$ as in (\ref{Phidef}). We write $A_r(g,n, g',n';t)$ for the corresponding 
kernel. Writing out 
$(\Phi^{-1} \circ \bA \circ \Phi\, \psi_r)(g,n)$ one finds 
\be
A_r(g,n,g',n')\ =\  A(\phi^{-1}(g,n),\phi^{-1}(g',n'))
\ =\ A(g' g^{-1} q^\up, g' g^{-1}n, q^\up, n')\,.   
\label{TrT}
\end{equation}
In the last step we used the invariance of $\bA$, 
i.e.~$\ell_{\cM}(g) \circ\bA  = \bA \circ \ell_{\cM}(g)$ for all $g \in G$.
Since $\bA_r^t = \Phi^{-1} \circ \bA^t \circ \Phi$ for $t \in \N$, also all 
iterated kernels will be related by (\ref{TrT}). 

For the reasons explained in Section 2.3 the kernel 
$A_r$ is most useful. From the last expression in (\ref{TrT}) one sees 
explicitly that $A_r$ only depends on the right invariant combination 
$g' g^{-1}$, consistent with $\rho(g) \circ \bA_r = \bA_r \circ \rho(g)$ 
for the operator. Further $A_r$ is manifestly constant on the left equivalence  
classes $(k g, k n) \sim (g,n)$ and $(k' g', k' n') \sim (g', n')$ with 
$(k,k') \in K \times K$. We set 
\be 
\cA(g{g'}^{-1}, n,n';t) := A_r(g,n,g',n';t)\,,
\label{Tr1}
\end{equation}
and note
\ba 
&& \cA(g,n,n') = \cA(g^{-1},n',n)\,,
\nonum
&& \cA(kg,kn,n';t) = \cA(g,n,n';t) = \cA(gk^{-1},n,kn';t)\,,\quad k \in K\,. 
\label{Tr2}
\end{eqnarray}
We may thus interpret $\bA_r$ either as an integral operator acting on 
$L^2(\cM_r)$ or on $L^2(G\times\cN)$; in the latter interpretation it 
automatically projects onto the $\ell_r(K)$ invariant subspace (with 
$\ell_r$ defined in (\ref{Lreps})). This means 
that the nonzero spectrum of $\bA$ lies automatically in that subspace. 
For convenience, we will therefore work with this interpretation 
of $\bA_r$. 

To simplify the notation we drop the subscript $r$ in the following
and write $\bA$ for $\bA_r$ etc. 
Due to the properties (\ref{Tr1}), (\ref{Tr2}) $\bA$ 
respects the fiber decomposition (\ref{LMdecomp}) in the following 
sense: 
\begin{lemma}
Let $\Sigma$ be a measurable subset of $\widehat 
G_r$, $\Sigma^c$ its complement and let $\cH$ and $\cH^c$ be the 
corresponding subspaces of $L^2(G)$ (which are orthogonal complements of 
each other):
\ba
\cH\is \int^{\oplus}_{\Sigma}\! d\nu(\sigma)\ \cL_{\sigma} \otimes 
\check{\cL}_{\check{\sigma}} \otimes L^2(\cN)\,,  
\nonum
\cH^c\is \int^{\oplus}_{\Sigma^c}\! d\nu(\sigma)\ \cL_{\sigma} \otimes
\check{\cL}_{\check{\sigma}} \otimes L^2(\cN)\,. 
\end{eqnarray}
Then $\bA\cH\subset\cH$ and $\bA\cH^c\subset\cH^c$.
\end{lemma}
\begin{proof} Let $\psi\in\cH,\ \psi_c\in\cH^c$ and consider 
\be 
s(g):= (\psi_c,\bA\rho(g)\psi) = (\rho(g)^{-1}\psi_c,\bA\psi)\, .
\end{equation}
The first expression shows that the Fourier transform $\widehat s$ of $s$ 
is supported in $\Sigma$. On the other hand, the second expression 
shows that the Fourier transform is supported in $\Sigma^c$. This is 
possible only if $s$ vanishes identically. Putting $g=e$, the lemma follows. 
\end{proof}

%%%%%%%%%%%%%%%%%%%%%%%%%%%%%%%%%%%%%%%%%%%%%%%%%%%%%%%%%%%%%%%%%%%
\newsubsection{Fiber decomposition of invariant selfadjoint operators}

In view of Lemma 3.1 the operator $\bA$ should also map the fiber spaces 
$\cL^2_{\sigma}(\cM_r)$ in (\ref{LMdecomp2}) onto itself. To give precise 
meaning to this statement we have to construct fiber operators 
$\bA(\sigma)$ from $\bA$. In a first step one applies the expansion 
(\ref{psi1}) to the kernel $\cA(g,n,n')$ 
\ba 
&& \cA(g, n,n') = \int_{\widehat{G}_r}d\nu(\sigma) 
\Tr\Big[\widehat{\cA}(\sigma,n',n)  
\pi_{\sigma}(g)\Big]\,.
\nonum
&& \widehat{\cA}(\sigma,n',n) 
: = \int_G \! dg\, \cA(g,n,n') \pi_{\sigma}(g^{-1})\,. 
\label{Tr3}
\end{eqnarray}
The swapped order in the $(n,n')$ arguments was chosen in order to have 
\be 
\widehat{\cA}(\sigma,k' n', k^{-1} n)_{\kappa_1 s_1,\kappa_2 s_2} 
= \sum_{s s'} r_{\kappa_1}(k')_{s_1 s} \, 
\widehat{\cA}(\sigma,n',n)_{\kappa_1 s,\kappa_2 s'} \,
r_{\kappa_2}(k)_{s' s_2}\,,
\label{Tr4}
\end{equation}
where the indices $\kappa, s$ etc.~refer to the $K$-adapted basis 
(\ref{adaptb}). Note further the hermiticity and diagonal $K$ invariance 
imply
\be
\widehat{\cA}(\sigma,n',n) = 
\widehat{\cA}(\sigma,n,n')^\dagger\,,
\quad 
\Tr_{V_\kappa}\widehat \cA(\sigma,kn',kn) = \Tr_{V_\kappa}\widehat
\cA(\sigma,n',n)\,. 
\label{Tr5}
\end{equation}
Since by (\ref{Abound}) for a.e.~$n,n'$ the kernel $\cA$ is in 
$L^1(G)$, its Fourier transform $\widehat{\cA}(\sigma,n,n')$ is a 
compact operator on $\cL_\sigma$ for 
a.e.~$n,n'$, see \cite{folland}, Theorem 7.6. 
Moreover (\ref{Abound}) implies the following bound on 
$\widehat\cA(\sigma,n,n')$
\be
\sup_n \int\! dn' \Vert \widehat \cA(\sigma,n,n') \Vert\le K_A\ ,
\label{hatabound}
\end{equation}
where $\Vert\, .\,\Vert$ denotes the operator norm on $\cL_\sigma$.
If we momentarily introduce the spaces
\be
L^p_{\sigma}(\cN):=\bigl\{\psi:\cN \ra \cL_\sigma\,\big\vert\int\!dn 
\, \left(\psi(n), \psi(n)\right)_{\sigma}^{p/2} <\infty\bigr\}\ ,  
\label{Lspdef}
\end{equation}
it follows first that $\widehat\cA(\sigma,n,n')$ defines a bounded linear 
operator from $L^1_{\sigma}(\cN)$ to $L^1_{\sigma}(\cN)$ 
as well as from $L^\infty_{\sigma}(\cN)$ to $L^\infty_{\sigma}(\cN)$; 
by complex interpolation it follows then essentially as in \cite{Lax} 
p.173 ff that $\widehat\cA(\sigma,n,n')$ defines a bounded linear
operator from $L^p_{\sigma}(\cN)$ to $L^p_{\sigma}(\cN)$ for all 
$p\in[1,\infty]$, where the $p = \infty$ norm is given by 
$\sup_n \sqrt{ (\psi(n), \psi(n))_{\sigma} }$.

Furthermore the projection property of $\bA$ onto the $\ell_r(K)$ invariant 
subspace translates into an equivariance property analogous to 
(\ref{LMred1}):
\be
\widehat\cA(\sigma,n',k^{-1}n)\pi_\sigma(k^{-1})= \widehat\cA(\sigma,n',n)
= \pi_{\sigma}(k) \widehat{\cA}(\sigma,k^{-1}n',n)\,,\quad 
{\rm for}\quad k\in K\,,
\label{Tequiv}
\end{equation}
using (\ref{psi2}), (\ref{Tr2}), (\ref{Tr3}). 
In the following we wish to convert the action of $\bA$ on 
$L^2(\cM_r)$ into an action $\bA(\sigma)$ on the fibers $\int^{\oplus} 
\!d\nu(\sigma) \, L_{\sigma}^2(\cM_r)$ such that 
$\widehat{\bA \psi}(\sigma,n) = \bA(\sigma) 
\widehat{\psi}(\sigma,n)$ holds. To this end we 
note    
\be 
(\bA \psi)(g,n) = \int\! dg' dn'\, 
\cA(g{g'}^{-1}, n,n') \psi(g',n')\,.
\label{TTsigma2}
\end{equation}
From the definitions (\ref{psi1}) one readily computes 
\be 
\widehat{\bA \psi}(\sigma,n) = 
\int \! dn'\,\widehat{\psi}(\sigma, n') \, 
\widehat{\cA}(\sigma,n',n)\,.
\label{TTsigma3}
\end{equation}
Note that $\widehat{\cA}$ acts from the right on $\widehat{\psi}$; 
it acts only on the factor $\check\cL_{\check\sigma}\otimes L^2(\cN)$. If 
one views $\widehat{\psi}$ as a matrix, the `column' index is unaffected.

We can now define the fiber operator $\bA(\sigma)$ by this action from the 
right, Eq.~(\ref{TTsigma3}),  on the elements of $L^2_{\sigma 
\check\sigma}(\cN)$ introduced in (\ref{LNdef1}):
\be 
\bA(\sigma)\widehat\psi(\sigma,n):= \widehat{\bA \psi}(\sigma,n)\ ; 
\label{fiberop}
\end{equation}
the fiber operators $\bA(\sigma)$ then form a measurable field of operators
such that
\be
\bA=\int_{\widehat G_r}^\oplus\! d\nu(\sigma)\, \bA(\sigma) \,.
\end{equation}
The spectrum of $\bA(\sigma)$ has infinite multiplicity because
of the spectator role of the first tensor factor: we have 
\be
\bA(\sigma)=\1\otimes \check\bA_{\check\sigma}\,,
\label{fiberop2}
\end{equation} 
where $\check\bA_{\check\sigma}$ acts on `from the right'. This means that 
$\check\bA_{\check\sigma}$ maps the space of bounded linear forms on 
$\cL_\sigma$ with values in $L^2(\cN)$ into itself, or equivalently, it is 
a map from $\check\cL_{\check\sigma}\otimes L^2(\cN)$ into itself. Thus 
$\check\bA_{\check\sigma}$ is the (partial) dual of an operator 
$\bA_\sigma$ mapping $\cL_\sigma\otimes L^2(\cN)$ into itself.
We will see that in many cases the spectrum of $\bA_\sigma$ has only 
finite multiplicity. As will become clear below, however, both 
$\bA(\sigma)$ and $\bA_{\sigma}$ in general mix the different 
$\kappa$ channels in the decomposition (\ref{LMred9}).

Since $\bA$ is a bounded operator, it follows 
from general considerations (see \cite{folland} (7.30)) that 
$\bA(\sigma): L^2_{\sigma\check{\sigma}}(\cN) \ra
 L^2_{\sigma\check{\sigma}}(\cN)$ is a bounded operator 
for almost all $\sigma$ (see also Proposition 3.4 below).
Similarly $\bA_\sigma$ is a bounded operator  
$L_{\sigma}^2(\cN) \ra L_{\sigma}^2(\cN)$ for almost all 
$\sigma$, with $L^2_{\sigma}(\cN)$ as in (\ref{LNdef}a) of 
Definition 2.2. On this space the left action $\ell_r(K)$ corresponds to
$f(n)\mapsto \pi_\sigma(k) f(k^{-1})$; the singlet subspace of 
$L^2_{\sigma}(\cN)$ under this action is just $L^2_{\sigma}(\cN)_0$
as defined in (\ref{LNdef}b). 

Whenever $\bA(\sigma)$ and hence $\bA_{\sigma}$ are bounded 
operators we call $\sigma \in \widehat{G}_r$ {\it nonexceptional}. 
This is the case for all $\sigma$ in the discrete part $\widehat{G}_d$ 
of the restricted dual (see Proposition 3.4 below) and for 
almost all $\sigma$ in the remainder $\widehat{G}_r\backslash 
\widehat{G}_d$. Whenever $\sigma$ does not occur as an 
integration variable we will assume it to be non-exceptional.   

Let us now describe the operator $\bA_\sigma$ explicitly. 
In terms of the operator valued kernel $\widehat\cA$, its action is  
given by
\be
\label{Tsigma1}
(\bA_\sigma f)(n)=\int\!dn'\, \widehat\cA(\sigma,n',n)^\dagger f(n')
=\int\!dn' \,\widehat \cA(\sigma,n,n') f(n')\, .
\end{equation}
The equivariance property carries over to $\bA_\sigma$:
\be
\label{Tsigma2}
(\bA_\sigma f)(kn)\ = \pi_{\sigma}(k)(\bA_\sigma f)(n)\,, 
\end{equation}
or, in the $K$-adapted basis $(\bA_{\sigma} f)(kn)_{\kappa s} = 
\sum_s r_{\kappa}(k)_{s s'}(\bA_{\sigma} f)(n)_{\kappa s'}$.
This is true whether or not $f$ satisfies the corresponding equivariance 
property; it is a reflection of the fact that $\bA$ projects onto the 
$\ell_r(K)$ invariant subspace of $L^2(G\times \cN)$. 

\addtocounter{proposition}{2}
\begin{proposition}\label{adecompose} 
Let $\bA$ be a standard invariant selfadjoint operator on 
$L^2(\cM_r)$ in the sense of Definition 3.1. Then: 
\begin{itemize}
\itemsep -3pt
\item[(a)] The operator ${\bf A}$ has a decomposition 
${\bf A} = \int^{\oplus} \! d\nu(\sigma)(\1 \otimes 
\check\bA_{\check\sigma})$, respecting the fibers in Proposition 2.2. 
Further ${\bf A}_{\sigma}$ is well-defined as a bounded linear and 
selfadjoint operator on $L^2_{\sigma}(\cN)$ for all $\sigma \in 
\widehat{G}_d$ and almost all $\sigma \in  \widehat{G}_r\backslash 
\widehat{G}_d$ (called non-exceptional). The associated 
norms satisfy 
\be
{\rm ess} \sup_\sigma\Vert {\bf A}_\sigma\Vert\ = \Vert \bf A \Vert\ .
\label{asigmabound} 
\end{equation} 
\item[(b)] $\lb \in$ ess-Spec$(\bA)$=Spec$(\bA)$ iff for all $\eps >0$ the 
set $\{ \sigma \in \widehat{G}_r \,|\, \mbox{Spec}(\bA(\sigma)) \cap 
(\lb - \eps, \lb + \eps) \neq \emptyset \}$ has positive measure, where
$\bA(\sigma) = \1 \otimes \check{\bA}_{\sigma}$.
Further $\lb$ is an eigenvalue of $\bA$ iff the set 
$\{ \sigma \in \widehat{G}_r \,|\, \lb \;\mbox{is an eigenvalue of}\;
\bA(\sigma) \}$ has positive measure. 
\end{itemize}
\end{proposition} 
\begin{proof}
(a) The decomposition has been just explained; the boundedness a.e. as 
well as (\ref{asigmabound}) follow from the general theory of direct 
integral decompositions of operators (see \cite{folland} or 
\cite{RS4}, Theorem XIII.83). For the sake of completeness we sketch the 
argument:

For all $\psi\in L^2(\cM_r)$ one has from (\ref{psi1}) and (\ref{fiberop}) 
\be 
({\bf A} \psi)(g,n) = \int_{\widehat{G}_r} \! d\nu(\sigma) \,
\Tr\ [{\bf A}(\sigma) \widehat{\psi}(\sigma,n) \pi_{\sigma}(g)]\,.
\label{Aprop1}
\end{equation}
The norm of ${\bf A} \psi$ is related to that of 
$\widehat{{\bf A}\psi}(\sigma,n)$ by the Parseval identity 
(\ref{Planchpsi})  
\be 
({\bf A} \psi, {\bf A} \psi) = 
\int_{\widehat G_r} \! d\nu(\sigma) \,
({\bf A}(\sigma) \widehat{\psi}(\sigma),{\bf A}(\sigma) 
\widehat{\psi}(\sigma))_{\sigma\check\sigma}\ ,
\label{Aprop2}
\end{equation}
with $(\;,\;)_{\sigma\check{\sigma}}$ defined in (\ref{LNdef1}).
Since ${\bf A}$ is a bounded operator the left hand side 
of (\ref{Aprop2}) is finite, hence the integrand 
on the right hand side is finite $\nu$-almost everywhere. The 
bound (\ref{asigmabound}) follows by choosing sequences of $\psi$ such 
that their Fourier transforms $\widehat \psi(\sigma,n)$ are becoming 
concentrated at a particular value $\sigma$ (see \cite{folland}, 
Proposition 
7.33). Further $\bA(\sigma)$ is symmetric and has a unique selfadjoint 
extension for a.e.~$\sigma$, which we denote by 
the same symbol. Since ${\bA}(\sigma) = \1 \otimes \check\bA_{\check\sigma}$, 
the same holds for the fiber operators ${\bf A}_{\sigma}$.

(b) The statements about the spectrum in general as well as the 
eigenvalues follow as in Theorem XIII.85 of \cite{RS4}. 
\end{proof}

From our analysis we have obtained the following: The Hilbert space 
$L^2(\cM_r)$ is resolved into the direct integral $\int^\oplus 
d\nu(\sigma)\cL_\sigma\otimes\check\cL_{\check\sigma}\otimes L^2(\cN)$. 
Likewise $\bA$ can be 
resolved into a direct integral $\int^\oplus\bA(\sigma)d\nu(\sigma)=
\int^\oplus\1_{\cL_\sigma}\otimes \check \bA_{\check\sigma}$. The 
fiber operators $\bA_\sigma$ are ($\nu$ a.e.) selfadjoint and therefore 
have a spectral resolution, which leads to a direct integral 
decomposition of the spaces $L^2_\sigma(\cN)$ as well as of $\bA_\sigma$
and $\bA(\sigma)$:
\ba
L^2_\sigma(\cN)_0&=& \int^\oplus d\mu_\sigma(\lb) \cL^2_{\lb\sigma}(\cN)\,,
\nonum
\bA_\sigma&=&\int^\oplus d\mu_\sigma(\lb)\lb\ \1_{\cL^2_{\lb\sigma}(\cN)}\,,
\label{globdecomp1}
\end{eqnarray}
as well as
\ba
\cL^2_\sigma(\cM_r)&=&\int^\oplus 
d\mu_\sigma(\lb)\, \cL_{\sigma}\otimes \cL^2_{\lb\check\sigma}(\cN)\,,
\nonum
\bA(\sigma)&=&\int^\oplus d\mu_\sigma(\lb)\lb\ 
\1_{\cL_{\sigma} \otimes \cL^2_{\lb \check \sigma}(\cN)}\,.
\label{globdecomp2}
\end{eqnarray}
Combining these decompositions, we have 
\ba
L^2(\cM_r)=\int^\oplus d\nu(\sigma) d\mu_\sigma(\lb)\, 
\cL_{\sigma}\otimes \cL^2_{\lb\check\sigma}(\cN)\,, 
\nonum 
\bA=\int^\oplus d\nu(\sigma)d\mu_\sigma(\lb)\lb
\,\1_{\cL_{\sigma} \otimes \cL^2_{\lb \check\sigma}(\cN)}\, .
\end{eqnarray}
So we have identified the measure $d\mu(\lb,\sigma)$ in (\ref{nuclear1}) 
as $d\nu(\sigma)\ d\mu_\sigma(\lb)$.

%%%%%%%%%%%%%%%%%%%%%%%%%%%%%%%%%%%%%%%%%%%%%%%%%%%%%%%%%%%%%%%%%%%%
\newsubsection{Relating the spectral problems of $\bA$ and 
$\bA_{\sigma}$: $\bA_\sigma$ compact}

Our goal is to analyze concretely how the spectral resolution of the 
fiber operators $\bA_\sigma$ relates to the spectral problem of $\bA$, 
including the construction of generalized eigenvectors. At first we 
consider an important special case in which the kernel of $\bA$ is 
`almost' square integrable implying that the fiber operators 
$\bA_\sigma$ are compact. Explicitly we require: 
\begin{itemize}
\item[{(C)}] $\int\! dg dn dn'\, |\cA(g,n,n')|^2 < \infty$, or equivalently 
$\int_{Q^{\nu-1}}\! dq \int dm |A(q^{\uparrow}, q, m)|^2  < \infty$.
\end{itemize}
Here $A(m,m')$ is the kernel of the original operator acting 
on $L^2(\cM)$ while $\cA$ is the kernel of its image under the 
isometry in Proposition 2.1. Of course since $A(m,m')$ is invariant it 
can never be square integrable in the proper sense, see 
Proposition 3.1. However (C) implies that the fiber operators are 
Hilbert-Schmidt for almost all $\sigma$. 

\begin{lemma}\label{Ascompact}
Let $\bA$ be a standard invariant selfadjoint operator as in Definition 
3.1 whose kernel in addition satifies condition (C). 
Then the fiber operators ${\bf A}_{\sigma}$ and $\bA(\sigma)$ are 
Hilbert-Schmidt for all $\sigma \in \widehat{G}_d$ and for almost all
$\sigma \in \widehat{G}_c$. These $\sigma \in \widehat{G}_r$ are called 
non-exceptional.
\end{lemma}

\begin{proof} The Parseval identity (\ref{Planchpsi}) applied to 
$\cA(g,n,n')$
gives
\be
\int\!dg |\cA(g,n,n')|^2 =
\int_{\widehat{G}_r} \!d\nu(\sigma) \, {\rm Tr}[
\widehat{\cA}(\sigma,n,n')^\dagger \widehat{\cA}(\sigma,n,n')]\,.
\label{AfiberHS2}
\end{equation}
The integral over $n,n'$ of the left hand side is finite by assumption;
that of the right hand side can be expressed in terms of
the Hilbert-Schmidt norm of the integral operator $\bA(\sigma)$
and gives $\int d\nu(\sigma) \Vert \bA(\sigma)\Vert^2_2$.
Thus $\bA(\sigma)$ must have finite Hilbert-Schmidt norm for almost all 
$\sigma \in \widehat{G}_r$, as asserted. The equivalence to the second 
condition in (C) can be seen from (\ref{TrT}). In detail
\ba
\label{AfiberHS3}
&& \int\! dg dn dn' |\cA(g,n,n')|^2
= \int\! dg dn dn' |A(g q^{\uparrow}, g n, q_0, n')|^2
\\
&&  = \int \! d\gamma_Q(q_1)  dn dn'
|A(q_1, g_s(q_1) n, q^{\uparrow}, n')|^2
= \int_{Q^{\nu-1}} \!\!dq \int\! dm |A(q^{\uparrow}, q, m)|^2 \,.
\nonumber
\end{eqnarray}
In the second equality we used that the fact that the section $g_s: Q \ra
G$ provides a one-to-one correspondence  between points in $Q$ and
right $K$ orbits in $G$. By definition $(q_1, g_s(q_1) n) = (q_1, q_2,
\ldots q_{\nu}) = m$ and the $d\gamma_Q(q_1)  dn =
d\gamma(m)$ integral just defines the iteration of the modulus of the kernel.
\end{proof}

The spectral problem for the fiber operators $\bA_\sigma$ and 
$\bA(\sigma)$ is now trivial: these operators have discrete spectrum 
except for 0, which is an accumulation point of eigenvalues of finite 
multiplicity. All eigenvectors lie in the respective $L^2$ spaces, 
i.e.~in $L^2_{\sigma}(\cN)$ for $\bA_{\sigma}$ and in 
$L^2_{\sigma\check\sigma}(\cN)$ for $\bA(\sigma)$. 
Since $\bA(\sigma)$ is of the form $\1 \otimes 
\check{\bA}_{\check\sigma}$ a dense set of eigenvectors 
$F : \cN \ra \cL_{\sigma} \otimes \check{\cL}_{\check\sigma}$   
exists such that $F(n)$ is trace class (not just Hilbert-Schmidt 
as in the definition of $L^2_{\sigma\check\sigma}(\cN)$. 
The goal in the following is to `lift' these normalizable 
eigenvectors of the fiber operators to non-normalizable 
$\sigma$-equivariant eigenfunctions of $\bA$. To this end the 
action of $\bA$ 
has to be extended to functions outside of $L^2(\cM_r)$. Our
Definition 3.1 ensures that $\bA$ can be extended naturally as an 
operator from $L^\infty(\cM_r)$ to itself, bounded in the sup norm.
In order to relate the eigenvectors of the fiber operators to the 
eigenfunctions of $\bA$ (which will no longer be square integrable), we 
need a generalization of the Fourier transformation. 

In preparation we introduce two pairs of Banach spaces as follows:

\begin{definition} 
\begin{eqnarray*}
&& \nspace L^{1,2}(G\times \cN):= L^1(G)\otimes L^2(\cN)\,,
\quad \;\;
\Vert \phi\Vert^{1,2}:= \Big[ \int\! dn 
\big(\int\! dg |\phi(g,n)| \big)^2 \Big]^{1/2}\,,
\\
&& \nspace L^{\infty,2}(G\times \cN):= L^\infty(G)\otimes 
L^2(\cN)\,,
\quad
\Vert t\Vert^{\infty,2}:= \Big[ \int\! dn 
\big( \sup_g |t(g,n)| \big)^2 \Big]^{1/2}\,.
\label{LGNdef}
\end{eqnarray*}
Further, for $\sigma \in \widehat{G}_r$
\be 
L^2(\cN, {\cC}(\cL_{\sigma}))  
:=\Big\{ C : \cN \ra {\cC}(\cL_{\sigma}) \,\Big|\, 
\int \! dn \Vert C(n) \Vert^2 < \infty \Big\}\,,
\label{L2NC}
\end{equation}
where $\cC(\cL_{\sigma})$ is the space of compact operators  on 
$\cL_{\sigma}$ and $\Vert C(n) \Vert$ is the operator norm. 
Similarly
\be 
L^2(\cN, {\cal J}_1(\cL_{\sigma}))  
:=\Big\{ F : \cN \ra {\cal J}_1(\cL_{\sigma}) \,\Big|\, 
\int \! dn \Vert F(n) \Vert_1^2 < \infty \Big\}\,,
\label{L2NJ}
\end{equation}
where  ${\cal J}_1(\cL_{\sigma})$ is the space of trace class 
operators on $\cL_{\sigma}$ and $\Vert F(n) \Vert_1 = {\rm Tr}[|F(n)|]$ 
is the trace norm of $F(n)$. 
\end{definition}

Denoting by $'$ the topological duals one has 
\begin{subeqnarray}
L^{\infty,2}(\cN \times G) \is  L^{1,2}(G \times \cN)' 
\\
L^2(\cN, {\cal J}_1(\cL_{\sigma})) \is L^2(\cN, 
{\cC}(\cL_{\sigma}))'\,.
\label{LNduals} 
\end{subeqnarray} 
Of course the spaces are not reflexive, the double duals are 
much larger than the original spaces. The relations (\ref{LNduals}) 
follow from the well-known facts $L^1(G)' = L^{\infty}(G)$ 
and $\cC(\cL_{\sigma})' = {\cal J}_1(\cL_{\sigma})$, see 
e.g.~\cite{RS}, Theorem VI.26, for the latter. 
Concretely one has $|(t,\phi)| \leq  \Vert t\Vert^{\infty,2} 
\Vert \phi \Vert^{1,2}$, $t \in L^{\infty,2}$, $\phi \in L^{1,2}$. 
Thus $(t,\,\cdot\,)$ defines a linear bounded functional on 
$L^{1,2}$; moreover every such functional is of that form. 
Similarly,   
\be 
(F,C)_{\sigma\check\sigma} = \int \! dn {\rm Tr}[F(n)^{\dagger} C(n)] 
\leq \Big[\int \! dn \Vert F(n)\Vert_1^2 \Big]^{1/2} 
\Big[\int \! dn \Vert C(n)\Vert^2 \Big]^{1/2}\,,
\end{equation} 
so that $(F, \,\cdot\,)_{\sigma\check\sigma}$ defines a linear 
bounded functional on $L^2(\cN, \cC(\cL_{\sigma}))$; moreover 
every such functional is of this form. 

The two pairs of spaces (\ref{LNduals}a,b) are related by the generalization 
of the Fourier transform we are aiming for. To see this we introduce

\begin{definition} A function $t = t_{\sigma} \in L^{\infty,2}(G \times \cN)$  
is called $\sigma$-equivariant for some $\sigma \in \widehat{G}_r$,
if there exists a $\ell_{\sigma} \in L^2(\cN, \cC(\cL_{\sigma}))'$  
such that 
\be 
(t_{\sigma}, \phi) = \ell_{\sigma}(\widehat{\phi}(\sigma))\quad 
\mbox{for all} \;\;\phi \in L^{1,2}(G \times \cN)\,,
\label{tell}
\end{equation}  
where $\widehat{\phi}(\sigma,n)$ is the Fourier transform (\ref{psi1})
of $\phi$ (which is $\cC(\cL_{\sigma})$-valued for 
$\phi(\,\cdot\,,n) \in L^1(G)$). 
Such  $t_{\sigma}$ intertwine the right $G$ action $\rho$ 
on $L^{\infty}(G)$ with that of $\pi_{\sigma}$, i.e. 
\be
(\rho(g^{-1}) t_{\sigma},\phi)= \ell_{\sigma}( 
\pi_{\sigma}(g)\widehat\phi)\ ,
\end{equation}
 for all $g \in G$ and $\phi\in L^{1,2}(G 
\times \cN)$.   
The $\sigma$-equivariant subspace of $L^{\infty,2}(G \times \cN)$ is 
denoted by $\cL_{\sigma}^{\infty,2}(G \times \cN)$. 
\end{definition}

\begin{proposition} \label{sigmaeq}
Every $\sigma$-equivariant $t_{\sigma} 
\in L^{\infty,2}(G \times \cN)$ is of the form 
\be
t^F_\sigma(g,n)= \Tr[F(n)\pi_\sigma(g)]\,,
\label{Omsigma}
\end{equation}
for a unique $F \in L^2(\cN, {\cal J}_1(\cL_{\sigma}))$. 
Equivalently the map $t_{\sigma}:L^2(\cN, {\cal J}_1(\cL_{\sigma})) 
\ra \cL_\sigma^{\infty,2}(G \times \cN)$, $F \mapsto 
t_{\sigma}^F$, is a bounded injection. The 
intertwining relation becomes $t_{\sigma}^{\pi_{\sigma}(g_0) F}(g,n)= 
t_{\sigma}^F(gg_0, n)$. A substitute for the Parseval relation is  
\be 
(t_{\sigma}^F,\phi) = (F, \widehat{\phi}(\sigma))_{\sigma\check\sigma}\,,
\label{Parseval}
\end{equation}
for all $\phi\in L^{1,2}(G\times\cN)$.
\end{proposition}

\begin{proof}
We begin by showing that the map $F(n) \mapsto t^F_{\sigma}(\,\cdot\,,n)$ 
is injective for almost all $n\in\cN$.   

Here we can appeal to a theorem due to Glimm \cite{glimm,folland} that implies 
that for a type I group (which we have here) the closure in operator norm 
of the algebra generated by $\pi_\sigma (g),\ g\in G$ consists of all 
compact operators on $\cL_\sigma$. Then if $ t_{\sigma}^F(\phi)=0$ for 
all $\phi\in \cD(G)$ also $\Tr[F\pi_\sigma(g)]=0$ for all $g\in G$ and by 
the above mentioned theorem $\Tr[FA]=0$ for all compact $A$, which 
implies $F=0$.

A slightly more down to earth argument goes as follows: Assume 
again that $\Tr F\pi_\sigma(\phi)=0$ for all $\phi\in \cD(G)$. Replacing 
$\phi$ by $E_\kappa*\phi*E_{\kappa'}$ we see that $\Tr[P_\kappa F 
P_{\kappa'}\pi_\sigma(\phi)]=0$ for all $\phi\in \cD(G)$ and $\kappa,\kappa'$.
Note that  ${\rm Ran} P_\kappa=m_\kappa V_\kappa$ and ${\rm Ran} 
P_\kappa'= m_{\kappa'} V_{\kappa'}$ have finite dimension.  
By a theorem of Kadison \cite{kadison}, the norm closure of the algebra 
generated by $\pi_\sigma (g),\ g\in G$, contains operators that map any 
finite set of linearly independent vectors into any other given finite 
set of vectors of the same cardinality. This means that for any linear 
map $L_{\kappa',\kappa}$ from $m_\kappa V_\kappa$ to  
$m_{\kappa'} V_{\kappa'}$ there is an element $A$ in that norm closure 
such that $L_{\kappa',\kappa}= P_{\kappa'} A P_{\kappa}$. Hence 
$\Tr[FL_{\kappa',\kappa}]=0$ for any such linear map, which implies $F=0$.

To complete the proof, let $\ell_{\sigma} \in 
L^2(\cN, \cC(\cL_{\sigma}))'$ be the element associated with 
$t_{\sigma}$ via (\ref{tell}). On account of (\ref{LNduals}b) there exist some 
$F \in L^2(\cN, {\cal J}_1(\cL_{\sigma}))$ such that 
$\ell_{\sigma}(C) = (F^{\dagger},C)_{\sigma\check\sigma}$ for 
all $C \in L^2(\cN, \cC(\cL_{\sigma}))$. This holds in particular for 
$C(\sigma,n) = \widehat{\phi}(\sigma,n)$ for $\phi \in L^1(G) \otimes 
L^2(\cN)$.
%On the other hand by 
%Glimm's theorem all of $L^2(\cN, \cC(\cL_{\sigma}))$ arises 
%as the image of $L^1(G) \otimes L^2(\cN)$ under $\pi_{\sigma}$. 
%We can thus always assume $C$ to be of the form 
%$C(\sigma,n) = \widehat{\phi}(\sigma,n)$ for some $\phi \in 
%L^1(G) \otimes L^2(\cN)$. 
A simple computation then gives 
$\ell_{\sigma}(\widehat{\phi}(\sigma)) = (t_{\sigma}^F, \phi)$, 
with $t_{\sigma}^F$ as in (\ref{Omsigma}). The uniqueness of $F$ 
follows the above injectivity result. For the  norm of $t_{\sigma}^F$ one 
has $\Vert t_{\sigma}^F \Vert^{\infty,2} \leq 
(\int \! dn \Vert F(n) \Vert_1^2 )^{1/2}$. 
\end{proof} 

We add some remarks. Proposition \ref{sigmaeq} allows to give the space 
$\cL_\sigma^{\infty,2}(G\times \cN)$ the structure of a pre-Hilbert space 
by defining a scalar product
\be
(t^F_\sigma, t^{F'}_\sigma):= \int dn \Tr[F(n)^\dagger F'(n)]\ ;
\label{sigmascalar} 
\end{equation}
and of course by completion this gives rise to a Hilbert space, which can 
be identified with $L^2(\cN,\cJ_2(\cL_\sigma))$, where $\cJ_2(\cL_\sigma)$ 
is the space of Hilbert-Schmidt operators on $\cL_\sigma$.

Comparing (\ref{Omsigma}) with (\ref{psi1}) 
one sees that $F$ can be viewed as the unique Fourier transform 
of $t = t_{\sigma}$, but in a fixed $\sigma$ fiber (it is a `$\delta$ 
function' in $\widehat G_r$). 
The equivariance property (\ref{tell}) characterizes the fiber. 
Equivalently the $\sigma$-equivariant subspace of 
$L^{\infty,2}(G \times \cN)$ can be identified with 
$L^2(\cN, {\cal J}_1(\cL_{\sigma}))$ by the bounded injection
$t_{\sigma}$. The map $t_{\sigma}$ can also be turned into an 
isometry by transporting the norm to its image.

Since $L^2(\cN, {\cal J}_1(\cL_{\sigma}))$ carries the 
representation $\pi_{\sigma}\times \pi_{\check\sigma},\,
\sigma \in \widehat{G}_r$, of $G \times G$, where 
$(\pi_{\sigma} \times \pi_{\check\sigma})(g_0,g_1)F(n) = 
\pi_{\sigma}(g_0) F(n) \pi_{\sigma}(g_1^{-1})$,
the intertwining property in Proposition 3.5 in 
principle generalizes to $(\pi_{\sigma} \times 
\pi_{\check\sigma})(g_0,g_1)\,t^F_{\sigma}(g,n) 
= t^F_{\sigma}(g_1^{-1} g g_0, n)$. However since we eventually 
are interested in functions on $\cM_r = (G \times \cN)/d(K)$,
only left actions with $g_1 = k \in K$ are useful.    

Whenever $F(n)$ satisfies the 
equivariance condition (\ref{LMred1}) the function $t^F_{\sigma}(g,n)$ 
projects to one on $\cM_r$, i.e.~$t^F_{\sigma}(kg,kn)=
t^F_{\sigma}(g,n)$ for $k\in K$. We denote the left $K$ invariant 
subspace of $L^{\infty,2}(G \times \cN)$ by $L^{\infty,2}(\cM_r)$ 
and the subspace of $L^2(\cN, {\cal J}_1(\cL_{\sigma}))$ whose 
elements satisfy  (\ref{LMred1}) by  $L^2(\cN, {\cal J}_1(\cL_{\sigma}))_0$. 
Proposition 3.5 evidently remains valid as an injective identification 
of $L^2(\cN, {\cal J}_1(\cL_{\sigma}))_0$ with the $\sigma$-equivariant 
subspace of $L^{\infty,2}(\cM_r)$, for which we write 
$\cL_{\sigma}^{\infty,2}(\cM_r)$.   
\smallskip

To proceed we now show that in addition to intertwining the group 
actions, the map $t_{\sigma}$ also intertwines the action of $\bA$ 
with that of $\bA(\sigma)$. 
Indeed, using {\it only} the definitions
(\ref{Tr3}), (\ref{TTsigma2}), (\ref{Omsigma}) one computes
\ba
[\bA t^F_{\sigma}](g,n) \is  \int\! dn'\, \Tr  
\big[ F(n') \ \widehat \cA(\sigma,n',n) \,\pi_{\sigma}(g)\big]
\nonum
\is \Tr\big[ (\bA(\sigma) F)(n)\pi_{\sigma}(g) \big] = 
t_{\sigma}^{\bA(\sigma)F}(g,n)\,.
\label{TTsigma}
\end{eqnarray}
If the Fourier transform were well-defined on these functions
the relation (\ref{TTsigma}) would amount to
$\widehat{ \bA t_{\sigma}^F} = \bA(\sigma) F$, just as in  
(\ref{TTsigma3}), (\ref{fiberop}). The point of (\ref{TTsigma})
is that it remains valid for $\bA$ acting on 
$\cL^{\infty,2}_{\sigma}(G \times \cN)$
and $\bA(\sigma)$ acting on $L^2_{\sigma\check\sigma}(\cN)$.

Since $n \mapsto F(n)$ takes values in the trace class operators and
the kernel $\widehat{\cA}$ acts from the right one can expand
$F(n)$ into rank 1 operators of the form
\be
F(n) = \sum_{i=1}^\infty v_i \otimes \check{f}_i(n)\,,
\label{rank1a}
\end{equation}
with $v_i \in \cL_{\sigma}$ and $f_i: \cN \ra \cL_{\sigma}$
in $L^2_{\sigma}(\cN)$. Mostly it is therefore the restriction 
of $t_{\sigma}$ to rank $1$ operators $F(n) = v \otimes \check{f}(n)$
that is needed and it is convenient to introduce a separate notation 
for it. We define 
\ba
\label{Xdef}
&& \tau_{v\sigma} : L^2_{\sigma}(\cN) \ra
\cL_{\sigma}^{\infty,2}(G\times \cN)\,,\quad f \mapsto \tau_{v\sigma}(f) \,,
\\[2mm]
&& \tau_{v\sigma}(f)(g,n) := t_{\sigma}^{v \otimes \check{f}}(g,n)
= \Tr\big[\!\left(v\otimes
\check f(n)\right) \pi_\sigma(g)\big]
=(f(n), \pi_\sigma(g) v)_{\sigma}\,.
\nonumber
\end{eqnarray}
Note that the map is antilinear in $f$ but linear in $\check f$,
and that $t_{\sigma}^F(g,n) = \sum_i \tau_{v_i \sigma}(f_i)(g,n)$, 
for $F$ of the form (\ref{rank1a}). 
Whenever $f$ is $K$ equivariant the image function is 
left $K$ invariant, i.e.~(\ref{Xdef}) also defines a map 
$\tau_{v\sigma}: L_{\sigma}^2(\cN)_0 \ra \cL_{\sigma}^{\infty,2}(\cM_r)$. 
It is straightforward to see that the map is bounded in the norm 
$\Vert \,\cdot\,\Vert^{\infty,2}$. It is thus a bounded injection and can 
be made into an isometry onto its image by transferring the norm to the image.
The representation $\pi_{\sigma}$ acts on rank 1 valued operators merely
by rotating the reference vector: $v \otimes \check{f}(n) \mapsto
(\pi_{\sigma}(g_0)
v) \otimes \check f(n)$, so that
$\tau_{v\sigma}(f)(gg_0,n) =
\tau_{\pi_{\sigma}(g_0) v,\sigma}(f)(g,n)$. 
Further $\tau_{v\sigma}$ interwines the action of
$\bA$ and $\bA_{\sigma}$: either from (\ref{TTsigma}) or directly  
from (\ref{TTsigma2}), (\ref{Tsigma1}) one finds
\be
(\bA \tau_{v\sigma}(f))(g,n)=(\bA_\sigma f_{\sigma}(n),
\pi_\sigma(g)v)_{\sigma} =
(\tau_{v\sigma}(\bA_{\sigma}f))(g,n) \,.
\label{factor2}   
\end{equation}

This intertwining relation now allows to reduce the spectral problem
for ${\bf A}$ to the simpler spectral problems of the
fiber operators $\bA_{\sigma}$ in the decomposition ${\bf A} =
\int^{\oplus} \!d\nu(\sigma) (\1 \otimes {\check\bA}_{\check\sigma})$.
We write
\begin{subeqnarray}
\label{ECdef}
&& \cE_{\lb,\sigma}({\bf A}) =
\Big\{ \Omega_{\sigma} \in \cL_{\sigma}^{\infty,2}(\cM_r) \,
\Big|\, \Omega_{\sigma}\;\mbox{is eigenfunction of}
\\[-4mm] 
&& \makebox[47mm]{} \;\;\;\;{\bf A}\;\;
\mbox{with spectral value} \;\;\lb \in {\rm Spec}({\bf A})\Big\}\,,
\nonum
&& \cE_{\lb}({\bf A}_{\sigma}) =
\Big\{\chi_{\sigma} \in L^2_{\sigma}(\cN)_0 \,
\Big|\, \chi_{\sigma}\;\mbox{is eigenvector of}
\\[-4mm]
&& \makebox[47mm]{} \;\;\;\;{\bf A}_{\sigma}\;\;
\mbox{with eigenvalue} \;\;\lb \in \mbox{Spec}({\bf 
A}_{\sigma})\Big\}\,,
\nonumber
\end{subeqnarray}
and similarly for $\cE_{\lb}(\check\bA_{\check\sigma})\subset
L^2_{\check\sigma}(\cN)$ and
$\cE_{\lb}(\bA(\sigma)) \subset L^2(\cN, {\cal J}_1(\cL_{\sigma}))_0$.

The space $\cE_{\lb,\sigma}(\bA)$ inherits the pre-Hilbert space 
structure from $\cL_\sigma^{\infty,2}(\cM_r)$ and we denote its completion 
by $\overline{\cE_{\lb,\sigma}(\bA)}$, whereas $\cE_{\lb}(\bA_\sigma)$ is 
already a Hilbert space.

The relation of these spectral problems is described by
\begin{proposition}
\label{relacomp}
Let $\bA$ be a standard  invariant selfadjoint operator on $L^2(\cM_r)$ 
in the 
sense of Definition 3.1 and assume that its kernel satisfies in addition 
condition (C). Then the fiber operators $\bA(\sigma)$, $\bA_{\sigma}$ 
are Hilbert-Schmidt for all non-exceptional $\sigma \in \widehat{G}_r$ 
by Lemma 3.2. Further for these $\sigma \in \widehat{G}_r$  
\begin{itemize}
\item[(a)] For each $v\in\cL_\sigma$ the map $\tau_{v\sigma}: 
\cE_{\lb}(\bA_{\sigma}) \ra
\cE_{\lb,\sigma}(\bA)$ is a bounded injection. 
\item[(b)] If ${\bA}_{\sigma}$, $\sigma \in \widehat{G}_d$,  
has an eigenvector (normalizable eigenfunction) for some $\lb$,
the associated generalized eigenfunction in (a) of ${\bf A}$
is normalizable (i.e.~$\lb$ is an eigenvalue of $\bA$ of infinite 
multiplicity).
\item[(c)] A complete orthonormal set of eigenfunctions of $\bA(\sigma)$  
in $\cE_\lb(\bA(\sigma))$ is given by
\be
\Big\{e_i\otimes\check\chi_{\lb\sigma r}\Big\}\ ;
\label{tensorbasis}
\end{equation}
a complete orthonormal set of eigenfunctions $\bA$ in
$\overline{\cE_{\lb,\sigma}(\bA)}$ is given by
\be 
\Big\{\tau_{e_i \sigma}(\chi_{\lb\sigma r})(g,n) \Big\}\ ,  
\label{Omcompact}
\end{equation}
where $\chi_{\lb\sigma r}$ runs through an orthonormal basis of 
$\cE_\lb(\bA_\sigma)$ and $e_i$ runs through an orthonormal basis of 
$\cL_\sigma$. The eigenfunctions (\ref{Omcompact}) vanish pointwise for 
$g\to\infty$ (i.e.~as $g$ leaves any compact subset of $G$).  

$\overline{\cE_{\lb,\sigma}(\bA)}$ and
$\cE_{\lb}(\bA(\sigma))$ are isometric as Hilbert spaces.
In particular the spectrum of ${\bA}(\sigma)$ is contained in the
spectrum of ${\bA}$ for almost all $\sigma \in \widehat{G}_r$.
\item[(d)] An orthonormal basis of $L^2_\sigma(\cN)_0$ is given by 
\be
\Big\{\chi_{\lb\sigma r} \Big\}\ ,
\label{Omcompact1}
\end{equation}
where $\chi_{\lb\sigma r}$ runs through an orthonormal basis of 
$\cE_\lb(\bA_\sigma)$, and $\lb$ runs through the eigenvalues of 
$\bA_\sigma$.
An orthonormal basis of $L^2(\cN,\cJ_2(\cL_\sigma))_0$ is given 
by
\be
\Big\{e_i\otimes\check \chi_{\lb\sigma r}\Big\}\ ,
\label{Omcompact2}
\end{equation}
where in addition $e_i$ runs through an orthonormal basis of 
$\cL_\sigma$.
\item[(e)] The set of all eigenfunctions 
$\bigcup_{\lb\sigma}\cE_{\lb,\sigma}(\bA)$ is complete in the sense that 
for each $\phi\in (L^2\cap L^1)(\cM_r)$ the following Parseval relation 
holds:
\be
(\phi,\phi)= \int 
d\nu(\sigma)d\mu_\sigma(\lb)
\sum_{i,r}\int dn \bigl|\bigl(\chi_{\lb\sigma r}(n),
\widehat\phi(\sigma,n)e_i\bigr)_\sigma\bigr|^2\ ,
\end{equation}
where the measure $d\mu_\sigma(\lb)$ introduced in (\ref{globdecomp1}) 
here is just the counting measure of the eigenvalues $\lb$ of 
$\bA_\sigma$.
\end{itemize}
\end{proposition}
\begin{proof}
(a) Since $\bA_\sigma$ is selfadjoint, we can assume that $\lb$ is 
real. Clearly, if $\chi \in \cE_{\lb}(\bA_{\sigma})$,
i.e.~$\bA_{\sigma}\chi
= \lb \chi$, then $\bA \tau_{v\sigma}(\chi) =
\tau_{v\sigma}(\bA_{\sigma}\chi) =
\lb \tau_{v\sigma}(\chi)$,
i.e.~$\tau_{v\sigma}(\chi) \in \cE_{\lb,\sigma}(\bA)$.
By Proposition 3.5 the map $\cE_\lb(\bA_\sigma)\ni\chi \mapsto 
\tau_{v\sigma}(\chi)\in\cE_{\lb,\sigma}(\bA)$ is injective.

(b) Recall from \cite{Dixmier1}, Theorem 14.3.3, that the coefficients
of a square integrable representation $\pi_{\sigma}$,
$\sigma \in \widehat{G}_d$, satisfy, using the $K$-adapted
basis (\ref{adaptb}) 
\be
\int\! dg \,\pi_{\sigma}(g)^*_{\kappa_1 s_1,\kappa_2 s_2}
\pi_{\sigma}(g)_{\kappa_3 s_3,\kappa_4 s_4} =
d_{\sigma}^{-1} \, \delta_{\kappa_1 \kappa_3}
\delta_{s_1 s_3} \delta_{\kappa_2 \kappa_4} \delta_{s_2 s_4}\,,
\label{pidiscrete}
\end{equation}
where $d_{\sigma}$ is the formal degree of $\pi_{\sigma}$.
Let $\chi \in L^2_{\sigma}(\cN)$
be the normalizable eigenfunction of ${\bA}_{\sigma}$
for some $\sigma \in \widehat{G}_d$, and consider the
associated generalized eigenfunction $\tau_{v \sigma}(\chi)$
of $\bA$. Using (\ref{pidiscrete}) one can verify by direct
computation that
\be
d_{\sigma}\! \int \!dg dn\,|\tau_{v\sigma}(\chi)(g,n)|^2 =   
(v,v)_{\sigma}(\chi,\chi)_{\sigma}\,,
\label{Aprop8}
\end{equation}
holds. It follows that $\tau_{v\sigma}(\chi)$ is actually
normalizable,
i.e.~an element of $L^2(\cM_r) \cap \cL^{\infty,2}_{\sigma}(\cM_r)$.
For the spectral values this implies: if $\lb \in
$ d-Spec$({\bf A}_{\sigma})$ then it is an eigenvalue of infinite
multiplicity of $\bA$.

(c) (\ref{tensorbasis}) follows directly from the completeness of the 
eigenfunctions of $\bA_\sigma$ and the relation between the operators 
$\bA(\sigma)$ and $\bA_\sigma$. The second statement follows from the fact 
that all the functions in $\overline{\cE_{\lb,\sigma}(\bA)}$  are 
$\sigma$-equivariant and hence by Proposition 3.5 of the form 
$\Omega_\sigma(g,n)=\Tr[F(n)\pi_\sigma(g))]$. The eigenfunction condition 
translates into the condition that $F$ is an eigenvector of $\bA(\sigma)$ 
with eigenvalue $\lb$; a complete set of such eigenvectors is given by
(\ref{tensorbasis}), as we have just seen. By construction the map 
$F\mapsto t^F_\sigma$ preserves the scalar product, which implies the 
orthonormality of the set (\ref{Omcompact}). So $t^F_\sigma$ provides an 
isometry between $\overline{\cE_{\lb,\sigma}(\bA)}$ and 
$\cE_{\lb}(\bA(\sigma))$. The vanishing at infinity in $G$ of the 
eigenfunctions (\ref{tensorbasis}) follows from the Howe-Moore theorem,
see \cite{Zimmer}, Theorem 2.2.20.

(d) This follows from the spectral theorem applied to $\bA_\sigma$ and
$\bA(\sigma)$ together with part (c). 

(e) Since $\phi\in L^2\cap L^1$, its partial Fourier transform 
$\widehat\phi(\sigma,n)$ is a.e.~a Hilbert-Schmidt operator in 
$\cL_\sigma$. The statement then follows from the completeness of the 
fiber eigenvectors (part (a)) and the Plancherel theorem of Appendix A1.
\end{proof}

We repeat a cautioning remark made in Section 2.2: the fact that 
$\lb_0\in {\rm Spec}(\bA_{\sigma_0})$ for some $\sigma_0$ and hence 
$\cE_{\lb_0,\sigma_0}(\bA)$ is nontrivial does in itself {\it not} imply that 
$\lb_0\in {\rm Spec}(\bA)$. In fact there are explicit 
counterexamples. By Proposition \ref{adecompose}b one needs a 
set $\Sigma$ of nonzero Plancherel measure, such that for each neighborhood 
$U$ of $\lb_0$ we have $\bigcup_{\sigma\in \Sigma} 
{\rm Spec}(\bA_\sigma)\cap U\ne \emptyset$. 

In the case of compact fiber operators $\bA_\sigma$ we have however obtained 
the following relation between the spectral problems of $\bA$ and $\bA_\sigma$: 
for every spectral value $|\lb|\le\Vert \bA\Vert$ of $\bA$ there is an 
eigenvalue of $\bA_\sigma$. The corresponding normalizable eigenvector 
$\chi$ of $\bA_{\sigma}$ generates a $G$ orbit of $\cL_\sigma^{\infty,2}$ 
eigenfunctions of $\bA$ via $\tau_{v\sigma}(\chi)$, and {\it all} 
$\sigma$-equivariant eigenfunctions of $\bA$ arise in that way by taking 
linear combinations in $\chi$ and the reference vectors $v \in 
\cL_{\sigma}$. In addition there may be `fake' solutions of $\bA \Omega = 
\lb \Omega$, where $\lb \neq {\rm Spec}(\bA)$. The value $|\lb|=\Vert 
A\Vert$ of course always belongs to the spectrum of $\bA$.

%%%%%%%%%%%%%%%%%%%%%%%%%%%%%%%%%%%%%%%%%%%%%%%%%%%%%%%%%%%%%%%%%%%%

\newsubsection{Relating the spectral problems of $\bA$ and
$\bA_{\sigma}$: $\bA_\sigma$ not necessarily compact}
\vspace{-3mm} 

The discussion becomes a little more complicated in the case of not 
necessarily compact fiber operators. These naturally arise if the 
original kernel is not `almost' square integrable as in 
condition (C). In this section we require instead the 
weaker conditions  
\begin{itemize}
\itemsep 3pt
\itemindent 10pt
\item[{(C1)}]
$\sup_m \int \! dm' |A(m,m')|^2 < \infty$\,,
\item[{(C2)}] There is an invariant multiplication operator $M$
given by a $d(G)$ invariant function in $L^2(\cM)$ also denoted by 
$M$, such that $M^{-1} \bA M$ defines a bounded (invariant) operator on 
$L^2(\cM)$.
%\item[{(C3)}] There is a left $K$ invariant function $M_G\in L^2(G)$ such 
%that $M^{-1}M_G^{-1}\bA M_G M$ defines a bounded operator on $L^2(\cM)$.  
\end{itemize}
For a positive kernel $A(m,m') >0$ the condition (C1) amounts to 
$\sup_m A(m,m;2) < \infty$, which is weaker than the integrability 
condition in Lemma 3.1 (at least as far as decay properties of $A$ are 
concerned). The multiplication operator acts by $(M\psi)(m) = 
M(m) \psi(m)$, for $\psi \in L^2(\cM)$. Invariance $M(d(g)m) = M(m)$, 
for all $g \in G$, means that $M$ maps into multiplication by a function 
$\tilde M$ on $\cN$, given by 
\be
\tilde M(n):= M(q^\up,n)
\end{equation}
under the isometry $\Phi^{-1}: L^2(\cM) \ra L^2(\cM_r)$ in (\ref{Phidef}). 
We drop the tilde and continue to write $M \in L^2(\cN)$ for this function 
and note that the kernel of $M^{-1} \bA M$, viewed as a bounded operator 
on $L^2(\cM_r)$ is given by $M(n)^{-1} \cA(g{g'}^{-1},n,n') M(n')$. 
The counterpart of Lemma 3.2 is now given by 

\begin{lemma} Under the conditions (C1), (C2) the fiber operators 
$\bA_{\sigma}$ have for all $\sigma \in \widehat{G}_d$ and for 
allmost all $\sigma \in \widehat{G}_c$ the properties   
\begin{itemize}
\itemsep 3pt 
\itemindent 10pt
\item[(C1')]
$\sup_n \int dn' \Vert \widehat\cA(\sigma,n,n')\Vert^2_2 = 
:C_A<\infty$, where $\Vert\,.\,\Vert_2$ denotes the Hilbert-Schmidt norm 
for operators on $\cL_\sigma$.
\item[(C2')]
there is a function $M\in L^2(\cN)$ such that $M^{-1} \bA_\sigma M$
defines a bounded operator on $L^2_{\sigma}(\cN)$, where $M$ maps each 
$L^2_\sigma(\cN)$ into itself via $(M\phi)(n):= M(n)\phi(n)$. 
\end{itemize}
\end{lemma} 
The goal in the following will be to first gain control over the 
eigenfunctions $\bA_{\sigma}$ (a step which was trivial in Section 3.3) 
and then to lift them to $\pi_{\sigma}$ equivariant generalized eigenfunctions 
of $\bA$. 
\begin{proof} 
(C1') follows from (C1) by applying (\ref{AfiberHS2}). (C2') is obvious.
\end{proof}

For the first step we will make use of the Gel'fand-Maurin theory of 
generalized eigenvectors and eigen\-spaces, in particular of a version due 
to Berezanskii as described (and proven) in Maurin \cite{maurin}. 
Berezanskii's construction, applied to the present situation, requires to 
set up for almost each $\sigma\in\widehat G_r$ a triplet
\be
\Phi_\sigma(\cN)\subset L^2_{\sigma}(\cN) \subset \Phi_\sigma'(\cN)\ ,
\end{equation}
in which $\Phi_\sigma = \Phi_{\sigma}(\cN)$ is the domain of an unbounded, 
densely defined closed operator $B_\sigma$ whose inverse is Hilbert-Schmidt 
and $\Phi'_{\sigma} = \Phi'_{\sigma}(\cN)$ is the topological dual of 
$\Phi_{\sigma}$. Choosing 
\be 
B_\sigma:= M^{-1} \bA_\sigma^{-1}\ ,
\label{Ber1}
\end{equation}
we compute 
\be
{\rm Tr}_{L_\sigma^2(\cN)}\left[(B_\sigma^{-1})^\dagger (B_\sigma^{-1}\right]
\le \int\! dn dn' |M(n)M(n')|\, \Vert\widehat \cA(\sigma,n,n')\Vert^2_2
\le\Vert M\Vert_2^2\, C_A\ ,
\label{Ber2}
\end{equation}
where in the last step we used property (C1)' and the fact that therefore
$\Vert \widehat\cA(\sigma,n,n')\Vert^2_2$ is the kernel of a bounded 
operator on $L_{\sigma}^2(\cN)$ with norm $\le C_A$, see \cite{Lax}, p.176. 
This shows that $B_\sigma^{-1}$ is indeed Hilbert-Schmidt. Property (C2)' now guarantees 
that $\bA_\sigma$ maps $\Phi_\sigma$ into itself. Indeed from  
$\bA_{\sigma} M L^2_{\sigma}(\cN) = M L^2_{\sigma}(\cN)$ one sees that 
$M L^2_{\sigma}(\cN)$ coincides both with the range of $B_{\sigma}^{-1} 
= \bA_{\sigma} M$ and with the domain of $B_{\sigma}$.

$\Phi_\sigma$ is a pre-Hilbert space with the scalar product 
\be
(\phi, \psi)_B:= (B_\sigma\phi, 
B_\sigma\psi)_{\sigma}+(\phi,\psi)_{\sigma}\ ; 
\label{Ber3}
\end{equation}
we denote its completion with respect to the norm $\Vert 
\phi_{\sigma}\Vert_B:= \sqrt{(\phi,\phi)_B}$  by $\overline\Phi_{\sigma}\subset 
L_\sigma^2(\cN)$. The dual of $\bA_\sigma$ acting on  $\Phi'_\sigma$ will 
also be denoted by $\bA_{\sigma}$. Berezanskii's theorem guarantees 
that $\bA_\sigma$ has a complete set of eigendistributions in $\Phi_\sigma'$. 
We now proceed to show that in the situation at hand these are in fact 
almost everywhere defined {\it functions} from $\cN$ with values in 
$\cL_{\sigma}$ (which are of course not square integrable). More precisely we 
have 

\begin{proposition}\label{fibereig}
Under the conditions (C1) and (C2) above, for almost all $\sigma \in  
\widehat{G}_r$ the fiber operator $\bA_\sigma$ has a complete set of 
generalized eigenfunctions  $\chi_{\lb\sigma r}$, where $r=1,\ldots 
g(\lb)$ with $g(\lb)\in \N\cup\{\infty\}$ denotes the 
multiplicity of the spectral value $\lb$. The generalized eigenfunctions 
satisfy for almost all $\lb \in {\rm Spec}(\bA_{\sigma})$ 
\begin{itemize}
\itemsep 3pt
\item[(a)] $\bA_\sigma\chi_{\lb \sigma}=\lb\,\chi_{\lb\sigma}$, 
\item[(b)] $M\chi_{\lb \sigma}\in L^2_\sigma(\cN)$. 
\item[(c)] For $\phi\in M L^2_\sigma(\cN)_0$ we have the Parseval 
(completeness) relation
\be
\int\! dn\, (\phi(n),\phi(n))_\sigma =\int\! dn \int\! d\mu_\sigma(\lb) 
\sum_{r=1}^{g(\lb)}
\Bigl|\bigl(\chi_{\lb\sigma r}(n),\phi(n)\bigr)_\sigma\Bigr|^2\ .
\end{equation}
\end{itemize}
We denote the space of these eigenfunctions by $\cE_{\lb}(\bA_{\sigma})$. 
\end{proposition}
\begin{proof}
We first show that (the dual of) $\bA_\sigma$ maps $\Phi'_\sigma$ into 
almost everywhere defined functions from $\cN$ with values in 
$\cL_{\sigma}$. More precisely, for any $\chi\in\Phi'_\sigma$ 
\be
M^\dagger \bA_\sigma\chi=(B_\sigma^{-1})^\dagger\chi\in L^2_\sigma(\cN)\, .
\label{Ber4}
\end{equation}
Since $\Phi_{\sigma}$ is a pre-Hilbert space, by the Riesz representation 
theorem each element $\chi$ of $\Phi'_{\sigma}$ can be represented by an 
element in $\overline{\Phi}_{\sigma}$. This means that there is a 
$\psi_\chi\in \overline{\Phi}_{\sigma} \subset L^2_\sigma(\cN)$ such that 
for all $\phi \in \Phi_{\sigma}$ 
\be
|(\chi,\phi)_{\sigma}|=|(\psi_\chi,\phi)_B|\le \Vert\psi_\chi\Vert_B  
\Vert\phi\Vert_B\, .
\label{Ber5}
\end{equation}
Replacing $\phi$ by $B_\sigma^{-1}\phi$, and using 
$\Vert B_{\sigma}^{-1} \phi\Vert_B^2 = (\phi,\phi)_{\sigma} + 
(B_{\sigma}^{-1}\phi, B^{-1}_{\sigma} \phi)_{\sigma} \leq 
(1 + \Vert B_{\sigma}^{-1} \Vert^2) \Vert \phi \Vert^2_{\sigma}$, gives  
\be 
|(\chi,B_\sigma^{-1}\phi)_{\sigma}|\le \Vert \psi_\chi\Vert_B \Vert 
B_\sigma^{-1}\phi\Vert_B \le
{\rm const} \Vert\psi_\chi\Vert_B \, \Vert\phi\Vert_{\sigma}\, .
\label{Ber6}
\end{equation}
This shows that $(\chi,B_\sigma^{-1}\phi)_{\sigma}$ defines a bounded linear 
functional on the dense domain $\Phi_{\sigma} \subset L^2_\sigma(\cN)$ 
and hence is given by an element of $L_\sigma^2(\cN)$, as claimed
in (\ref{Ber4}). Now (a) and (c) follow from  Berezanskii's theorem
and (b) follows from (a) and (\ref{Ber4}).   
\end{proof}
Consistency requires that the dual $\bA'_{\sigma}$ of $\bA_{\sigma}$ 
(with a separate notation only at this point) is such that $M 
\bA'_{\sigma} M^{-1}$ is a bounded operator 
on $L_{\sigma}^2(\cN)$, which is however just the relation dual to (C2)'. 

Let us add that it is not possible to apply this method directly 
to obtain the $\sigma$-equivariant generalized eigenfunctions   
of $\bA$ itself. As long as $\bA$ and $M$ are invariant the product 
$\bA M$ will never be Hilbert-Schmidt, as the infinite group 
volume is once more overcounted. One would have to replace the invariant 
multiplication operator $M$ by an operator of multiplication by a 
(necessarily noninvariant) function in $L^2(\cM_r)$, which would make the 
strategy at least cumbersome.

We rather proceed as in Section 3.3. In order to avoid having to 
introduce further notation we continue to use the spaces in 
Definition 3.2 but indicate the modified notion of square 
integrability by a mnemonic prefactor $M$ or $M^{-1}$. For example
\ba 
&& F \in M^{-1} L^2(\cN, {\cal J}_1(\cL_{\sigma})) 
\quad {\rm if} \quad \int \! dn |M(n)|^2 \Vert F(n)\Vert_1^2 < 
\infty\,,
\nonum
&& t \in M^{-1} L^{\infty,2}(G \times \cN) 
\quad \;\;\;\,{\rm if} \quad \int \! dn |M(n)|^2 \sup_g |t(g,n)|^2 < 
\infty\,.
\label{Ber7}
\end{eqnarray}
The $\sigma$-equivariant subspace of $M^{-1} L^{\infty,2}(G \times \cN)$
is denoted by $M^{-1} \cL_{\sigma}^{\infty,2}(G \times \cN)$, etc. 
We define $t_{\sigma}^F(g,n):= \Tr[F(n)\pi_\sigma(g)]$ as in 
(\ref{Omsigma}). With these modified spaces Proposition 3.5 carries 
over, and states that every element of $M^{-1} \cL_{\sigma}^{\infty,2}(G \times \cN)$, 
is of the form  $t_{\sigma}^F(g,n)$ with a unique $F \in  
M^{-1} L^2(\cN, {\cal J}_1(\cL_{\sigma}))$. 
The intertwining property for the $G$ actions is manifest and ${\bf 
A}t_{\sigma}^F = t_{\sigma}^{\bA(\sigma) F}$ carries over from 
(\ref{TTsigma}). Since $F(n)$ is trace class we can expand it as in 
(\ref{rank1a}) 
\be 
F(n) = \sum_i v_i \otimes \check{\chi}_i(n) \,,\sspace 
v_i \in \cL_{\sigma}\,,\;\;\chi_i \in M^{-1} L_{\sigma}^2(\cN)\,,
\label{Ber8}
\end{equation}
and define 
\ba 
&& \tau_{v \sigma}: M^{-1} L_{\sigma}^2(\cN) \rra 
M^{-1} \cL_{\sigma}^{\infty,2}(\cM_r)\,,
\nonum
&& \chi \mapsto \tau_{v\sigma}(\chi)\,,\sspace 
\tau_{v\sigma}(\chi)(g,n) = (\chi(n), \pi_{\sigma}(g)v)_{\sigma}\,,
\label{Ber9}
\end{eqnarray}
where again the spaces and their norms are defined in the obvious way.

In parallel to Eq.~(\ref{ECdef}) we write
\begin{subeqnarray}
\label{EMdef}
&& \cE_{\lb,\sigma}({\bf A}) = 
\Big\{ \Omega_{\sigma} \in M^{-1}\cL_{\sigma}^{\infty,2}(\cM_r) \,
\Big|\, \Omega_{\sigma}\;\mbox{is eigenfunction of}
\\[-4mm] 
&& \makebox[47mm]{} \;\;\;\;{\bf A}\;\;
\mbox{with spectral value} \;\;\lb \in {\rm Spec}({\bf A})\Big\}\,,
\nonum
&& \cE_{\lb}({\bf A}_{\sigma}) = 
\Big\{ \chi_{\sigma} \in M^{-1} L^2_{\sigma}(\cN)_0 \,
\Big|\, \chi_{\sigma}\;\mbox{is eigenfunction of}
\\[-4mm] 
&& \makebox[47mm]{} \;\;\;\;{\bf A}_{\sigma}\;\;
\mbox{with spectral value} \;\;\lb \in {\rm Spec}({\bf 
A}_{\sigma})\Big\}\,,
\nonumber
\end{subeqnarray}
and similarly for $\cE_{\lb}(\check\bA_{\check\sigma})\subset  
M^{-1} L^2_{\check\sigma}(\cN)_0$ and 
$\cE_{\lb}(\bA(\sigma)) \subset M^{-1} L^2(\cN, {\cal J}_1(\cL_{\sigma}))_0$. 

We obtain the following counterpart of Proposition \ref{relacomp}:
\begin{proposition}
\label{relanoncomp}
Let $\bA$ and $\bA(\sigma)$, $\bA_{\sigma}$ with $\sigma \in
\widehat{G}_r$ non-exceptional as in Proposition 3.4, and assume that 
$\bA$ satisfies conditions (C1), (C2). Then
\begin{itemize}
\item[(a)] $\tau_{v \sigma}: \cE_{\lb}(\bA_{\sigma}) \ra
\cE_{\lb,\sigma}(\bA)$ is a bounded injection.
\item[(b)] A complete set of generalized eigenfunctions of $\bA(\sigma)$ 
in $\cE_\lb(\bA(\sigma)$ is given by 
\be
\Big\{e_i\otimes\check\chi_{\lb\sigma r}\Big\}\ ;
\label{nctensorbasis}
\end{equation}
a complete set of generalized eigenfunctions of $\bA$ in
$\cE_{\lb,\sigma}(\bA)$ is given by
\be
\Big\{\tau_{e_i \sigma}(\chi_{\lb\sigma r})(g,n) \Big\}\ ,
\label{Omnoncompact}
\end{equation}
where $\chi_{\lb\sigma r}$ runs through the complete set of Propostion 
\ref{fibereig} and $e_i$ runs through an orthonormal basis of $\cL_\sigma$.
The eigenfunctions (\ref{Omnoncompact}) vanish at infinity in $G$. 
$\cE_{\lb,\sigma}(\bA)$ and $\cE_{\lb}(\bA(\sigma))$ are 
homeomorphic as Banach spaces, provided one uses on the image the norm 
transported from the preimage. In particular the spectrum of 
${\bA}(\sigma)$ is contained in the spectrum of ${\bA}$ for almost all 
$\sigma \in \widehat{G}_r$.  
\item[(c)]
A complete set of of generalized eigenfunction spanning 
$L^2_\sigma(\cN)_0$ is given by
\be
\Big\{\chi_{\lb\sigma r} \Big\}\ ,
\label{Omnoncomp1}
\end{equation}
where $\chi_{\lb\sigma r}$ runs through the complete set above 
and $\lb$ runs through the eigenvalues of $\bA_\sigma$.  A complete set of 
eigenfunctions spanning $L^2(\cN,\cJ_2(\cL_\sigma))_0$ is given by
\be
\Big\{e_i\otimes\check \chi_{\lb\sigma r}\Big\}\ ,
\label{Omnoncomp2}
\end{equation}
where in addition $e_i$ runs through an orthonormal basis of
$\cL_\sigma$.
\item[(d)] The set of all eigenfunctions
$\bigcup_{\lb\sigma}\cE_{\lb,\sigma}(\bA)$ is complete in the sense that
for each $\phi\in (L^2\cap L^1)(\cM_r)$ the following Parseval relation
holds:
\be
(\phi,\phi)= \int
d\nu(\sigma)d\mu_\sigma(\lb)
\sum_{i,r}\int dn \bigl|\bigl(\chi_{\lb\sigma r}(n),
\widehat\phi(\sigma,n)e_i\bigr)_\sigma\bigr|^2\ ,
\end{equation}
with the measure $d\mu_\sigma(\lb)$ introduced in (\ref{globdecomp1}).
\end{itemize}
\end{proposition}
\begin{proof} The proof parallels that of Proposition \ref{relacomp}. The 
only change is that the completeness of the set of eigenvectors has to be 
replaced by the completeness relation of Proposition \ref{fibereig} (c).
\end{proof}

Symbolically we can summarize the content of Propositions 3.6b and 
3.8b in the isometry 
\be 
\cE_{\lb ,\sigma}(\bA) \simeq \cE_{\lb}(\bA(\sigma))\,.
\label{Eiso}
\end{equation}
That is, under the conditions (C) or (C1),(C2) the 
space of eigenfunctions of the fiber operator $\bA(\sigma)$ is 
naturally isometric to the space of $\sigma$-equivariant eigenfunctions 
of $\bA$. 
When $\bA = \bt$ is a transfer operator and the spectral 
value is $\lb = \Vert \bt \Vert$, the generalized eigenspace
$\cE_{\Vert \bt \Vert,\sigma}(\bt) = \cG_{\sigma}(\bt)$ is the
$\sigma$-equivariant piece of ground state sector of $\bt$.

%%%%%%%%%%%%%%%%%%%%%%%%%%%%%%%%%%%%%%%%%%%%%%%%%%%%%%%%%%%%%%%%%%%%%%%%%%%%%%%%%%%%
\newsubsection{Absence of the discrete series}

A priori the spectral value of $\lb$ of the operator ${\bf A}$ is of 
course unrelated to the type $\sigma$ of (irreducible) representation 
with respect to which a generalized eigenfunction transforms. 
The goal of the Section 4 will be to show that for 
a transfer operator $\bt$ and the spectral value $\lb = \Vert \bt \Vert$,
the possible representations $\sigma \in \widehat{G}_r$ are 
severely constrained. A first flavor of the correlation between the 
spectral value of $\bA$ and the type $\sigma$ 
of irreducible representations allowed can be obtained from the  

\begin{corollary} 
Let $\bA$ be as in Section 3.3 or 3.4 and $\bA(\sigma) = \1 \otimes 
\check{\bA}_{\check\sigma}$ one of its fiber operators. Under any one of 
the following conditions the eigenspaces $\cE_{\lb,\sigma}(\bA) \simeq 
\cE_{\lb}(\bA(\sigma))$ for $\sigma \in \widehat{G}_d$, the discrete 
series of $G$,  are empty: 
\begin{itemize}
\item[(a)] $\cE_{\lb}(\bA(\sigma))$ contains a left $K$ singlet,
i.e.~a generalized eigenfunction $F$ obeying $F(kn) = F(n)$, for all $k \in K$. 
\item[(b)] $\cE_{\lb,\sigma}(\bA)$ contains a right $K$ singlet, 
i.e.~a generalized eigenfunction $\Omega$ obeying $\Omega(gk,n) = 
\Omega(g,n)$, for all $k \in K$.   
\item[(c)] $\bA = \bt$ is a transfer operator, $\lb=\Vert \bt \Vert$ 
and it is an eigenvalue of $\bt(\sigma)$.
\end{itemize}
\end{corollary}
\begin{proof} 
(a),(b) For definiteness let us consider the setting in Section 3.3, for 
the one in Section 3.4 only notational changes are required. 
Via the isometry $F \ra t_{\sigma}^F$ the assumptions in (a) and (b) 
amount to ${\rm Tr}[F(kn) \pi_{\sigma}(g)] =  {\rm Tr}[F(n) \pi_{\sigma}(g)]$
and   ${\rm Tr}[F(n) \pi_{\sigma}(gk)] =  {\rm Tr}[F(n) \pi_{\sigma}(g)]$,
respectively, for all $k \in K$, $g \in G$ and almost all $n \in \cN$. 
In terms of $P_0$, the projector onto the $\kappa =0$ singlet subspace in
$\cL_{\sigma}$, this becomes ${\rm Tr}[F(n) \pi_{\sigma}(g)] =  
{\rm Tr}[F(n) P_0 \pi_{\sigma}(g)]$
and ${\rm Tr}[F(n) \pi_{\sigma}(g)] =  {\rm Tr}[F(n) \pi_{\sigma}(g)P_0]$,
respectively. However these can be nonvanishing functions only if 
the projections $P_0 \pi_{\sigma}(g)$ or $\pi_{\sigma}(g) P_0$ 
are nontrivial, which requires $\pi_{\sigma}$ to be of $K$-type 1.  
All $K$-type 1 representations are however non-discrete 
(in fact spherical principal series) representations.   

(c) The proof is by contradiction. Let $\sigma \in \widehat{G}_d$ 
be given and suppose that the point spectrum of $\bt_{\sigma}$
is nonempty and contains $\Vert \bt \Vert$. The spectral radius equals
$\Vert \bt \Vert$ and the normalizable eigenfunction $\chi_{\sigma}$ 
of $\bt_{\sigma}$ has eigenvalue $\Vert \bt \Vert$. By  
Prop.\ref{relacomp}(b) the associated eigenfunction 
$\tau_{v\sigma}(\chi_\sigma)$ of $\bt$ is normalizable. This 
contradicts Prop.~\ref{spectrum}(b). 
\end{proof} 

Part (c) of the corollary is useful because in many cases all the fiber 
operators $\bt_{\sigma}$ are compact (see Section 3.3). Then 
the nonzero spectrum of $\bt_\sigma$ is discrete
for all nonexceptional $\sigma$. Parts (a), (b) are useful, because often 
one has independent reasons to expect that a ground state should exist, 
which is at least a singlet under a maximal compact subgroup $K$ of $G$.    
In all cases the corollary entails that all discrete 
series representations are ruled out as candidates for the 
representation carried by the ground state sector.  
The remaining part of the restricted dual corresponds to 
noncompact Cartan subalgebras (see Appendix A.2) and the associated 
irreducible representations are cuspidial 
principal series. To avoid having  
to discuss the restricted dual $\widehat{G}_r$ of generic 
linear reductive Lie groups, we focus from now on on 
the case $G = {\rm SO}_0(1,N)$. Its representation theory shows
all the typical complications (existence of discrete series, 
in particular) and one can reasonably expect that 
the subsequent results will generalize to other Lie groups.

%%%%%%%%%%%%%%%%%%%%%%%%%%%%%%%%%%%%%%%%%%%%%%%%%%%%%%%%%%%%%%%%%%%%%%%%%%
\newpage
\newsection{The structure of the ground state sector}
 
For definiteness we consider from now on the case $G = {\rm SO}_0(1,N)$.
The restricted dual then is a disjoint union 
$\widehat{G}_r = \widehat{G}_d \cup \widehat{G}_p$, 
of a set $\widehat{G}_d$ describing discrete UIR (unitary irreducible 
series representations) and a set $\widehat{G}_p$ describing principal 
series UIR. A UIR $\sigma_d \in  \widehat{G}_d$ is parameterized 
a tuple of integers $\pm s \in \N,\,\xi \in \widehat{M}_s \subset 
\widehat{M}$, a UIR $\sigma_p \in \widehat{G}_p$ is labeled
by a real parameter $\om \geq 0$ and again by a tuple of integers 
$\xi \in \widehat{M}$. Here $\widehat{M}$ is the dual of the subgroup 
${\rm SO}(N\!-\!2)$. We refer to appendices A and B for a brief survey of 
the relevant aspects of the representation theory of linear reductive Lie 
groups in general, and of ${\rm SO}_0(1,N)$ in particular. 

Recall from Section 3 that the space of $\sigma$-equivariant 
ground states $\cG_{\sigma}(\bt)$ can  be isometrically identified 
with the eigenspace $\cE_{\Vert \bt \Vert}(\bt(\sigma))$ of the 
fiber operator $\bt(\sigma) = \1 \otimes \check{\bt}_{\check\sigma}$.
The goal in the following is to successively rule out more and more 
UIRs $\sigma \in \widehat{G}_r$ for which the fiber spaces in
$\cG_{\sigma}(\bt)$ can be nonempty. This will lead to 
Theorems 1.1 and 1.2, as announced in the introduction. 

\setcounter{proposition}{0}
\setcounter{corollary}{0}
\setcounter{lemma}{0}

%%%%%%%%%%%%%%%%%%%%%%%%%%%%%%%%%%%%%%%%%%%%%%%%%%%%%%%%%%%%%%%%%%%%%%%%%%%%%%
\newsubsection{ $\cG_{\sigma}(\bt)$ is empty for all but one principal 
series representation} 

We begin by studying in more detail the restricted transfer operators 
$\bt_{\om \xi} := \bt_{\sigma}$, where we now write the label $\sigma$ as 
$(\om, \xi)$ corresponding to a member of the principal series (see 
Appendix B). We also denote by $\widehat M$ the subset of $\widehat G_r$ 
corresponding to the principal series. 

Here the realization of the model 
space $\cL_{\sigma}$ as $L^2_{\xi}(K)$ and the realization of the 
infinite sums over $\kappa$ in terms of $k$-integrations is useful. See 
Appendix A.8. We first show that in this realization the function space 
$L^2_{\sigma}(\cN)_0$ in (\ref{LNdef}), for which we now write 
$L_{\xi}^2(\cN)$, 
takes the following form 
\ba
\label{princNK1}
&& f \in L^2(\cN \times K, V_{\xi})\;,
\nonum
&& f(k_0\, n, k) = f(n, k k_0) \,,\quad \;\quad k_0 \in K\,,
\\[2mm]
&& f(n, m k) = r_{\xi}(m) f(n,k)\,,\quad m \in M\,.
\nonumber
\end{eqnarray}
For fixed $n \in \cN$ the second equation is just the 
defining relation for $L^2_{\xi}(K)$, see (\ref{prep1}). 
The first relation is the realization of the equivariance condition 
under $\ell_r(K)$ in the compact model of the principal representation.  
Indeed the right action $f(n,k) \mapsto f(n,k) \pi_{\om \xi}(g)$ 
can be read off from Eq.~(\ref{kernelaction}b) and is given by 
\be 
f(n,k) \pi_{\om \xi}(g) =  (\delta \nu)^{-1}( a(k g^{-1}))\,
f(n, g^{-1}[k]) \,.
\label{princNK2}
\end{equation}
For $g = k_0 \in K$ the right hand side of (\ref{princNK2}) reduces to 
$f(n, kk_0^{-1})$ so that the equivariance condition assumes the form 
given in (\ref{princNK1}). As a check one can verify the left $K$ 
invariance 
\be 
f(k_0 n, k) \pi_{\om\xi}(k_0 g) = 
f(n, k) \pi_{\om\xi}(g)\,,\quad k_0 \in K\,,
\label{princNK3}
\end{equation}
using $a(k g^{-1} k_0^{-1}) = a(k g^{-1})$ and $k(k g^{-1} k_0^{-1}) = 
k(k g^{-1}) k_0^{-1}$ via (\ref{Iwaaction}).

The fiber operator $\bt_{\om \xi}$ will now act on the space of functions
(\ref{princNK1}) as an integral operator with kernel 
\ba
&& \cT_{\om \xi}(n,n';k,k') := \int_G \! dg \, \cT(g,n',n) [\pi_{\om \xi}(g)](k',k) 
\nonum 
&& \quad = \int_P\!dp\, \cT({k'}^{-1} p k, n',n) \, 
(\delta^{-1} \nu)(p) \,r_{\xi}(m(p)) = 
\cT_{\om \xi}(kn, k'n'; e, e) \,,
\label{princT1}
\end{eqnarray}
using (\ref{pxi2}). Explicitly $\nu = \nu_{\om}$ and $\delta$ are
given by (\ref{Phomeos}).
From (\ref{Tr2}) and (\ref{pxi2}) one verifies the symmetries 
\be
\cT_{\om \xi}(k_0 n, n'; k k_0^{-1}, k') = \cT_{\om \xi}(n, n'; k, k') =
\cT_{\om \xi}(n, k_0 n'; k, k' k_0^{-1}) \,,
\end{equation}
which ensure that $\bt_{\om \xi}$ acts consistently on the 
function space (\ref{princNK1}). For fixed arguments the kernel (\ref{princT1}) 
is a linear map on $V_{\xi}$ whose matrix elements obey   
\be 
\Big| \bra v , \cT_{\om \xi}(n,n';k,k') \,v'\ket_{V_{\xi}} \Big|\leq 
\cT_{0,0}(n,n';k,k') \sup_{m \in M} |\bra v, r_{\xi}(m) v'\ket_{V_{\xi}}|\,,
\label{princT2}
\end{equation}
for all $v,v' \in V_{\xi}$. The inequality is strict unless $\om =0$ 
and $\xi =0$.  Here we wrote $\xi =0$ for the trivial representation of 
$M$ and $\om =0$ refers to the $\om \ra 0$ limit of the
principal series. The bound (\ref{princT2}) is a manifest consequence 
of (\ref{princT1}) and the unitarity of $\nu:A \ra U(1)$.  
With these preparations we can show the crucial

\begin{proposition}\label{tnorm} 
(a) The operators $\bt_{\om \xi}:L^2_{\xi}(\cN) \ra L^2_{\xi}(\cN)$ 
are bounded for all (not almost all) $\om \geq 0$, $\xi \in \widehat{M}$.
Their norms $\Vert\bt_{\om \xi}\Vert$ 
are continuous functions of $\om$ and obey
\be 
\Vert\bt_{\om \xi}\Vert \le \Vert \bt_{0 0}\Vert\quad \mbox{for all} \;\;
\xi \in \widehat{M}\;\; \mbox{and} \;\;\om \geq 0\,,
\label{Tomnorm}
\end{equation}
where the inequality is strict unless $\om =0$ and $\xi =0$.\\
(b) $\bt_{0 0}$ is a transfer operator in the sense of Definition 2.1 uniquely  
associated with $\bt$. 
\end{proposition}
\begin{proof}
(a) For the boundedness we show that the kernel of $\bt_{\om\xi}$ 
obeys the stronger condition
\be 
\sup_{\Vert v\Vert=\Vert v'\Vert=1}\,\,\sup_{n,k} \int \!dn' dk' \Big|
\bra v, \cT_{\om\xi}(n,n',k,k') v'\ket_{V_{\xi}} \Big| < \infty\,.
\label{Tombound1}
\end{equation} 
Clearly the expression under the first sup is bounded by 
\be 
\sup_n \int \!dg dn'\, \cT(g,n',n) \;
\sup_{k, g_0} \int \!dk' \big|\bra v, \pi_{\om,\xi}(g_0)(k,k') 
v'\ket_{V_{\xi}}\big|\,.
\label{Tombound2}
\end{equation}
The first factor is bounded by the constant $K_{\cT}$ in 
(\ref{Abound}). To estimate the second factor, let $\phi_l$ 
be a sequence of positive normalized functions on $G$ 
approximating a delta distribution centered around $g_0 \in G$. 
From (\ref{pxi2}) 
\be 
\big|\bra v, \pi_{\om\xi}(\phi_l)(k,k') v'\ket_{V_{\xi}} \big|
\leq \int_P \!dp\, \delta(p)^{-1} \phi_l(k^{-1} p k') 
\; \sup_m \big| \bra v, r_{\xi}(m) v'\ket_{V_{\xi}}\big|\,.
\label{Tombound3}
\end{equation}
The second factor is a finite constant and it suffices to 
consider the $dk'$ integral of the first factor.  
Recall from Appendix A that $P=NAM$, $dp = d(na) dm 
= \delta(a)^2 dn da dm$. Parameterizing $k'$ as $k' = 
k'(\vec{s}) m'$, with $m' \in M$, $k'(\vec{s}) \in 
K/M$ (see Appendix B) one has 
\be 
\int_K \!dk' \!\int_P d(na) dm \,\delta(a)^{-1} \phi_n(k^{-1} nam k') = 
\int_M\! dm'\! \int_G dg \,\delta(a(g)) \phi_n(k^{-1} g m') \leq 1\,,       
\label{Tombound4}
\end{equation} 
using $dg = dn da dm dS(\vec{s})$ from (\ref{NAKmeasure}), (\ref{MSmeasure}), 
and the fact that $\delta(a(g)) \leq 1$. This gives (\ref{Tombound1}). 
From here it follows (see Definition 2.1) that $\bt_{\om\xi}$ 
defines a bounded operator from $L^p_{\xi}(\cN)$ to $L^p_{\xi}(\cN)$,
for all $1 \leq p\leq \infty$. For $p=2$ this is the assertion,
for $p \neq 2$ it gives an alternative (though less general) 
proof of the extension discussed after Eq.~(\ref{Lspdef}).

To see the continuity of the norms, recall 
\be 
\Vert\bt_{\om \xi}\Vert = \sup_f |(f, \bt_{\om \xi} f)|\,,
\quad f \in L^2_{\xi}(\cN)\,,\;\; (f,f) =1\,.
\label{Tomnorm1}
\end{equation}
We may assume that the supremum is reached on a sequence 
$(f_l)_{l \in \N}$ of normalized functions in $L^2_{\xi}(\cN)$,
in which case it suffices to show that $\lim_{l \ra \infty} 
(f_l, \bt_{\om\xi} f_l)$ is continuous in $\om$. 
Each matrix element $(f_l, \bt_{\om\xi} f_l)$ is manifestly 
a continuous function of $\om$. To show that this remains 
true for the limit $l \ra \infty$ we establish a uniform bound:
using (\ref{princNK1}), (\ref{princT1}) one has for the matrix elements 
\ba 
(f_l, \bt_{\om\xi} f_l) \is \int\! dn dn' dk dk' \! \int_P \!dp\,
\cT({k'}^{-1} p k, n',n) (\delta^{-1} \nu)(p) 
\nonum
&& \times \bra f_l(n,k), r_{\xi}(m(p)) f_l(n',k') \ket_{V_{\xi}}\,, 
\label{Tomnorm2}
\end{eqnarray}   
where $\om$ enters only through the phase $\nu(p)=\nu_{\om}(p)$, see
(\ref{Phomeos}).  
The modulus of these matrix elements is easily seen to have the 
$l$ and $\om$ independent upper bound 
\be 
\sup_{n,k,n',k'} \int_P\!dp\,\cT({k'}^{-1} p k, n',n) \,
\delta(p)^{-1} < \infty\,.
\end{equation}
By dominated convergence one can take the $l \ra \infty$ limit inside 
the integrals, after which continuity in $\om$ is manifest.

Finally, to verify (\ref{Tomnorm}) we return to (\ref{Tomnorm1}). 
From (\ref{princT2}) the modulus of the matrix elements can be bounded by 
\ba 
&& \big| (f,\bt_{\om\xi} f)\big| \leq  
\int \! dk dn dk' dn' \Big | 
\bra f(n,k) , \cT_{\om \xi}(n,n'; k,k') f(n',k')\ket_{V_{\xi}} \Big|
\nonum
&& \quad \leq \int \! dk dn dk' dn' \, \cT_{0,0}(n,n';k,k')\,
\Vert f(n,k)\Vert_{V_{\xi}}  \Vert f(n',k')\Vert_{V_{\xi}}\,,  
\label{Tomnorm3}
\end{eqnarray}
where the second inequality is strict unless $\om =0$ and $\xi =0$. 
Here both norms are pointwise nonnegative functions in $L^2_{0}(\cN)$.
In fact one gets a norm preserving map from $L^2_{\xi}(\cN)$ to 
a subset of $L^2_0(\cN)$ by mapping $f(n,k)$ to 
$f_0(n,k):= \Vert f(n,k)\Vert_{V_{\xi}}$. In particular 
$f$ has unit norm in $L^2_{\xi}(\cN)$ if and only if $f_0$ 
has unit norm in $L^2_0(\cN)$.

On the other hand to get the norm of $\bt_{00}$ the 
supremum of $(f_0, \bt_{00} f_0)$ over all normalized 
$f_0 \in L^2_0(\cN)$ has to be taken. These are complex valued
functions but by the integral representation (\ref{princT1})
$\cT_{00}(n,n',k,k')\ge 0$ and the supremum is reached on 
the subset of (all) nonnegative functions in $L^2_0(\cN)$. Among those 
are the ones in the range of the map $L^2_{\xi}(\cN) \ra 
L^2_0(\cN),\; f \mapsto \Vert f\Vert_{V_{\xi}}$, described above.  

So one has for $(\xi,\om)\neq(0,0)$
\be
\Vert\bt_{\om\xi}\Vert 
\leq  \sup_{f \in L^2_\xi(\cN)} | (f, \bt_{\om \xi} f)|
< \sup_f (|f| , \bt_{00} |f|)   
\leq  \sup_{f_0{\in L^2_0(\cN)}} (f_0, \bt_{00} f_0) = 
\Vert\bt_{00}\Vert\,.
\label{Tomnorm4}
\end{equation}
(b) From part (a) we know that $\bt_{00}$ is a 
bounded linear integral operator with spectrum contained in 
$[0, \Vert \bt \Vert]$. The integral representation (\ref{princT1}) 
shows that its kernel is strictly positive almost everwhere. 
Thus $\bt_{00}$ is also positivity improving and by (\ref{Tombound1}) it is a 
transfer operator in the sense of Definition 2.1. 
\end{proof}

As a corollary we obtain part (b) of Theorem 1.1 announced in the 
introduction. 

\begin{corollary} $\cG_{\sigma}(\bt)$ is non-empty for at most 
one principal series representation -- the limit of the 
spherical principal representation, $\sigma 
= (\om=0,\xi=0)$. Further $\cG_{00}(\bt)$ can be isometrically 
identified with the ground state sector $\cG(\bt_{00})$ 
of the transfer operator $\bt_{00}$. 
\end{corollary}

We add some comments on the discrete series. 
Recall from (\ref{asigmabound}) 
\be 
\Vert \bt \Vert = {\rm ess} \sup_{\sigma \in \widehat{G}_r} 
\Vert\bt_{\sigma}\Vert\,,
\label{Tomnorm5}
\end{equation}
and that a discrete series UIR is labeled by a tuple of 
integers $\pm s \in \N$, $\xi \in \widehat{M}_s \subset \widehat{M}$.  

\begin{remark} We expect that a preferred class of 
discrete series representations exists such that a bound similar 
to (\ref{Tomnorm}) holds. 
If there are discrete series UIR for which $\Vert \bt_{\pm s\xi}\Vert=
\Vert \bt\Vert$, a natural conjecture is that they have the minimal 
possible $M$ weight for a given $s \in \N$, i.e.~$\xi =
\xi_s:= (|s|,\ldots,|s|)$. This requires that the  
norms $\Vert\bt_{\pm s \xi}\Vert$ of the operators 
$\bt_{\pm s \xi}$ obey 
\be 
\Vert\bt_{\pm s \xi}\Vert \le \Vert \bt_{\pm s, \xi_s}
\Vert\quad \mbox{for all} \;\; \xi \in \widehat{M_s}\,,
\label{Tsnorm}
\end{equation}
where the inequality is strict unless $\xi =\xi_s:= (|s|,\ldots,|s|)$.
\end{remark}

A verification would require a concrete model for the discrete series 
(like that of the principal series described in Appendix A.8) 
producing a counterpart of Eq.~(\ref{princT1}). We leave this for 
a future investigation. Here we proceed by identifying simple 
conditions under which discrete series representations are 
ruled out as candidates for the representation carried by
$\cG_{\sigma}(\bt)$. A set of such conditions has already been 
identified in corollary 3.1. We proceed by studying the 
role of positivity.

%%%%%%%%%%%%%%%%%%%%%%%%%%%%%%%%%%%%%%%%%%%%%%%%%%%%%%%%%%%%%%%%%%%%%%%%%%%%%%%%%%%

\newsubsection{The role of strict positivity} 

One property which one expects from a ground 
state is ``not to have nodes''. This means the ground state 
wave function can be chosen strictly positive a.e. In this 
section we show that, somewhat surprisingly, the positivity 
requirement uniquely determines the representation 
carried by $\sigma$-equivariant eigenfunctions. 
   
As before we denote by $\pi_{00}$ the limiting  representation of the 
spherical principal series $\pi_{\om 0}$, $\om \geq 0$, of 
$G={\rm SO}_0(1,N)$, 
and by $\bA_{00}$ the corresponding fiber operator in the 
decomposition of $\bA$ (see Section 3). Further we set 
\be 
\cP = \{ \Omega \in L^{\infty,2}(\cM_r) \;\mbox{or}\;
\Omega \in M^{-1} L^{\infty,2}(\cM_r)\,|\,\Omega(g,n) > 0 
\;\mbox{a.e.}\}\,,
\label{positive}
\end{equation}
for the settings in Sections 3.3 or 3.4, respectively.

\begin{proposition}\label{pos}
$\cE_{\lb,\sigma}(\bA) \cap \cP = \{ 0 \}$ unless 
$\pi_\sigma = \pi_{00}$. Further the intersection is 
isometric to $\cE_{\Vert \bA \Vert}(\bA_{00})$ and 
is spanned by a.e.~strictly positive functions.  
\end{proposition}

\begin{proof} Recall from Section 3 that the generalized eigenspaces 
$\cE_{\lb,\sigma}(\bA)$ invariant under $\pi_{\sigma}$ are generated by 
`factorized' functions of the form $\sum_i \tau_{v_i 
\sigma}(\chi_i)(g,n)$.
Here 
\be 
\tau_{v \sigma}(\chi)(g,n)  = (\chi(n), \pi_{\sigma}(g) v)_{\sigma} = 
\sum_{\kappa s} \chi(n)_{\kappa s}^* (\pi_{\sigma}(g) v)_{\kappa s}\,,  
\label{posi1}   
\end{equation}
with $v \in \cL_{\sigma}$ and $\chi \in \cE_{\lb}(\bA_{\sigma})$ 
a generalized eigenfunction of ${\bf A}_{\sigma}$. By assumption at least 
one such function is strictly 
positive a.e.~on $\cM_r$, for which we write $\Omega(g,n)
= \sum_i (\chi_i(n), \pi_{\sigma}(g)v_i)_{\sigma}$. Replacing $g$ with 
$kg$, $k \in K$, and averaging over $K$ must still produce an 
a.e.~strictly positive function. On the other hand, evaluated in the 
$K$ adapted basis the average reads
\be 
\int_K\!dk\,\Omega(kg,n) = \sum_i \chi_i(n)^*_{00} 
P_0 \pi_{\sigma}(g) v_i >0\,.
\label{posi2}
\end{equation}    
Importantly the $K$ average cannot vanish, so at least one of the 
projections $P_0 \pi_{\sigma}(g) v_i$ must be nonzero. This means 
$\cL_\sigma$ must contain a $K$-invariant vector, namely 
$P_0 \pi_{\sigma}(g)v_i$ for some $i$, and $\pi_{\sigma}$ is by 
definition of $K$-type 1. But 
for the UIR of ${\rm SO}_0(1,N)$ this is the case only for 
$\pi_{\om,\xi=0}$, see Appendix B. Discrete series representations 
and non-spherical principal series representations are ruled out.   
Thus, for $\cE_{\lb,\sigma}(\bA) \cap \cP$ 
to be non-empty the corresponding UIR $\pi_{\sigma}$ must be a 
member of the spherical principal series, $\pi_{\sigma} = \pi_{\om 0}$,
for some $\om \geq 0$.

It remains to show that only the $\om \ra 0$ limit can occur. To 
this end we replace $g$ in (\ref{posi2}) by $gk$ and average 
over $K$. Using the fact that $\pi_{\om,0}(g)_{00,00}$ evaluates 
to the Legendre function $E_{\om,00}(q_0)$, $q = g^{-1} q^{\uparrow} =
(q_0, \vec{q})$, in (\ref{EPrel}) one gets the condition  
\be 
\int_{K\times K}\!dk dk'\,\Omega(kgk',n) = 
\sum_i \chi_i(n)^*_{00} (v_i)_{00}\, E_{\om,00}(q_0) >0\,,
\label{posi3}
\end{equation}
for almost all $n \in \cN$ and $q_0 > 1$. However, $E_{\om,00}$ is an 
oscillating function of its argument unless $\om=0$. This means 
(\ref{posi3}) cannot be satisfied for $\om \neq 0$.     
\end{proof}

We remark that the first part of the argument extends to all 
Lie groups of the type considered: $\cE_{\lb,\sigma}(\bA) \cap \cP 
= \{ 0 \}$ unless $\pi_{\sigma}$ is of $K$-type 1. 
The second part of the argument shows that $\pi_{\sigma}$ 
must have a $K$-spherical function $p(g) = \pi_{\sigma}(g)_{00,00}$ 
which is real and positive on all of $G$, i.e.~on $K\backslash G/K$.  
An interesting (apparently open) question is which nonamenable Lie 
groups have a UIR with a pointwise positive $K$-spherical function 
(not just one of positive type).

As a corollary we now obtain part (a) of Theorem 1.1 as anticipated 
in the introduction.

\begin{corollary} $\cG_{\sigma}(\bt)$ is empty for all but the 
principal series representations whenever one of the following holds: 
(i) $\cG_\sigma(\bt)$ contains a strictly positive function. 
(ii) $\cG_\sigma(\bt)$ contains a $K$-singlet. (iii) $\bt_{\sigma}$ 
is compact. 
\end{corollary}

\begin{proof} 
(i) is a special case of Proposition 4.2. (ii) and (iii) follow from 
Corollary 3.1.
\end{proof}

Combined parts (a) and (b) of Theorem 1.1 show that under broad 
conditions the representation $\sigma$ carried by the 
equivariant fibers $\cG_{\sigma}(\bt)$ of the ground state sector 
is {\it uniquely} determined and always the {\it same} for every 
invariant transfer operator, namely the limit of the spherical 
principal series. The existence of generalized ground states 
in $\cG_{00}(\bt) \simeq \cG(\bt_{00})$ is however not guaranteed from 
the outset. A simple but important special case where it is, 
is when all the fiber operators $\bt_{\sigma}$ are compact.

%%%%%%%%%%%%%%%%%%%%%%%%%%%%%%%%%%%%%%%%%%%%%%%%%%%%%%%%%%%%%%%%%%%%%%%
\newsubsection{$\cG_\sigma(\bt)$ for $\bt_{\sigma}$ compact}

Here we present the proof of Theorem 1.2, as anticipated in the 
introduction. We reformulate the theorem in a more precise way as follows:

{\bf Theorem 1.2.'} {\it Let $\bt$ be an invariant transfer operator 
in the sense of Definition 2.1 and 3.1, and assume that its 
kernel satisfies in addition the condition (C) in Section 3.3. 
Then all its ground state fibers $\cG_{\sigma}(\bt)$ but 
$\cG_{00}(\bt) \simeq \cE_{\Vert \bt \Vert}(\bt_{00})$ are empty. Further
$\bt_{00}$ is a transfer operator whose ground state 
sector $\cG(\bt_{00})=\cE_{\Vert \bt \Vert}(\bt_{00})$ is
nonempty and is spanned by a unique normalizable and a.e.~strictly positive 
function $\psi_0 \in L^2(\cN \times S^{N-1})\simeq L^2(\cN) \otimes
\cL_{00}$. Explicitly $\cG(\bt)$ is the linear hull of functions 
\be 
\Omega(g,n) = \int_{S^{N-1}} \!\!dS(p) \,
\frac{\psi_0(n,\vec{p})}{(q_0 - \vec{q} \cdot \vec{p})^{\frac{N-1}{2}}}\,,
\label{orbitcompact}
\end{equation}
where $q = g q^{\uparrow} = (q_0, \vec{q})$. 
}
\begin{proof} By Theorem 1.1 we know that 
at most $\cG(\bt_{00})$ is nonempty. All functions in 
$\cG(\bt_{00})$ are thus $M$-singlets and take values in 
$\C = V_{\xi =0}$. By Lemma 3.2 and Proposition 4.1  
$\bt_{00}$ is a Hilbert-Schmidt operator and a transfer 
operator in the sense of Definition 2.1. Its eigenspace 
$\cE_{\Vert \bt \Vert}(\bt_{00})$ coincides with 
its ground state sector $\cG(\bt_{00})$. By the generalized 
Perron-Frobenius theorem \cite{GlimmJaffe,RS4} it follows that  
the ground state sector of $\bt_{00}$ is one-dimensional 
(non-empty, in particular) and is spanned by a unique a.e.~strictly 
positive function $\psi_0(n,k) > 0$ on $\cN \times K$, with 
$\psi_0(k_0 n, k) = \psi_0(n,kk_0)$, namely the ground state of 
$\bt_{00}$. Note that via the realization (\ref{princNK1}) the result 
is applied to scalar valued functions on $\cN \times K$ rather 
than to $\cL_{00}$-valued functions on $\cN$. Since $\xi=0$ 
the function also obeys $\psi_0(n,mk) = \psi_0(n,k)$, for all 
$m \in M$,  and thus can be viewed as a function on $\cN \times 
S^{N-1}$, with $S^{N-1} = M\backslash K$. This is related to 
the fact that the $K$-content of the principal series representations 
$\pi_{\om,\xi}$ with $\xi =0$ equals $\widehat{K}_{\xi =0} 
= \{(0,\ldots, 0, \ell)\,,\; \ell \geq0\}$. These are the ${\rm SO}(N)$ 
irreducible representations which can be realized on the space 
of symmetric traceless tensors, or equivalently on the harmonic 
symmetric polynomials of degree $\ell$ on the sphere $S^{N-1}$.   
In the original model, where $\psi_0:\cN \ra \cL_{00}$, 
we relabel the components $\psi_0(n)_{\kappa s}$ in the 
$K$-adapted basis such that $\psi_0(n)_{\kappa s} = \psi_0(n)_{\ell m}$,
where $\psi_0(n)_{\ell m}$ are the components of $\psi_0(n,\vec{p})$ in 
a basis of spherical harmonics $Y_{\ell m}(\vec{p})$ on $S^{N-1}$.   
The matrix element $\tau_{v,00}(\psi_0)(g,n) = (\psi_0(n), \pi_{00}(g) v_0)$, 
then evaluates to a double sum over labels of the spherical harmonics.
This holds for any reference vector 
$v \in \cL_{00}$. Specifically we choose now for $v$ the 
unique $K$-invariant vector $v_0 \in \cL_{00}$, i.e.~the 
vector satisfying $\pi_{00}(k) v_0 = v_0$ for all $k \in K$.  
In the $K$-adapted basis of $\cL_{00}$ this vector has 
components $(v_0)_{\kappa s} = (v_0)_{00} \delta_{\kappa =0} 
\delta_{s=0}$. The matrix element then evaluates to 
\be 
\tau_{v_0,00}(\psi_0)(g,n) = (\psi_0(n), \pi_{00}(g) v_0) = 
\sum_{\ell m} \psi_0(n)_{\ell m} \pi_{00}(g)_{\ell m,00} (v_0)_{00}\,. 
\label{orbitcompact1} 
\end{equation}
The matrix elements $\pi_{00}(g)_{\ell m,00}$ are right $K$ invariant 
and reduce to functions on $Q = G/K$, with the identification $q = g
q^{\uparrow}$. In the case at hand these are the Legendre functions 
$E_{\om=0,\ell m}$ described in (\ref{Hdef}).  
Using the integral representation in (\ref{Hdef}) the sum  
can be replaced by an integral over $S^{N-1}$: 
$\sum_{\ell m} \psi_0(n)_{\ell m} E_{0,\ell m}(q) = \int\! dS(p)\, 
\eps_{0,p}(q) \psi_0(n,\vec{p})$, with $\psi_0(n,\vec{p}) = 
\sum_{\ell m} \psi_0(n)_{\ell m} Y_{\ell m}(\vec{p})$.  Inserting 
(\ref{Edef}) 
and omitting the overall constant $(v_0)_{00}$ gives the 
announced expression (\ref{orbitcompact}). 

It remains to show that the linear hull of these functions 
is dense in $\cG(\bt)$. We proceed in two steps. First we show 
that the choice of the $K$-invariant vector $v_0$ was inessential.
This is because on account of the irreducibility of $\cL_{00}$ 
every vector $v \in \cL_{00}$ lies in the closure of 
linear combinations of the form $\sum_i c_i \pi_{00}(g_i) v_0$.
Thus $\tau_{v, 00}(g,n)$ with $v \in \cL_{00}$ generic lies in 
the linear hull of the functions (\ref{orbitcompact1}). 
In a second step we appeal to Proposition 3.6, according 
to which linear combinations of the form (\ref{Omcompact}) 
(but now with a unique $\chi_{\lb = \Vert \bt \Vert, \sigma =00, r=1} =
\psi_0$) are dense in $\cE_{\lb = \Vert \bt 
\Vert,\sigma = (00)}(\bt) = \cG(\bt)$. 
\end{proof}

\begin{remark} Theorem 1.2' is a generalization of the result 
announced in Section 3.3 of \cite{O1Nmodel}; as a special case it 
contains the omitted proof of (a slightly corrected variant of) 
Theorem 2 in \cite{O1Nmodel}. Explicitly:   
Let $\T$ be the transfer operator of the hyperbolic 
nonlinear sigma-model (see Eqs. (2.3),(2.4) of \cite{O1Nmodel}). 
Then $\T^2$ satisfies the premise of Theorem 1.2.  
\end{remark}
\begin{proof} 
One only has to check that the condition (C) is 
satisfied. This follows from the Lemma in Section 3.3 of \cite{O1Nmodel}. 
\end{proof}

%%%%%%%%%%%%%%%%%%%%%%%%%%%%%%%%%%%%%%%%%%%%%%%%%%%%%%%%%%%%%%%%%%%%%%%%%%%%%%%%%%
\newsection{Conclusions and outlook} 
\setcounter{theorem}{0}

The set of ground states of a quantum mechanical many particle system 
-- equivalently, a statistical mechanics system on a finite lattice -- 
whose dynamics is invariant under a nonamenable unitary representation of 
a locally compact group has been shown here to have a remarkable structure: 
despite the invariance of the dynamics there are always infinitely 
many noninvariant and nonnormalizable ground states. 
This spontaneous symmetry breaking cannot be prevented by 
tuning a coupling or temperature parameter, and it takes place 
even in one and two dimensions and even for finite systems. The phenomenon 
is also not limited to a semiclassical regime. The linear space spanned 
by these 
ground states carries a distinguished unitary irreducible representation of 
the group, which for ${\rm SO}_0(1,N)$  has been identified as the 
limit of the spherical principal series. These properties hold 
for a large class of transfer operators or Hamiltonians. Compactness 
of the transfer operators restricted to the group invariant fibers 
is a sufficient but by no means necessary condition. The same is true
for the very existence of generalized ground states. 

As explained in Section 2.1 the well-known theory of generalized 
eigenfunctions due to Gel'fand \cite{gelfand} and their application to 
group representations due to Maurin \cite{maurin} is not sufficient to 
assure the existence of generalized ground states (neither as functions nor
as distributions). However, we can guarantee their existence as functions
in various situations considerably broader than the above one. These involve 
by neccessity a subsidiary condition on the transfer operator. One can take
\cite{NieSei3} (s1) $\bt_{00}$ is a compact operator. 
(s2) the integral kernel of $\bt$ has an 
extremizing configuration $m_*\in \cM$, in the sense that $T(m,m';t) 
\leq T(m_*,m_*,t)$ for all $m,m' \in \cM$, and possibly others. 
In these cases one has the following construction principle:

\begin{theorem} Let $\bt$ be as in Theorem 1.1 and such that the 
subsidiary condition (s1) or (s2) is satisfied. Set $\Omega_t(v) := 
\bt^t v/(v,\bt^t v)$ for $v \in L^2(\cM)$ and $t \in \N$. Then there 
exist (explicitly known) vectors $v_* \in L^2(\cM)$ and a sequence of 
integers $(t_j)_{j \in \N}$ such that the limit 
\be 
\Omega_{\infty} := w^*\!-\!\lim_{j \ra \infty} \Omega_{t_j}(v_*)\,,
\label{Thm2}
\end{equation}
exists and is an element of $\cG(\bt)$.  
\end{theorem}

The point of Theorem 5.1 is that even a weak limit has the tendency 
to either diverge or to vanish. It is only by a carefully tuned 
`renormalization' that a finite and nonzero limit can be obtained 
which is a generalized ground state of $\bt$. The choice of the 
sequence $t_j$ is presumably inessential, but the choice of 
specific vectors $v_* \in L^2(\cM)$ is crucial. Typically these `seed' vectors 
form a set of measure zero in $L^2(\cM)$ and their identification 
is a major part of the Theorem.

The results for these `finite' systems can be generalized in a 
number of ways. The choice of $\cM$ as a product manifold 
$\cM = Q^{\times \nu}$ is natural from the viewpoint of a quantum 
mechanical or statistical mechanics interpretation. One could also 
replace $\cM$ with the phase space of a dynamical system with 
constraints generating the Lie algebra of a nonamenable group,  
in which case $\cM$ could be any proper riemannian $G$ manifold 
(in the sense of \cite{michor}). We expect that the results of sections 
2 and 3 remain valid in this more general setting. The main 
reason for specializing to $G = {\rm SO}_0(1,N)$ in Section 4
was that these groups have split rank 1, so their restricted dual 
is relatively simple. Nevertheless we expect that counterparts 
of Theorem 1.1 exists for basically every noncompact semi-simple 
Lie group with a finite center: only a tiny subset of unitary 
irreducible representations are potential candidates for the 
representation carried by the ground state sector of a system 
with a $G$ invariant dynamics. 

Another extension is to supersymmetric multiparticle systems with 
nonamenable symmetries (see \cite{macf} for one-particle
systems). 
The transfer operators associated with graded homogeneous spaces
are relevant for the description of disordered systems \cite{efetov83, 
efetov97, ZirnJMP}. They are not symmetric with respect to a positive 
definite scalar product, but have a unique ground state in a certain 
sense. Restricted to a bosonic subspace transfer operators related to the 
ones considered here arise. This leads to the expectation that the lowest 
lying excitations of the original system would have a structure analogous 
to that of the ground state space found here. 

The analysis of the thermodynamic limit is an open problem. 
In the operator/Hilbert space formulation the thermodynamic limit 
can presumably at best be taken for specific lattices (hierarchical, 
Cayley trees, etc). For the hypercubic lattices directly relevant   
to lattice quantum field theories one needs to analyze the 
limit in terms of correlation functions, see \cite{ZirnSpencer}. 
Expectations are however 
that the structure of the ground state sector, if it changes, will   
`fragment' even more \cite{hchain,O1Nmodel}. In fact the existence of 
non-invariant expectation values in the thermodynamic limit 
is a more-or-less direct consequence of the non-amenability 
\cite{O1Nmodel}. This should lead to a general no-go theorem  
(inverse, nonconstructive Coleman theorem).

We believe that this circle of ideas will have a variety of 
applications, ranging from localization phenomena in solid state 
physics and quantum Kaluza-Klein theories to AdS duality   
and the ground states of quantum gravity. We hope to return to 
these applications elsewhere. 

Acknowledgments: We wish to thank M. Zirnbauer for calling our attention 
to the transfer operators on graded homogeneous spaces. E. S. 
would like to thank H. Saller for useful discussions about abstract 
harmonic analysis. 

%%%%%%%%%%%%%%%%%%%%%%%%%%%%%%%%%%%%%%%%%%%%%%%%%%%%%%%%%%%%%%%%%%%%%%%%%%%%%%%%%%
\newpage 
\setcounter{section}{0}
\appendix
\newappendix{Harmonic analysis on noncompact Lie groups}

In this appendix we summarize the results on the representation theory and 
the harmonic analysis of non-compact Lie groups needed in the 
main text. In appendix B we present more explicit results 
for ${\rm SO}(1,N)$.

{\it 1.~Basic setting:} The harmonic analysis on a noncompact Lie group 
is governed by Plancherel-type decompositions  
\be 
L^2(G) = \int^{\oplus}_{\widehat{G}_r} \! d\nu(\sigma) \, 
\cL_{\sigma}(G) \otimes \check{\cL}_{\check{\sigma}}(G)\,,
\sspace \rho \times \ell = \int^{\oplus}_{\widehat{G}_r} \! d\nu(\sigma) 
\,\pi_{\sigma} \otimes \pi_{\check{\sigma}}\,,
\label{Planchrho}
\end{equation}
where $d\nu$ is the Plancherel measure on $\widehat{G}$ 
which has support only on a proper subset $\widehat{G}_r 
\subset \widehat{G}$, the restricted dual of the group $G$. 
The  $\pi_{\sigma}$ are unitary continuous 
representations and $\cL_{\sigma}:= \cL_{\sigma}(G)$ is 
irreducible for $\nu$ almost all $\sigma \in \widehat{G}_r$. 
In the following we present the precise version of (\ref{Planchrho}) 
used in the main text and set the notation. 

We use the variant of (\ref{Planchrho}) valid for any linear connected 
reductive Lie group $G$ (linear meaning that 
$G$ is a closed subgroup of ${\rm GL}(N,\R)$ or ${\rm GL}(N,\C)$, 
reductive meaning that it is stable under conjugate transpose). 
Examples are ${\rm SO}(1,N)$, ${\rm SO}(N,\C)$, ${\rm SL}(N,\R)$, 
${\rm SU}(m,n)$, ${\rm Sp}(2N,\R)$. 
The Plancherel measure $d\nu(\sigma)$ is then a combination of a 
counting measure for discrete variables and a Lebesgue measure 
for continuous variables (see \cite{Herb2} and references therein); 
for groups $G$ having no compact Cartan subgroup the discrete part is 
absent (by a theorem of Harish-Chandra, see e.g.~\cite{Taylor}). This holds   
in particular for the complex connected simple Lie groups \cite{Mackey,
Varadarajan}. The spaces $\cL_{\sigma}$ are Hilbert spaces equipped 
with an inner product $(\,,\,)_{\sigma}$ depending on $\sigma$. As a 
consequence of the multiplicity theorems decribed below one can 
identify each $\cL_{\sigma}$ with a subspace of $L^2(K)$,
where $K$ is a maximal compact subgroup of $G$. The $\pi_{\sigma}$ are 
distributions over a space of test functions 
$\cD$ on $G$ which take values in the trace class operators on 
$\cL_{\sigma}$. For the class of Lie groups considered the
distributions arise by integration with respect to a $L^{\infty}(G)$ 
function $\pi_{\sigma}(g)$. Thus, for each $\phi \in \cD$ the 
integral $\pi_{\sigma}(\phi) := \int \!dg \phi(g) \pi_{\sigma}(g)$ 
(where $dg$ is Haar measure on $G$) is a trace class operator whose matrix 
elements in some orthonormal 
basis $e_i$ in $\cL_{\sigma}$ we denote by $\int \!dg \phi(g) 
\pi_{\sigma}(g)_{ij}$. A typical choice for $\cD$ is $\cC^{\infty}_c(G)$, 
the smooth compactly supported functions on $G$. 
For $\psi \in L^2(G) \cap L^1(G)$ the operator 
$\pi_{\sigma}(\psi)$ is Hilbert-Schmidt, for $\psi \in L^1(G)$ it is 
still compact \cite{glimm,folland}. 
The $\C$-valued functions $g \mapsto \pi_{\sigma}(g)_{ij}$ are called the 
coefficients of $\pi_{\sigma}$. They are continuous functions but 
in general do not have sufficient decay to be (square) integrable 
on $G$. This weak continuity is equivalent (\cite{folland}, p.68) to 
the continuity of the $\cL_{\sigma}$-valued functions.

The Plancherel theorem states that $\phi \in \cD$ can be expanded 
in terms of eigenfunctions as
\ba
\phi(g) \is  \int_{\widehat{G}_r} \! d\nu(\sigma)\, \sum_{i,j} 
\widehat{\phi}(\sigma)_{ij} \; \pi_{\sigma}(g)_{ij}\,,
\nonum
\widehat{\phi}(\sigma)_{ij} \is \int_G \!dg 
\, \phi(g)\,\pi_{\sigma}(g^{-1})_{ij}\,. 
\label{Planch1} 
\end{eqnarray}
The Plancherel-Parseval identity
\be 
\int_G \!dg\, \phi(g)^* \psi(g) =
\int_{\widehat{G}_r}\! d\nu(\sigma) \, 
\sum_{ij} \widehat{\phi}(\sigma)^*_{ij}
\,\widehat{\psi}(\sigma)_{ij} \,,
\label{Planch2}
\end{equation}
expresses the unitarity of the map $\phi \mapsto \widehat{\phi}(\sigma)$. 

The irreducible representations on which the Plancherel measure has 
support turn out to come in families parameterized by conjugacy classes 
of Cartan subgroups of $G$. The Cartan subgroups 
are of the form $H = T \times R$, where $T$ is compact and $R \simeq \R^d$ 
for some $d$. Its dual $\widehat{H}$ has discrete parameters coming 
from $\widehat{T}$ and continuous parameters coming from $\widehat{R}$. 
The labels $\sigma$ should be thought of as elements of $\widehat{H}$ and 
the Plancherel measure takes the form 
\be 
\int_{\widehat{G_r}} \!d\nu(\sigma) = \sum_H \int_{\widehat{H}} \!d\sigma 
\,\nu(H: \sigma)\,,
\label{Planch6}  
\end{equation}
where the sum extends over the (finite number of) conjugacy classes 
of Cartan subgroups of $G$ and the functions $\nu(H: \sigma)$ are 
known (\cite{Herb2} and references therein). A Cartan subgroup here is 
defined as a maximal abelian subalgebra all of whose elements are diagonalizable 
as matrices over the complex numbers. The representations $\sigma \in 
\widehat{G}_r$ associated with a compact Cartan subgroup are 
called discrete series representations, all others are called 
(on the basis of a theorem) cuspidial principal series. Among them 
are the principal series proper associated with a unique 
conjugacy class to be described below. For example ${\rm SL}(2,\R)$ 
has two conjugacy classes of Cartan subgroups, $H = {\rm SO}(2)$ 
and $H = \{ {\rm diag}(a,a^{-1})\,|\, a \in \R \backslash \{0\}\}$.  
In contrast, ${\rm SL}(n,\R)$ with $n>2$ has $[n/2]+1$ conjugacy 
classes of Cartan subgroups, none of which is compact. The 
case $G= {\rm SO}_0(1,N)$ will be discussed in more detail in 
Appendix B.  

{\it 2.~Cartan subgroups:} The structure of the Cartan 
subgroups $H$ entering (\ref{Planch6}) is best described in terms of their Lie 
algebra $\mathfrak{h}$. For the Lie groups considered there exists
an involution $\iota$ such that the set of its fixed points 
generates a maximal compact subgroup $K$ of $G$. (More precisely 
$(G_{\iota})_0 \subset K \subset G_{\iota}$, where $G_{\iota}$ 
is the fixed point set and $(G_{\iota})_0$ is its identity component).
The involution $\iota$ of $G$ induces a decomposition of the 
Lie algebra $\mathfrak{g} = \mathfrak{k} \oplus \mathfrak{q}$, 
where $\mathfrak{k}$ and $\mathfrak{q}$ are even and odd under 
(the differential of) $\iota$, respectively. The Lie algebras $\mathfrak{h}$
can be assumed to be invariant under $\iota$. 
They are of the form $\mathfrak{h} = \mathfrak{k}_0 + \mathfrak{a}_0$, 
where  $\mathfrak{k}_0 \subset \mathfrak{k}$,   $\mathfrak{a}_0 \subset 
\mathfrak{a}$. Here $\mathfrak{a}$ is the Lie algebra of the 
subgroup $A$ in the Iwasawa decomposition $G = NAK$. To $\mathfrak{k}_0,\, 
\mathfrak{a}_0$ one can associate a subgroup $P_0 = N_0 A_0 M_0$ of $G$ 
called a ``cuspidial parabolic subgroup''. Here $N_0$ is a nilpotent 
subgroup of $N$, $A_0$ is a subgroup of $A$, and $M_0$ is such that 
$A_0 M_0$ is the centralizer of $A_0$ in $G$. There is a systematic 
technique, called parabolic induction, which allows one to construct 
unitary representations of $G$ from those of $P_0$. Almost all 
of them are irreducible and provide the above ``cuspidial principal
series'' representations. For our purposes the most important one
is the principal series proper, which is associated with the 
(up to conjugacy) unique Cartan subgroup $H$ for which 
$\mathfrak{h} \cap \mathfrak{q} = \mathfrak{a}$, i.e.~for which 
$A_0$ is all of $A$. The associated subgroup $P =  NAM$ is called 
minimal parabolic and the construction of the associated 
principal series representations will be detailed later. 

The other extreme case is when $A_0 =\{e\}$ consists of the 
identity only, in which case $P_0$ is all of $G$. 
The associated representations are precisely the above 
discrete series representations. By a theorem of 
Harish-Chandra $G$ has discrete series representations if 
and only it has a compact Cartan subgroup. Equivalently (\cite{Taylor}, 
p.282)
\be 
{\rm rank}K = {\rm rank} M + \dim A\,,
\label{discser1}
\end{equation}
where ${\rm rank}K$ is the dimension of the maximal torus of $K$.   
For example all ${\rm SO}_0(p,q)$ groups with $pq$ even 
have discrete series, as do ${\rm Sp}(n,\R)$, ${\rm SU}(p,q)$, and 
${\rm SL}(2,\R)$. Groups which do not have discrete series are 
${\rm SL}(n,\R),\,n > 2$, and ${\rm SO}_0(p,q)$ with $pq$ odd. 
Several constructions of (all) discrete series representations are 
known, see \cite{Schmid}. In particular there is an elegant 
construction based on the kernel of the Dirac operator on homogeneous 
spinor bundles, see \cite{Partha,AtiSchmid}.  

The content of the theorem mentioned after (\ref{Planch6}) is 
that the restricted dual $\widehat{G}_r$ is exhausted by the 
above cuspidial principal series representations and the discrete 
series representations.

{\it 3.~Tensor product conventions:} Let $\cH$ be a separable Hilbert 
space with 
inner product $(\;,\;)$, linear in the right and anti-linear in the left
argument. The conjugate Hilbert space $\check{\cH}$ is the Hilbert 
space with underlying additive group identical to that of $\cH$ but 
with scalar multiplication defined by $(\lb, \check{v}) \mapsto 
\lb^* \check{v}$, for $\lb \in \C$, $\check{v} \in \check{\cH}$,
and inner product $(\check{v}, \check{w}) := (w,v)$. The Hilbert 
space $\check{\cH}$ can canonically be identified with the 
space of linear forms $\cL(\cH, \C)$ on $\cH$. Indeed by 
Riesz theorem $\cL(\cH, \C) \ni (v, \cdot) \mapsto \check{v} \in 
\check{\cH}$ is a linear isomorphism of complex vector spaces.
The map $\cH \ra \check{\cH}$ is anti-linear; it associates to
each $v \in \cH$ a linear form as usual, $\check{v} =(v,\cdot\,)$, 
$\check{v}(u) = (v,u)$. The orthonormality 
relations of a basis $e_i, i\in \N$, then amount to 
$\check{e}_j(e_i) = \delta_{ij}$. If $v_i = \check{e}_i(v)$ are the 
components of a vector $v = \sum_i v_i e_i$, then the associated 
linear form $\check{v}$ has the complex conjugate components, 
$\check{v} = \sum_i v_i^* \check{e}_i$.  Suppose that $\cH$ carries in 
addition a representation $\pi$ of some Lie group $G$. Then 
$\pi(g) v$ has components $(\pi(g) v)_i = \sum_j \pi(g)_{ij} v_j$, 
where $\pi(g)_{ij} := \check{e}_i(\pi(g) e_j)$ are the matrix elements 
of $\pi(g)$. Note that in these conventions $\pi(g)$ acts by 
`matrix multiplication' on the components $v_j = \check{e}_j(v)$. 
We define the conjugate representation $\check{\pi}$ on $\check{\cH}$ 
by $(\check{\pi}(g) \check{v})(u) := \check{v}(\pi(g^{-1}) u))$.
Since $( (\check{\pi}(g) \check{v})^{\scriptstyle{\vee}},u) = 
(\check{\pi}(g) 
\check{v})(u) = \check{v}(\pi(g^{-1})u) = (v,\pi(g^{-1})u) = 
(\pi(g) v,u)$, this amounts to $[\check{\pi}(g) \check{v}]^{\scriptstyle{\vee}}
= \pi(g) v$, or $\check{\pi}(g) \check{v} = [\pi(g) v]^{\scriptstyle{\vee}}$. 
In particular $\check{\pi}(g) \check{v}$ has components  
$(\check{\pi}(g) \check{v})_i = \sum_j \pi(g)^*_{ij} v_j^*$, if 
$v_j$ are the components of $v \in \cH$.  

For separable Hilbert spaces $\cH_1$ and $\cH_2$ with orthonormal 
bases $e_i,\,i\in \N$, and $f_j,\, j \in \N$, repectively, we define the 
tensor product $\cH_2 \otimes \check{\cH}_1$ as the Hilbert space 
spanned by $f_i\, \check{e}_j =: f_i \otimes \check{e}_i$
and completed with respect to the inner product  $(f_i \otimes \check{e}_j, 
f_k \otimes \check{e}_l)_2 = \delta_{ik} \delta_{jl}$. 
The tensor product $\cH_2 \otimes \check{\cH}_1$ can be canonically identified 
with the space $\cJ_2(\cH_1,\cH_2)$ of Hilbert-Schmidt operators   
$F: \cH_1 \ra \cH_2$, equipped with the inner product 
$(F,F')_2 = {\rm Tr}[F^{\dagger} F'] = \sum_{ij} F_{ji}^* F'_{ji}$,  
where $F = \sum_{ij} f_i F_{ij} \check{e}_j$ are the components of $F$.
The isometry $\cH_2 \otimes \check{\cH}_1 \ra \cJ_2(\cH_1,\cH_2)$ 
is simply given by the extension of $(\sum_i v_i f_i) \otimes 
(\sum_j w_j e_j)^{\scriptstyle{\vee}} \mapsto \sum_{ij} v_i w_j^* f_i\, 
\check{e}_j$. 
(This isometry was in fact already used before in the 
identification  $f_i\, \check{e}_j = f_i \otimes \check{e}_i$.) 

Suppose now that $\cH_1$, $\cH_2$ carry unitary representations 
$\pi_1,\,\pi_2$, respectively. Then $\cH_2 \otimes \check{\cH}_1$ 
carries a unitary representation $\pi_2 \times \check{\pi}_1$
of $G \times G$, the outer tensor product of $\pi_2$ and $\check{\pi}_1$.   
It is given by $(\pi_2 \times \check{\pi}_1)(g_2,g_1)(v_2 \otimes \check{v}_1) := 
\pi_2(g_2) v_2 \otimes \check{\pi}_1(g_1) \check{v}_1$. In the 
realization as Hilbert-Schmidt operators this means
\be 
(\pi_2 \times \check{\pi}_1)(g_2,g_1) F = \pi_2(g_2) F \pi_1(g_1)^{\dagger} 
= \pi_2(g_2) F \pi_1(g_1)^{-1}\,. 
\label{piHS}
\end{equation}
The outer tensor product of two unitary representations is 
irreducible if and only if both constituents are. 
The diagonal representation $(\pi_2 \otimes \check{\pi}_1)(g) := 
(\pi_2 \times \check{\pi}_1)(g,g)$ of $G$ is called the inner tensor 
product of $\pi_2$ and $\check{\pi}_1$; of course it is in general not 
irreducible if $\pi_2$ and $\check{\pi}_1$ are. 

%%%%%%%%%%%%%%%%%%%%%%%%%%%%%%%%%%%%%%%%%%%%%%%%%%%%%%%%%%%%%%%%%%%%%%

{\it 4.~Relation to amenability:} The support of the Plancherel 
measure in (\ref{Planchrho}) also reflects the amenability or 
nonamenability of the underlying group (see \cite{Pat} for the definition
of amenable topological groups). For a continuous 
unitary representation $\pi$ of a locally compact group $G$ 
the support of $\pi$ is the set $\sigma \in \widehat{G}$ weakly 
contained in $\pi$. Here $\pi_1$ is said to be weakly contained 
in $\pi_2$ if every function of positive type can be approximated,
uniformly on compact subsets of $G$, by finite sums of functions of 
positive type associated with $\pi_2$. Here functions of 
positive type can be identified with the diagonal matrix 
elements $g \mapsto (v, \pi(g) v)$ of a representation.  
By definition $\widehat{G}_r$, the reduced
dual of $G$, is the support of the (left or right) regular 
representation of $G$; see \cite{Dixmier1}, Definitions 18.1.7 and 18.3.1.
An amenable locally compact group $G$ is characterized by the property 
that $\widehat{G}_r = \widehat{G}$ (\cite{Dixmier1}, Prop.~18.3.6). 
In fact, whenever $\widehat{G}_r$ weakly contains a single finite dimensional 
continuous unitary representation $\widehat{G}_r = \widehat{G}$ follows 
(\cite{Dixmier1}, Prop.~18.3.6 and Addendum 18.9.5). For the group itself 
one has: 
a connected semisimple Lie group with a finite center is amenable 
if and only if it is compact (see \cite{Pat} or \cite{Zimmer}, 
Prop.~4.1.8). All the noncompact (linear reductive) 
Lie groups considered here are therefore nonamenable. It follows that 
$\widehat{G}_r$ is a proper subset of $\widehat{G}$ and that 
$\widehat{G}_r$ cannot contain any finite dimensional continuous 
unitary representation. In contrast the Euclidean group 
${\rm ISO}(N)$ considered in Appendix C is amenable.  

Sometimes also the concept of an amenable representation is useful,
which relates to (\ref{piHS}). Specifically one is interested 
in situations where $\pi_1 \otimes \pi_2$ 
contains the unit representation of $G$, i.e.~the singlet. A simple case 
is when $\pi$ is finite dimensional; then $\pi \otimes \check{\pi}$ always 
contains the unit representation (as $F = \1$ by (\ref{piHS}) is clearly an 
invariant tensor). More generally one has (\cite{BHV}, Prop. 
A.1.11 and Corollary A.1.12): Let $\pi_1,\, \pi_2$ be unitary 
representations of the Lie group $G$ on $\cH_1,\,\cH_2$. Then the 
following are equivalent: (i) $\pi_1 \otimes \pi_2$ 
contains the unit representation. (ii) there exists a finite 
dimensional representation $\pi$ which is a subrepresentation 
of both $\pi_1$ and $\pi_2$. Further, if $\pi_1$ is irreducible 
condition (ii) can be replaced by: (ii)' $\pi_1$ is 
finite dimensional and is contained in $\check{\pi}_2$.  
The unitary representation $\pi$ is called amenable if 
for $\pi_1 = \pi$, $\pi_2 = \check{\pi}$ any one of the
conditions (i)--(iii) is satisfied. This in turn can be 
shown to be equivalent to \cite{bekka}: 
a unitary representation $\pi$ of a Lie group $G$ on a 
Hilbert space $\cH$ is amenable if there exists a positive 
linear functional $\om$ over $\cB(\cH)$ (the $C^*$-algebra of 
bounded linear operators on $\cH$) 
such that $\om(\pi(g) T \pi(g)^{-1}) = \om(T)$, 
for all $g \in G$ and all $T \in \cB(\cH)$. 
Further a locally compact group is amenable iff every ${\pi\in\widehat G}$
is amenable \cite{bekka}. For a simple noncompact Lie group (connected 
with finite center and rank $> 1$) the only amenable representations are 
those containing the trivial one.

%%%%%%%%%%%%%%%%%%%%%%%%%%%%%%%%%%%%%%%%%%%%%%%%%%%%%%%%%%%%%%%%%%%%%%%%%%

{\it 5.~Characters:} The coefficient functions  $g \mapsto \pi_{\sigma}(g)_{ij}$ 
form a unitary irreducible representation, viz
\ba 
&& \pi_{\sigma}(g_1 g_2)_{ij} = \sum_k \pi_{\sigma}(g_1)_{ik} 
\,\pi_{\sigma}(g_2)_{kj} \,,
\nonum
&& \pi_{\sigma}(e)_{ij} = \delta_{ij}\;,\sspace 
\pi_{\sigma}(g^{-1})_{ij} = [\pi_{\sigma}(g)_{ji}]^*\,.
\label{pi1}
\end{eqnarray}
For a compact Lie group the coefficients also obey orthogonality 
and completeness relations essentially summarizing the content of 
the Plancherel (or Peter-Weyl) expansion. In the case of a noncompact 
Lie group these have no direct counterpart in that double sums over 
products of the matrix coefficients or traces diverge. Instead character 
functions and the associated spectral projectors provide the appropriate 
substitute for orthogonality and completeness relations.   

Characters are defined as follows \cite{Varadarajan}. 
Since $\widehat{\phi}(\sigma)$ is a 
trace class operator for all $\phi \in \cD$ the trace
\be 
{\bf \Theta}_{\sigma}(\phi) := \sum_i \widehat{\phi}(\sigma)_{ii} = 
\sum_i \int_G\! dg \,\phi(g) \pi_{\sigma}(g^{-1})_{ii}\,,
\label{trace1}
\end{equation}
is well-defined and independent of the choice of orthonormal basis
on $\cL_{\sigma}$. Thus ${\bf \Theta}_{\sigma} : \cD \ra \C$ is a 
distribution over $\cD$ for every unitary irreducible representation 
$\pi_{\sigma}$. It characterizes such a representation in that 
${\bf \Theta}_{\sigma_1} = {\bf \Theta}_{\sigma_2}$
holds if and only if the representations $\pi_{\sigma_1}$ and 
$\pi_{\sigma_2}$ are unitarily equivalent. A representation $\pi_{\sigma}$ 
is called tempered if the distribution ${\bf \Theta}_{\sigma}$ extends
continously to $\cS(G)$, the Schwartz space of functions $\phi 
\in \cC^{\infty}(G)$ such that $\phi$ and all its derivatives 
are square integrable on $G$.  The sum in (\ref{trace1}) can 
of course not be pulled inside the integral: as all eigenvalues 
of $\pi_{\sigma}(g)$ have modulus one the sum over 
$\pi_{\sigma}(g)_{ii}$ diverges. However for the class of Lie groups considered 
a regularity theorem ensures the existence of a locally integrable 
function $\Theta_{\sigma}$ such that 
\be 
{\bf \Theta}_{\sigma}(\phi) = \int_G \!dg \, 
\phi(g) \Theta_{\sigma}(g^{-1})\,\quad \mbox{for all} \;\; \phi \in 
\cD\,.
\label{trace2}
\end{equation} 
The function $\Theta_{\sigma}$ is constant on conjugacy classes in the 
sense that $\Theta_{\sigma}(g g' g^{-1}) = \Theta_{\sigma}(g')$ for all 
$g \in G$ and $g' \in G'$. Here $G'$ is a dense open subset of $G$
characterized by the fact that each of its elements lies in precisely one 
Cartan subgroup $H$ of $G$, see e.g.~\cite{Varadarajan}.  
The character function $\Theta_{\sigma}$ is also an 
eigenfunction of $Z(G)$, the abelian algebra of all bi-invariant 
differential operators on $G$. In terms of them the Plancherel expansion 
(\ref{Planch1}) can be rewritten as
\be 
\phi(g) = \int_{\widehat{G}_r}\!d\nu(\sigma) \, 
(\Theta_{\sigma}*\phi)(g)\,,\quad \phi \in \cD\,.
\label{Planch4}
\end{equation}
Here 
\be 
(\Theta_{\sigma}*\phi)(g) := \int_G \! dg'\, 
\Theta_{\sigma}(g{g'}^{-1}) \phi(g') = 
\sum_i   \int_G \! dg' \,\phi(g') \sum_j 
\pi_{\sigma}(g)_{ij} \pi_{\sigma}({g'}^{-1})_{ji}\,.
\label{Planch5}
\end{equation}
Due to the properties of $\Theta_{\sigma}$ the above tempered irreducible 
representations come in families parameterized
by conjugacy classes of Cartan subgroups of $G$.

The bi-invariant function $(g,g') \mapsto \Theta_{\sigma}(g {g'}^{-1})$ 
in (\ref{Planch5}) can be viewed as a `regularized' version of the formal double
sum that would arise by pulling the sum over $i$ inside the integral.   
There are two natural ways to achieve such a regularization. 

One is by performing averages over Borel sets in $\widehat{G}_r$,
which gives rise to spectral projectors: 
\begin{subeqnarray} 
\label{EGspec}
&& E_I(g{g'}^{-1}) := \int_I \!d\nu(\sigma)\, \Theta_{\sigma}(g{g'}^{-1}) \;,
\\
&& \int \!dg \,E_I(g_1g^{-1}) E_J(gg_2^{-1}) = 
E_{I \cap J}(g_1g_2^{-1})\,,
\\
&& E_{\widehat{G}_r}(g{g'}^{-1}) = 
\int_{\widehat{G_r}} \!d\nu(\sigma) \, \Theta_{\sigma}(g{g'}^{-1}) 
= \delta(g,g')\,.
\end{subeqnarray}

{\it 6.~K-finite functions:} 
A literal way to take the sum over $i$ in (\ref{Planch5}) inside the 
integral is by restricting the class of functions $\phi$ to the 
$K$-finite ones. A function $f \in \cC_c^{\infty}(G)$ is called
left (resp.~right) {\it $K$-finite} (\cite{Wallach} p.236) if the set 
$\{f(kg), k\in K\}$ (resp.~$\{f(gk), k\in K\}$) lies in a finite dimensional 
subspace of $\cC(G)$, the continuous functions on $G$. It is called 
(bi)-K-finite if both holds. Let $K \ni k \mapsto r_{\kappa}(k)$, 
$\kappa \in \widehat{K}$, be the unitary irreducible representation 
of $K$ with hightest weight $\kappa$ on a vector space $V_{\kappa}$,
$d_{\kappa}:= \dim V_{\kappa}$. Their characters $k \mapsto \chi_{\kappa}(k) 
:= {\rm Tr}[r_{\kappa}(k)]$ 
obey $\chi_{\kappa} * \chi_{\kappa'} = \delta_{\kappa,\kappa'} 
d_{\kappa}^{-1} \chi_{\kappa}$ (where we took $\kappa$ to label a unitary
equivalence class and $*$ denotes the convolution product with 
respect to $K$). For any finite subset $I \subset \widehat{K}$ 
then $E_I := \sum_{\kappa \in I} d_{\kappa} \chi_{\kappa}$ 
is the spectral projector; in particular $E_I * E_J = E_{I \cap J}$.
The Fourier expansion takes the form $f = \sum_{\kappa \in \widehat{K}} 
d_{\kappa} f*\chi_{\kappa}$ and converges in the $L^2(K)$ norm. 
One can view the characters    
as functions on $G$ with support on $K$ only and convolute functions 
in $\cC_c^{\infty}(G)$ with the projectors $E_I$ (with the convention 
$(f*h)(g) = \int \!dg' f(g {g'}^{-1}) h(g') = \int \!dg' f({g'}^{-1}) h(g' g)$). 
Then (\cite{Wallach}, p.237)
\be 
f \in \cC_c^{\infty}(G) \;\; \mbox{is}\;\; \left\{ \begin{array}{ll}
\mbox{left $K$-finite} & \quad \mbox{iff} \;\;\;\; E_I * f = f\,,\\
\mbox{right $K$-finite} & \quad \mbox{iff} \;\;\;\; f*E_I =f \,,
\end{array}
\right.\,
\label{Kfinite} 
\end{equation}
for some finite $I \subset \widehat{K}$. The function $f$ is 
bi-$K$-finite if $E_I * f * E_I = f$ holds.

A representation $\pi$ of $G$ on a Hilbert space $\cH$ is called $K$-finite 
(\cite{Wallach}, p.232) if its restriction to the compact subgroup $K$ 
is unitary and decomposes into a unitary sum of irreducibles $(r_{\kappa}, 
V_{\kappa})$, $\kappa \in \widehat{K}$, each occuring with finite multiplicity 
$m_{\kappa}$. That is, $\cH = \bigoplus_{\kappa \in \widehat{K}} 
m_{\kappa} V_{\kappa}$ as a representation of $K$. Let $P_{\kappa}: 
\cH \ra m_{\kappa} V_{\kappa}$ be the orthogonal projection. 
Note that an alternative characterization of a $K$-finite 
representation $\pi$ is that $P_{\kappa}$ is an operator of finite rank on $\cH$ 
for all $\kappa \in \widehat{K}$. 
A vector $v \in \cH$ is called $K$-finite if $\pi(k) v, \, k\in K$, 
generates a finite dimensional subspace of $\cH$ (\cite{Lang}, p.25). 
Evidently this is the case iff $P_I v = v$ for some 
finite $I \subset \widehat{K}$. To study the relation between 
$K$-finite functions and $K$-finite vectors the following 
explicit realization of the projectors is useful:
\be 
P_{\kappa} v = d_{\kappa} \int_K \!dk \,\chi_{\kappa}(k^{-1}) \pi(k) v\;,
\quad v \in \cH\,.
\label{Pkappa}
\end{equation}
Consistency requires that $\pi(k) \circ P_{\kappa} = 
P_{\kappa} \circ \pi(k)$, $k \in K$, which indeed is a property 
of the right hand side of (\ref{Pkappa}) using that the 
character $\chi_{\kappa}$ is constant on $K$ conjugacy classes. 
To verify (\ref{Pkappa}) first note that the matrix elements of 
$\pi(k), \,k \in K$, are 
blockdiagonal in the basis $\{e_{\kappa, s},\,s=0,\ldots, m_{\kappa} d_{\kappa}\!-\!1,
\; \kappa \in \widehat{K}\}$, where $e_{\kappa, m + d_{\kappa}} = e_{\kappa,m}$,
$m =0,\ldots, d_{\kappa}\!-\!1$, is an orthonormal basis of $V_{\kappa}$. 
Explicitly 
\be 
(e_{\kappa' s'}, \pi(k) e_{\kappa s}) = \delta_{\kappa',\kappa} \;
r_{\kappa}(k)_{s's}\,.
\label{piblock}
\end{equation}
Any $v \in \cH$ can by assumption be expanded as $v = \sum_{\kappa,s} 
(e_{\kappa s} , v) \, e_{\kappa s}$ and $\pi(k)$ acts on 
$m_{\kappa} V_{\kappa}$ as the blockdiagonal matrix $r_{\kappa}(k)$.
Thus, to verify (\ref{Pkappa}) one only has to show that $(e_{\kappa' m'}, P_{\kappa} v) 
= \delta_{\kappa',\kappa} (e_{\kappa m'}, v)$. Taking the trace over $m_1=m_2$ in 
\be 
\int_K \! dk \,r_{\kappa}(k^{-1})_{m_1 m_2} \, r_{\kappa'}(k)_{m_3 m_4} 
= \frac{1}{d_{\kappa}} \delta_{\kappa \kappa'} \delta_{m_1 m_4} \delta_{m_2 m_3}\,,
\label{rortho}
\end{equation}
(see e.g.~\cite{BR}, p.170) this readily follows. 
 
The interest of these constructions lies in fact that any $K$-finite
representation is a direct orthogonal sum of unitary irreducible 
representations (\cite{Wallach}, Lemma 8.6.22). The irreducible 
representions are of course also $K$ finite and for them 
explicit bounds on the multiplicities are available. For the purposes here
two results are relevant: Let $G$ be a semi-simple linear connected 
real Lie group and $K$ a maximal compact subgroup. Then $\pi_{\sigma}$ is 
$K$-finite for any $\sigma \in \widehat{G}$ and the irreducible representation 
$\kappa \in \widehat{K}$  occurs in $\pi_{\sigma}|_K$ at most with 
multiplicity $d_{\kappa}$ (see \cite{Dixmier1}, p.331). A theorem by 
Harish-Chandra (\cite{Warner}, p.319) states that essentially the same is 
true for any 
connected semi-simple Lie group with finite center (with a technically 
slightly different notion of irreducibility). On account of these 
multiplicity bounds one can identify  $\cL_{\sigma}$ 
as a vector space with $\bigoplus_{\kappa} m_{\kappa} V_{\kappa}$ 
and since the multiplicities $m_{\kappa} \leq d_{\kappa}$ 
do not exceed those in the decomposition of $L^2(K)$ 
(which equal $d_{\kappa}$) one can identify each $\cL_{\sigma}$ 
with a subspace of $L^2(K)$. In certain cases the upper bound 
on the multiplicities is even sharper. In the case $G = {\rm SO}_0(1,N)$, 
$K = {\rm SO}(N)$ we focus on, each $\kappa \in \widehat{K}$ can occur at 
most once in $\pi_{\sigma}|_K$, by a result due to Dixmier 
\cite{Dixmier2}. Another case when this happens is for $G= {\rm SL}(2,\C)$ 
and $K = {\rm SU}(2)$, see \cite{Warner}, p.317, where also the general 
conditions for $m_{\kappa} \leq 1$ are discussed.

Let now $\pi$ be a $K$-finite representation and write $\pi(\psi) 
= \int \!dg \,\psi(g) \pi(g)$ for $\psi \in \cC_c^{\infty}(G)$. Using 
Eq.~(\ref{Pkappa}) one readily verifies the following relations 
\be 
\pi(E_{\kappa}*\psi) = \pi(\psi) P_{\kappa} 
\sspace 
\pi(\psi*E_{\kappa}) = P_{\kappa} \pi(\psi)\,,
\label{EPkappa}
\end{equation}
and similarly for the two-sided projections. The same holds for 
the Fourier coefficients $\int \!dg\, \psi(g) \pi(g^{-1}) = 
(\pi* \psi)(e)$. For an irreducible representation $\pi_{\sigma},\,
\sigma \in \widehat{G}$, and $\psi \in L^1(G) \cap L^2(G)$, the 
relations (\ref{EPkappa}) then imply the following decomposition 
of the Hilbert-Schmidt operator $\widehat{\psi}(\sigma) : 
\cL_{\sigma} \ra \cL_{\sigma}$: 
\ba 
\widehat{\psi}(\sigma) \is \sum_{\kappa_2, s_2, \kappa_1 s_1} 
\!\!e_{\kappa_2 s_2}\, \widehat{\psi}(\sigma)_{\kappa_2 s_2,\kappa_1 s_1} 
\,\check{e}_{\kappa_1 s_1} 
\nonum
\is \sum_{\kappa_2, \kappa_1} P_{\kappa_2} \widehat{\psi}(\sigma) P_{\kappa_1} 
= \sum_{\kappa_2, \kappa_1} 
(E_{\kappa_1}\! \!* \!\!\psi \!* \!\!E_{\kappa_2})^{\widehat{\;\;\;\,}}(\sigma)\,.
\label{FTkappadecomp}
\end{eqnarray} 
For left $K$-finite functions the sum over $\kappa_1$ is finite, 
for right $K$-finite functions the one over $\kappa_2$ is, and 
for bi-$K$ finite functions both sums are finite. 
Here $\check{e}_{\kappa s}$ is the basis dual to $e_{\kappa s}$, $s=0, \ldots,
m_{\kappa} \dim V_{\kappa} \! - \!1$. In particular 
$\widehat{\psi}(\sigma)_{\kappa_2 s_2,\kappa_1 s_1} = 
\check{e}_{\kappa_2 s_2}(\widehat{\psi}(\sigma) e_{\kappa_1 s_1}) = 
(e_{\kappa_2 s_2}, \widehat{\psi}(\sigma) e_{\kappa_1 s_1})_{\sigma}$.

%%%%%%%%%%%%%%%%%%%%%%%%%%%%%%%%%%%%%%%%%%%%%%%%%%%%%%%%%%%%%%%%%%%%%%%%

{\it 7.~Harmonic analysis on $G/K$:} If instead of Fourier analyzing 
functions in $L^2(G)$ one is only interested in $L^2(G/K)$ functions, 
where $Q \simeq G/K$ is a symmetric space of noncompact type, the 
harmonic analysis simplifies considerably. In group theoretical 
terms it amounts to the decomposition of the quasiregular representation 
$\ell_1$ of $G$ on $L^2(Q)$. We resume the 
specifications and notations of Section A.1; in particular $K$ 
is a maximal compact subgroup  of $G$, so that $Q$ is an indecomposable 
Riemannian symmetric space. The key simplification is 
that for the harmonic analysis on $G/K$ only a subset of the 
principal series representations proper is needed (see \cite{Helgason1} and 
\cite{Helgason2}, Section IV.7). 
Recall that the principal series representations proper are 
those induced by the minimal parabolic subgroup $P = NAM$,
where $G = NAK$ is the Iwasawa decomposition of $G$ and 
$M$ is the centralizer of $A$ in $K$. The inducing construction 
will be described below. The upshot is that the principal 
series representations $\pi_{\nu,\xi}$ are labeled by 
a character $\nu: A \ra U(1)$ of $A$ and by a unitary 
(finite dimensional) irreducible representation $\xi \in 
\widehat{M}$ of $M$. The principal series representations   
associated with the singlet $\xi =0$ of $M$ are called 
the {\it spherical principal series} representations
(or minimal or class 1 principal series; see e.g.~\cite{Warner}, Vol.1, 
p.462). These representations $\pi_{\nu,0}$ are thus labeled by 
elements of $\widehat{A}$ only, which can be identified with a 
subset of $\R^{\dim A}$. We write $\widehat{Q}$ for this subset 
and label the characters $\nu = \nu_{\om}$ and the representations 
$\pi_{\om,0} := \pi_{\nu_{\om},0}$ by points in $\om \in 
\widehat{Q}$. The abstract definition of $\widehat{G/K}$ 
as the subset of $\widehat{G}_r$ needed for the harmonic 
analysis on $G/K$ thus turns into the bijection $\widehat{G/K} 
\simeq \widehat{Q} \subset \R^{\dim A}$. Moreover the 
Plancherel measure $d\nu(\sigma)|_{\widehat{G/K}}$ 
restricted to the spherical principal series is absolutely continuous 
with respect to the Lebesgue measure $d\om$ on $\R^{\dim A}$,
\be 
d\nu(\sigma)\Big|_{\widehat{G/K}} = \frac{d\om}{|c(\om)|^2}\,.
\label{cfunction}
\end{equation}
Here $c(\om)$ is the Harish-Chandra $c$-function, for which 
an explicit formula in terms of the structure of $G/K$  is 
known. Concerning the $K$ content of the $\pi_{\om, 0}$ representations,
it is known that all of them contain the $K$-singlet with 
multiplicity $1$ (\cite{Helgason1}, p.414). 
As a consequence the representations $\pi_{\om, 0}$ can be set into 
one-to-one correspondence to $K$-spherical functions. A continuous 
function $p$ on $G$ is called $K$-spherical (\cite{Helgason2}, 
p.357) if it satisfies 
\be 
\int \! dk \,p(g_1 k g_2) = p(g_1) p(g_2)\,.
\label{Kspherical}
\end{equation}
This implies that $p$ is $K$-bi-invariant and normalized $p(e) =1$. 

One can explicate this structure by writing out the coefficients of
$\pi_{\om,0}$ in the $K$-adapted basis of Section 2.4 using 
(\ref{piblock}). An equivalent characterization of the matrix elements in 
(\ref{piblock}) is then 
\be 
\pi_{\om,0}(k_1 g k_2)_{\kappa s, \kappa' s'} = 
\sum_{s_1, s_2} r_{\kappa}(k_1)_{s s_1} \,
\pi_{\om,0}(g)_{\kappa s_1, \kappa' s_2} \, r_{\kappa}(k_2)_{s_2 s'}\,.
\label{Dix3}  
\end{equation} 
In particular one sees: $\pi_{\om,0}(g)_{00,\kappa s}$ is left $K$-invariant, 
$\pi_{\om,0}(g)_{\kappa s,00}$ is right $K$-invariant, and 
$\pi_{\om,0}(g)_{00,00}$ is $K$-spherical. 
The right K-invariant functions $g \mapsto [\pi_{\om,0}(g)]_{\kappa s,00}$ 
are sufficient for the harmonic analysis on the right coset space $G/K$. 
To see this observe that for a left $K$-invariant function ($\phi(gk) 
= \phi(g)$, for all $k \in K$) the decomposition (\ref{Planch1}) specializes to 
\ba 
\widehat{\phi}(\sigma)_{\kappa s, \kappa' s'} 
\is \int_{G/K} \!\!d\gamma_G(gK) \,\phi(gK) 
\int_K\! d\gamma_K(k)\, \pi_{\sigma}(k^{-1}g^{-1})_{\kappa s, \kappa' s'} = 
\widehat{\phi}(\sigma)_{00,\kappa s} \delta_{\kappa',0} \delta_{s',0}\,,
\nonum
\phi(gK) \is \int_{\widehat{G/K}} \! d\nu(\sigma) \, \sum_{\kappa,s} 
\widehat{\phi}(\sigma)_{00,\kappa s} \, 
\pi_{\sigma}(gK)_{\kappa s,00}\,. 
\label{GKPlanch1}
\end{eqnarray}
Here we wrote $\widehat{G/K}$ for the subset of representations in 
$\widehat{G}_r$ for which the matrix elements $\widehat{\phi}(\sigma)_{\kappa s,00}$ 
are nonzero. According to the above results it consists of spherical 
principal series representations only, and there is a bijection $
\widehat{G/K} \simeq \widehat{Q} \subset \R^{\dim A}$ to a subset  
of $\R^{\dim A}$. Combined with (\ref{cfunction}) this allows for 
a very explicit description of the harmonic analysis on $G/K$. 
Via the Iwasawa decomposition the section $g_s(q)$ 
provides a one-to-one correspondence between points $q \in Q$ and 
right $K$-orbits (recall that $g_s(q)$ equals 
$n a$ viewed as a function of $q$). We define
\be 
E_{\om, \kappa s}(q) := [\pi_{\om,0}(g_s(q))]_{0\, 0,\kappa s}\,.
\label{Epi2}
\end{equation}
The functions $E_{\om, \kappa s}(q)$ are equivariant with respect to 
$\pi_{\om,0}$, i.e.
\be 
E_{\om, \kappa s}(g^{-1} q) = \sum_{\kappa' s'} 
E_{\om, \kappa' s'}(q)\, \pi_{\om,0}(g)_{\kappa' s',\kappa s}\,.
\label{Epi4}
\end{equation} 
The spectral synthesis formulas then assume the form
\ba 
\psi(q) \is \int_{\widehat{Q}} \!\frac{d\om}{|c(\om)|^2}\, 
\sum_{\kappa s} \widehat{\psi}(\om)_{\kappa s} E_{\om,\kappa s}(q)\,,
\nonum
\widehat{\psi}(\om)_{\kappa s} \is \int_Q \! d\gamma_Q(q)\, 
\psi(q) E_{\om,\kappa s}(q)\,.
\label{Epi5}
\end{eqnarray} 
Finally we mention that in the fiber decomposition
\be 
L^2(G/K) \simeq \int^{\oplus} \!\frac{d\om}{|c(\om)|^2}\, \cL_{\om}\,,
\sspace \ell_1 \simeq \int^{\oplus} \frac{d\om}{|c(\om)|^2}\, \pi_{\om,0}\,,
\label{GKharmonic}
\end{equation}
all fiber spaces $\cL_{\om}$ are isometric to $L^2(K/M)$. This can be 
understood from the fact that generic (nonspherical) principal series 
representations can be modeled on $L^2(K)$ (see below); so for the 
$M$ singlets this gives model spaces isometric to $L^2(K/M)$.

%%%%%%%%%%%%%%%%%%%%%%%%%%%%%%%%%%%%%%%%%%%%%%%%%%%%%%%%%%%%%%%%%%%%%%%%%%%

{\it 8.~Principal series:} As is clear from the preceeding discussion 
the principal series of unitary irreducible representations is 
at the core of the harmonic analysis for noncompact Lie groups.  
For complex (noncompact semisimple connected) Lie groups (with a finite 
center) it suffices in fact for the harmonic analysis. Since we will 
need a number of results that occur in its construction we present here 
a concise but in principle selfcontained summary thereof. For definiteness we 
consider again the class of linear reductive Lie groups, though 
everything goes through also for arbitary non-compact semi-simple Lie 
groups with a finite center.

The principal series arise as special cases of so-called 
{\it multiplier representations} defined as follows. Consider 
a diffentiable manifold $\cM$ carrying a differentiable $G$ action 
$\cM \ni m \mapsto g.m \in \cM$. For a finite dimensional 
vector space with inner product $\bra \;,\;\ket_V$ let $L^2(\cM, V)$ be 
the Hilbert space of functions $f: \cM \ra V$ square integrable 
with respect to 
\ba 
&& (f_1, f_2) := \int d\gamma(m) \bra f_1(m), f_2(m) \ket_V\;,
\nonum
&& \bra f_1(m), f_2(m) \ket_V  := \sum_s f_1(m)^*_s \,f_2(m)_s\,.
\label{multirep1}
\end{eqnarray}
Let further $G \times \cM \ni (g,m) \mapsto \chi(g,m)$, 
$\chi(g,m) : V \ra V$, be a cocycle satisfying 
\be 
\chi(g_1 g_2, m) = \chi(g_1, m) \chi(g_2, g_1.m)\,,\quad 
\chi(e,m) = \1\,.
\label{multirep2}
\end{equation}
Set 
\be 
[\pi(g) f](m) := \chi(g,m) f(g.m) 
\sqrt{\frac{d\gamma_g(m)}{d \gamma}} \,,
\label{multirep3}
\end{equation}
where $d\gamma_g = d(\gamma \circ g)$ is the translated measure 
and $d\gamma_g/d\gamma$ is the Radon-Nikodym derivative.       
The latter ensures both the representation property and 
the unitarity with respect to the inner product 
(\ref{multirep1})
\be 
\pi(g_1) \pi(g_2) = \pi(g_1 g_2) \,,
\quad (\pi(g_1) f_1, \pi(g_2) f_2) = (f_1, f_2) \,.
\label{multirep4}
\end{equation}
One now applies this construction principle to the group 
manifold $K$, where $K$ is the maximal compact subgroup 
of $G$. Via the Iwasawa decomposition $G = NAK = B K$ it 
carries a $G$-action induced by the right translations.
Explicitly let $g = n(g) a(g) k(g) = b(g) k(g)$ 
be the unique Iwasawa decomposition of some $g \in G$. 
Then 
\be 
B\backslash G \simeq K \,,\quad k_0 g = b(k_0g) k(k_0g) \,,
\quad k_0 \in K,
\label{Kmultirep1}
\end{equation}
that is, the coset space $B \backslash G$ of left equivalence classes 
$b g \sim g$, $b \in B$, is isometrically identified with 
$K$ by picking the representative $k(k_0 g) =: g[k_0]$ in the 
Iwasawa decomposition. For each $g \in G$ the map $k_0 \mapsto g[k_0]$ 
defines a diffeomorphism on $K$, which is the $G$-action 
on $K$ needed to define the multiplier representation (\ref{multirep3}). 
We write $d(g[k])$ for the translated bi-invariant Haar measure $dk$ 
in $K$ and $d(g[k])/dk$ for the corresponding Radon-Nikodym 
derivative. According to the general construction 
\be 
[\pi(g) f](k) = \chi(g,k) f(g[k]) \sqrt{ \frac{d(g[k])}{dk}}\,,
\label{Kmultirep2}
\end{equation}
defines a unitary representation on $L^2(K,V)$ for any 
cocycle $\chi(g,k): V \ra V$. It remains to compute the 
Radon-Nikodym derivative. It comes out to be $\Delta(b(kg))^{-1}$,
where $b \mapsto \Delta(b)$ is the (right) modular function of 
the (non-unimodular) subgroup $B = AN$. That is,
\be 
\int_K\! dk \, f(k) = \int_K \!dk\, f(g[k]) \Delta(b(kg))^{-1}\,,
\label{Kmultirep3}
\end{equation}
for all $f \in \cC_c(K)$. The proof is simple and instructive, so 
we present it here, also in order to set the conventions 
(see e.g. \cite{Lang}, p.44f and \cite{Varadarajan}, p.83f), with the 
opposite 
conventions). Let $db$ be the left invariant 
Haar measure on $B$ and $dg, \, dk$ the bi-invariant Haar measures 
on $G,\,K$, respectively. Then  
\be 
dg = db dk\, \quad \mbox{for functions of}\;\; g = bk\,,
\label{Bmeasure1}
\end{equation}  
where the order $bk$ (rather than $kb$) is important. 
Indeed, by definition the map $B \times K \ra G,\; 
(b,k) \mapsto bk$ is an isomorphism. There exists therefore
an analytical function $J: B \times K \ra \R^+$ such that 
$dg = J(b,k) db dk$. Since $dg = d(b_0 g k_0)$ for all 
$b_0 \in B$, $k_0 \in K$ it follows that 
$J(b_0 b, k k_0) = J(b,k)$. Thus $J$ must be a constant,
which by a change of normalization can be set to unity. 
Note that if functions of $kb$ were considered the 
counterpart of (\ref{Bmeasure1}) would read $dg = d_rb dk$, 
with $d_rb$ the right Haar measure on $B$. Our convention for the 
modular function is that of \cite{folland}, p.46ff]  
\be 
d(b b_0) = \Delta(b_0)\,db \,,
\label{Bmeasure2}
\end{equation}
so that $d_rb = \Delta(b^{-1}) db$. With these preparations 
at hand the verification of (\ref{Kmultirep3}) is straightforward. 
Let $E \in \cC_b(B)$ be a function such that $\int_B \!db E(b) = 1$. 
Given an arbitrary $f \in \cC_c(K)$ define a function 
$F \in \cC_c(G)$ by $g = bk \mapsto E(b) f(k)$. Then 
\be 
\int \! dk f(k) = \int_G \! dg F(g) = \int_G \! dg F(g g_0) 
= \int\! dk db F(bk g_0) \,,
\label{Kmultirep4}
\end{equation}
for all $g_0 \in G$. By (\ref{Kmultirep1}) $k g_0$ decomposes 
as $k g_0 = b(kg_0) g_0[k]$, so that $F(bk g_0) = 
E(bb(kg_0)) f(g_0[k])$. Inserting into (\ref{Kmultirep4}) and  
shifting the integration variable $b \ra b b(kg_0)^{-1}$ 
gives (\ref{Kmultirep3}).

To proceed we describe the action of $G$ on itself in terms 
of the Iwasawa decomposition $G = NAK$. Converting the 
relations in \cite{Varadarajan}, p.84 into the present conventions one 
has 
\ba 
\label{Iwaaction}
k(g_1 g_2) \is k(k(g_1) g_2) = g_2[k(g_1)]\,,
\nonum
a(g_1 g_2) \is a(g_1) a(k(g_1) g_2) \,, 
\\[1mm]
n(g_1 g_2) \is n(g_1)[ a(g_1) n(k(g_1) g_2) a(g_1)^{-1}]\,. 
\nonumber
\end{eqnarray} 
As a check note $n(g_1 g_2) a(g_1 g_2) k(g_1 g_2) = 
n(g_1) a(g_1) (nak)(k(g_1)g_2) = g_1 g_2$. In particular it 
follows that $a(g,k) := a(kg)$ is an $A$-valued cocycle,
\be 
a(g_1 g_2,k) = a(g_1, k) a(g_2 , g_1[k])\,,\quad 
a(e,k) =e\,.
\label{acocycle}
\end{equation}
We evaluate it on two types of continuous group homomorphisms. 
The first, $\delta:A \ra \R_+$, is obtained from 
\be 
d(a n a^{-1}) =: \delta(a)^2 dn \,,\sspace 
\Delta(an) = \delta(a)^2\,,
\label{Kmultirep5}
\end{equation}
where $g=nak$ is the Iwasawa decomposition. The first relation in 
(\ref{Kmultirep5}) defines $\delta$, the second readily follows from 
$db = da dn$ for functions of $b=an$ (see \cite{Varadarajan}, p.80, 
\cite{Lang}, p.39). 
The second group homomorphism $\nu: A \ra U(1)$ assigns to each 
$a \in A$ a complex phase $\nu(a) \in U(1)$.    

We define a representation of $G$ on $L^2(K,V)$ by 
\be 
[\pi_{\nu}(g)f](k) := \delta(a(g,k))^{-1} \nu (a(g,k)) f(g[k])\,.
\label{Kmultirep6}
\end{equation}
Since it is obtained by specialization of (\ref{Kmultirep2}) 
it is unitary. The following property of $\pi_{\nu}$ will allow 
one to restrict it to subspaces of $L^2(K,V)$ such that the 
resulting representations are equivalent to the principal series 
representations sought after. Namely: 

(i) The restriction of 
$\pi_{\nu}$ to $K$ is the right regular representation of $K$, 
i.e.~$\pi_{\nu}(k_0)f(k) = f(kk_0)$.  
(ii) $\pi_{\nu}$ commutes with the left regular representation 
of $M$
\be 
\ell(m) \circ \pi_{\nu}(g) = \pi_{\nu}(g) \circ \ell(m), 
\quad \mbox{for all} \;\;m \in M\,.
\label{Kmultirep7}
\end{equation}
Here $M$ is the centralizer of $A$ in $K$,
that is the subgroup $M \subset K$ whose elements commute with all 
elements of $A$. Property (i) follows from $a(k_0,k) =e$ and
$k_0[k] = kk_0$. Equation (\ref{Kmultirep6}) is equivalent to 
\be 
a(m g) = a(g) \,,\quad 
k(m g) = m k(g) \,,\quad \mbox{for all} \;\; m\in M\,.
\label{Kmultirep8}
\end{equation}
To verify (\ref{Kmultirep7}) it suffices to note that 
$n(mg) = m n(g) m^{-1}$ since $M$ normalizes $N$. 
Thus $m g = n(m g) a(m g) k(m g) = m n(g) a(m g) 
m^{-1} k(m g)$, which gives (\ref{Kmultirep7}).  

Let now $m \mapsto 
r_{\xi}(m)$, $\xi \in \widehat{M}$,  be an 
irreducible representation of $M$ on $V_{\xi}$ and consider 
\be 
L^2_{\xi}(K) = \{ f \in L^2(K,V_{\xi}) \,|\,
f(m k) = r_{\xi}(m) f(k)\,,\;m \in M\}\,.
\label{prep1}
\end{equation}
Eq.~(\ref{Kmultirep6}) implies that for $f \in L^2_{\xi}(K)$
one has $[\pi_{\nu}(g) f](m k) = r_{\xi}(m) [\pi_{\nu}(g) f](k)$. 
Thus $\pi_{\nu}: L^2_{\xi}(K) \ra L^2_{\xi}(K)$ and the 
restriction 
\be 
\pi_{\nu,\xi}(g) := \pi_{\nu}(g)\Big|_{L^2_{\xi}(K)}\,,
\label{prep2}
\end{equation}
is well defined. The subgroup $P := NAM$ is a (minimal) parabolic subgroup 
of $G$ and the representations (\ref{prep2}) are the 
{\it $P$-principal series} representations of $G$ in the so-called 
`compact model'. Explicitly 
\be 
[\pi_{\nu,\xi}(g)f](k) = \delta(a (k g))^{-1} 
\nu(a (k g)) f(g[k])\,,\quad g= nak\,.
\label{prep3}
\end{equation}
  
The standard definition of the $P$-principal series is 
by `parabolic induction'. For completeness we briefly recap 
this construction and verify that it is equivalent to 
(\ref{prep3}). First note that in the above notation 
$\chi_{\nu,\xi}(nam) = \nu(a) r_{\xi}(m)$ is a 
unitary representation of $P = NAM$ on $V_{\xi}$. 
Indeed, rearranging the Iwasawa components of $p_1,p_2 \in P$ 
one finds $p_1 p_2 = n(p_1) n(m(p_1) a(p_1) p_2) 
a(p_1) a(p_2) m(p_1) m(p_2)$,
using $a n(g) a^{-1} = n(ag)$ and $m n(g) m^{-1} = n(mg)$. 
The $P$-principal series is then defined as the representation 
of $G$ induced by $\chi_{\nu,\xi}$, ${\rm Ind}_P^G \,\chi_{\nu,\xi}$.
This means one considers the linear space of functions 
$F: G \ra V_{\xi}$ such that 
\ba 
&& F(p g) = \Delta(p)^{-1/2} \chi_{\nu,\xi}(p) F(g)\,,
\nonum
&& (F_1,F_2):=\int_K \! dk \bra F_1(k), F_2(k)\ket_{V_{\xi}}\,, 
\label{prep4}
\end{eqnarray}
which upon completion with respect to the norm given forms a 
Hilbert space $\cH_{\chi_{\nu,\xi}}$. Here $\Delta$ is the modular 
function of $P= NAM$, 
which (since $M$ is compact) coincides with the modular function of 
$B=NA$ and hence is given by $\Delta(nam) = \delta(a)^2$. 
Explicitly the covariance equation in (\ref{prep4}) thus reads
\be 
F(nam g) = \delta(a)^{-1} \nu(a) r_{\xi}(m) \,F(g)\,.
\label{prep5}
\end{equation}
The induced representation then is defined as the restriction 
of the right regular representation to $\cH_{\chi_{\nu,\xi}}$,
\be 
[{\rm Ind}_P^G(g_0) F](g) := F(g g_0)\,,\quad 
F \in \cH_{\chi_{\nu,\xi}}\,.
\label{prep6}
\end{equation}
The unitarity follows from (\ref{Kmultirep3}) applied to 
$f(k) = \bra F_1(g_0[k]), F_2(g_0[k]) \ket_{V_{\xi}}$.  
The equivalence to the `compact model' (\ref{prep2}), (\ref{prep3})
comes about as follows. Via 
\be 
F(nak) = \delta(a)^{-1} \nu(a) f(k)\,,
\label{prep7}
\end{equation}
one can set up a correspondence between functions $F \in \cC_c(G)$ 
and $f \in \cC_c(K)$. Moreover $F \in \cH_{\chi_{\nu,\xi}}$ if and 
only if $f \in L^2_{\nu,\xi}(K)$. To see this one rewrites the 
argument $nam g$ in (\ref{prep5}) as $n a m g = 
n n(amg) a a(g)\, m k(g)$; in this form the equivariance 
properties are directly mapped onto each other. The square 
integrability in both function spaces is the same, as 
$F|_K =f$ in (\ref{prep7}). Finally the correspondence 
(\ref{prep7}) maps the representations onto each other: 
We use the relation $k g_0 = (na)(kg_0) g_0[k]$ to 
rewrite $gg_0 = nak g_0$ as 
\be 
gg_0 = n n(ak g_0) a a(k g_0) g_0[k]\,.
\label{prep8}
\end{equation}
From here one easily verifies $[{\rm Ind}_P^G(g_0)F](g) 
= F(gg_0) = [\pi_{\nu,\xi}(g_0) f](k)$, for $g = nak$. 
Taken together, the restriction map $F \ra F|_K$ intertwines 
the unitary representations ${\rm Ind}_P^G \chi_{\nu,\xi}$ on
$\cH_{\chi_{\nu,\xi}}$ and $\pi_{\nu,\xi}$ on $L^2_{\xi}(K)$.

In summary, the principal series representations of a 
semisimple Lie group are parameterised by a unitary 
character $\nu: A \ra U(1)$, i.e.~$\nu \in \widehat{A}$,  
and by an element $\xi \in \widehat{M}$, where $M$ is the 
centralizer of $A$ in $K$ and $G = NAK$ is the Iwasawa decomposition. 
From the viewpoint of the Plancherel measure (\ref{Planch6}) they 
account for the conjugacy classes of Cartan subgroups of the form 
$H = T \times R$, where the noncompact part $R$ is isomorphic to 
$\widehat{A}$ (see App.A.1.2), while the compact part $T$ comes 
from the Cartan subgroup $H_M$ of $M$. Schematically $H = 
\widehat{A} \times H_M$, $\sigma = (\nu,\xi)$. The construction 
does not ensure irreducibility in itself; typically however 
principal series representations are irreducible. By Kostant's 
theorem (\cite{Warner}, Thm. 5.5.2.3) this is the case whenever $G$ is a 
semisimple connected Lie group with finite center and $P$ is a 
minimal parabolic subgroup. In particular this holds for 
all ${\rm SO}_0(1,N)$, $N \leq 3$, and for the two-fold 
simply connected covering ${\rm Spin}(1,N)$ of ${\rm SO}_0(1,N)$,
with $N$ odd. In rare cases a principal series representation may 
fail to be irreducible; however it then decomposes into 
a direct sum of irreducible representations, the number of which 
cannot exceed the order of the Weyl group (see \cite{Warner}, Vol.1, 
Corr.~5.5.2.2, p.461).   

The action of $\pi_{\nu,\xi}$ on $L^2_{\xi}(K)$ and its 
matrix elements can be described fairly explicitly. To 
this end set $\pi_{\nu}(\phi) := \int_G \! dg \phi(g) \pi_{\nu}(g)$,
for $\phi \in \cC_c(G)$. Then 
\ba 
[\pi_{\nu}(\phi)f](k_0) \is \int\! dk\, \pi_{\nu}(\phi)(k_0,k) f(k)\,,
\nonum
\pi_{\nu}(\phi)(k_0,k) \is \int\! d(na) \,
\phi(k_0^{-1} na k) (\delta^{-1} \nu)(a) \,,
\label{pkernel1}   
\end{eqnarray}
that is, $\pi_{\nu}(\phi)$ acts as an integral operator with the given 
kernel. If the function $\phi$ has support only on $K$ the kernel 
reduces to $\pi_{\nu}(\phi)(k_0,k) = \phi(k_0^{-1} k)$. Both 
statements follow directly from (\ref{prep3}) taking into account that 
$g[k_0] = k(k_0 g)$ and $dg = d(na) dk$ for functions of $g=nak$.   

Since $\pi_{\nu}$ commutes with the left regular representation of $M$,
see Eq.~(\ref{Kmultirep6}), the kernel obeys 
\be 
\pi_{\nu}(\phi)(mk_0,mk) = \pi_{\nu}(\phi)(k_0,k) \,,\quad m \in M\,.
\label{pkernel2}
\end{equation}
Using the fact that for any $m \in M$ the map $n \mapsto m^{-1} n m$ 
is a diffeomorphism of $N$ onto itself with unit Jacobian, $dn = 
d(m^{-1} n m)$, the invariance (\ref{pkernel2}) is also readily verified 
directly. Both the trace ${\rm Tr}[\pi_{\nu}(\phi)] = 
\int_K \!dk \pi_{\nu}(\phi)(k,k)$ and the Hilbert-Schmidt norm 
$\Vert \pi_{\nu}(\phi) \Vert^2_2 = {\rm Tr}[\pi_{\nu}(\phi)^* 
\pi_{\nu}(\phi)]$ are finite for $\phi \in \cC_c^{\infty}(G)$. 
This remains true upon restriction to $L^2_{\xi}(K)$ where 
$\pi_{\nu\xi}(\phi)$ basically gives the Fourier coefficients 
entering the Plancherel decomposition. In detail let 
\be 
[P_{\xi}f](k) := \int_M \!dm \,r_{\xi}(m^{-1}) f(mk)\,,
\quad [P_{\xi}f](m k) = r_{\xi}(m) [P_{\xi}f](k)\,,
\label{pxi1}
\end{equation}
be the projector from $L^2(K,V_{\xi})$ to $L^2_{\xi}(K)$. Then 
$\pi_{\nu\xi}(\phi) 
:= \pi_{\nu}(\phi) P_{\xi} =  P_{\xi}   \pi_{\nu}(\phi) = 
\widehat{\phi}(\nu,\xi)^{\dagger}$, acts as an  
integral operator with matrix valued kernel 
\ba 
\pi_{\nu\xi}(\phi)(k_0,k) \is 
\int_M \! dm \, \pi_{\nu}(\phi)(k_0, m k) r_{\xi}(m) 
\nonum
\is 
\int_P \! dp \phi(k_0^{-1} p k) (\delta^{-1} \nu)(p) r_{\xi}(m(p)) \,,   
\label{pxi2}
\end{eqnarray}
where $P = NAM$, $dp = d(na) dm$ and $m(n'a'm') = m'$. Note that 
the $\ell(M)$ invariance (\ref{pkernel2}) of the kernel $\pi_{\nu}(\phi)(k_0,k)$ 
has turned into a covariance 
\be 
\pi_{\nu\xi}(\phi)(m k_0, m k) = r_{\xi}(m)\,  \pi_{\nu\xi}(\phi)(k_0,k)\, 
r_{\xi}(m)^{-1}\,.
\label{pxi3}
\end{equation}

As noted before, for $\phi \in \cC_{c}^{\infty}(G)$ the kernel (\ref{pxi2}) 
defines a trace class operator on $L^2_{\xi}(K)$. Both the left and the 
right action of $\pi_{\nu \xi}(g)$ on $\pi_{\nu,\xi}(\phi)$ produces again 
a trace class operator whose kernel is readily worked out. One 
finds
\begin{subeqnarray}
\label{kernelaction} 
&& \mbox{kernel of $\pi_{\nu,\xi}(g)\pi_{\nu\xi}(\phi)$} = 
\pi_{\nu,\xi}(\ell(g)\phi)(k,k')
\nonum 
&& = (\delta^{-1} \nu)(a(kg)) \,\pi_{\nu, \xi}(\phi)(g[k], k')\,,
\\
&& \mbox{kernel of $\pi_{\nu,\xi}(\phi) \pi_{\nu \xi}(g)$} = 
\pi_{\nu,\xi}(\rho(g^{-1})\phi)(k,k')
\nonum 
&& = (\delta \nu)^{-1}(a(k'g^{-1})) \,\pi_{\nu, \xi}(\phi)(k, g^{-1}[k']))\,.
\end{subeqnarray}   
Note that (\ref{kernelaction}b) is the formal adjoint of (\ref{kernelaction}a),
as it should.

%%%%%%%%%%%%%%%%%%%%%%%%%%%%%%%%%%%%%%%%%%%%%%%%%%%%%%%%%%%%%%%%%%%%%%%%%

\newpage 
\newappendix{The restricted dual of SO(1,N)}

Here we explicate some of the general results in appendix A for the 
case ${\rm SO}_0(1,N)$. These groups have split rank 1, i.e.~the 
subgroup $A$ in the Iwasasa decomposition is one-dimensional. 
As a consequence the restricted dual is exhausted by 
the principal series proper and the discrete series; cuspidial 
principal series are absent. Further all unitary irreducible 
representations are multiplicity free, which gives rise to very 
explicit descriptions of their ${\rm SO}(N)$ content. 
Finally one can get explicit expressions for the coefficients
of the spherical principal series in terms of simple special functions 
(generalized Legendre functions).

{\it 1.~Group decompositions and orbits:} 
The groups ${\rm SO}_0(1,N)$ are generalizations of the Lorentz 
group ($N=3$) and the de-Sitter group ($N=4$). They are connected 
and locally compact; the two-fold simply connected covering group 
of ${\rm SO}_0(1,N)$ is ${\rm Spin}(1,N)$. The natural action
of ${\rm SO}_0(1,N)$ on $\R^{1,N}$ decomposes it into 6 types of 
orbits: the origin $\{x = 0\}$, the two-sheeted hyperboloids 
$\{ x \cdot x =r,\; \pm x^0 > 0\}$,
the one-sheeted hyperboloids (de-Sitter spaces) $\{x \cdot x = -r \}$ 
(with $r > 0$ in both cases), and finally 
the cones $\{x \cdot x =0,\, \pm x^0 > 0\}$.

Both for the detailed description of these orbits and for the 
representation theory the Iwasawa decomposition $G = NAK$ is 
instrumental. For $G={\rm SO}_0(1,N)$ 
it takes the following form: $K \simeq {\rm SO}(N)$ is the 
isotropy group of $q^{\uparrow} = (1,0,\ldots,0)$. $A$ is the one-dimensional 
subgroup generated by $a(\th),\, \th \in \R$, and $N$ is the $N\!-\!1$ dimensional
subgroup generated by $n(t), \,t = (t_1, \ldots , t_{N-1})^T \in \R^{N-1}$,
where 
\be
a(\th) = \left( 
\begin{array}{cc|c} \ch\th & \sh \th & \mbox{} \\[1mm]
\sh \th & \ch \th  & \mbox{} \\[2mm]
\hline 
\mbox{ } & \mbox{ } & \mbox{ } \\[-3mm] 
\mbox{ } & \mbox{ } & \mbox{ }\large{\1}_{N-1} 
\end{array} \right)\,,
\sspace 
n(t) = \left( 
\begin{array}{cc|c} 1+ \frac{1}{2} t^2 & -\frac{t^2}{2} & t^T \\[1mm]
\frac{t^2}{2} & 1- \frac{t^2}{2} & t^T \\[2mm]
\hline
t  & - t  &  \large{\1}_{N-1}  
\end{array} \right)\,.
\label{AN}
\end{equation}
with $t^2 := t_1^2 + \ldots + t_{N-1}^2$. Observe that 
$n(t) n(t') = n(t + t')$ and $a(\th) a(\th') = a(\th + \th')$, 
so that $N \simeq \R^{N-1}$ and $A \simeq \R$. Each element $g \in G$ 
admits a unique decomposition $g = n a k$ with $n \in N$, $a \in A$, 
and $k \in K$. Since $g^{-1} = k^{-1} a^{-1} n^{-1}$ the same 
holds for a decomposition with the subgroups oppositely ordered,
$G = K A N$. Let $\eta = {\rm diag}(1,-1,\ldots,-1)$ be the bilinear form 
on $\R^{1,N}$.

As before let $M$ denote the centralizer of $A$ in $K$. Clearly 
$M \simeq {\rm SO}(N\!-\!1)$, with ${\rm SO}(N-1)$ acting on the lower 
$(N\!-\!1)\times (N\!-\!1)$ block of the matrices. One has 
\be 
a(\th) n(t) a(\th)^{-1} = n(e^{\th} t)\,,\sspace 
m n(t) m^{-1} = n(mt)\,,\quad m \in M\,.
\label{ANrel}
\end{equation}
This shows that as subgroups $A N = N A$ and that $N M$ is the 
semidirect product of $N \simeq \R^{N-1}$ with $M$.

Explicit parameterizations of the various G-orbits in $\R^{1,N}$ 
can be obtained from the Iwasawa decomposition by letting it 
act on a reference vector of the orbit. For the upper 
sheet $\H_N = \{ q \cdot q =1\,,\; q_0 > 0\}$ of the two-sheeted 
hyperboloid we take $q^{\uparrow} = (1,0,\ldots , 0)$ as 
the reference vector. The action of $G$ via the $NAK$ decomposition 
then gives the `horispherical' coordinates on $\H_N$. Indeed,   
$n(t) a(\th) q^{\uparrow}$ parameterizes a unique 
point $q = (q_0, \ldots, q_N)$ in $\H_N$,  
\be 
q_0 = \ch \th + \frac{1}{2} t^2 e^{-\th}\,,
\quad 
q_1 = \sh \th + \frac{1}{2} t^2 e^{-\th}\,,
\quad q_i = e^{-\th} t_{i-1},\;\;i = 2,\ldots, N\,,
\end{equation} 
and $(\th, t_1,\ldots ,t_{N-1})$ are its horospherical coordinates. 
One can invert the transformation, in which case the product 
$n(t) a(\th)$ viewed as a function of $q$ gives back the section 
$g_s(q)$ of Eq.~(\ref{sectionHN}). The isometry 
$\H_N \simeq {\rm SO}(1,N)/{\rm SO}(N)$ is likewise manifest. 
Similar descriptions -- not needed here -- exist for the 
cone $\{x\cdot x =0, x^0 > 0\}$ and the one-sheeted hyperboloid
$\{ q \cdot q=-1\}$.   

Finally we note the relevant invariant measures. Let $dk$ denote the 
normalized Haar measure on $K = {\rm SO}(N)$. Set $da = d\th$ for 
$a = a(\th) \in A$ which gives Haar measure on $A$. Then 
\be 
d(na) = e^{-\th(N-1)} d\th \,dt_1\ldots dt_{N-1}\,, 
\label{NAmeasure}
\end{equation}
is the left invariant measure on $NA$. The left invariance 
$d(n(t_0) a(\th_0) na) = d(na)$ is easily checked from 
$a(\th_0) n(t) = n(e^{\th_0} t) a(\th_0)$. The Haar measure 
on $G$ in the $NAK$ Iwasawa decomposition is then given by 
\be 
dg = d(na) dk\,,\sspace g \in NAK\;.   
\label{NAKmeasure}
\end{equation}   
If elements $k \in K$ are decomposed according to $k = k(\vec{s}) m$, 
$m \in M$, $\vec{s} \in S^{N-1} \simeq K/M$, the normalized measures 
on $K,\,M$ and $S^{N-1}$ are related by 
\be 
dk = dS(\vec{s}) dm\,.
\label{MSmeasure}
\end{equation}

{\it 2.~Dual and restricted dual of ${\rm SO}_0(1,N)$:}
For $G= {\rm SO}_0(1,N)$ and ${\rm Spin}(1,N)$ the dual space (as a 
topological space) is known completely \cite{BB}. 
The lists in \cite{Ottoson}  and \cite{Hiraiirrep} are not quite complete and 
do not discuss the square integrability of the representations.  

To illustrate the relation to the restricted dual 
$\widehat{G}_r$ we briefly sketch these results here. All unitary irreducible 
representations (UIR) come from those of the Lie algebra. For general $N$ 
the relevant UIR of the Lie algebra ${\rm so}(1,N)$ have been classified 
by Ottoson \cite{Ottoson} and Schwarz \cite{Schwarz}.
Apart from the singlet $\pi_0$ there are three main types of UIR. For the 
harmonic analysis on $L^2({\rm SO}_0(1,N))$ only two of them are needed, 
the principal $\pi({\rm princ})$ and the discrete series $\pi({\rm disc})$.
In addition there are complementary series $\pi({\rm comp})$ (subdivided 
in \cite{Ottoson,Schwarz} into supplementary series and exceptional 
series). 
Each UIR is labeled by $r:={\rm rank}\,{\rm SO_0}(1,N) = [(N+1)/2]$ real 
parameters,
$(\xi_1, \ldots, \xi_{r-1}, s)$. The $\xi_i$ are nonnegative and are either 
all integers or all halfintegers and are ordered $\xi_1 \leq \xi_2 \leq 
\ldots \leq \xi_{r-1}$. The remaining parameter $s$ can be real or complex. 
Depending on the value 
of $s$ the parameters $\xi$ are constrained by further 
conditions, which together with the value of $s$ specify to
which series a given UIR belongs. For the principal series one 
has 
\be
\pi({\rm princ}): \;\;
s = i\om\,,\;\; \om \geq 0\,,\;\;\xi \in \widehat{{\rm SO}}(N\!-\!1)\,,
\label{Pseries}
\end{equation} 
where the explicit description of the weights $\xi$ is given in Eq.~(\ref{SOprinc1}) 
below. The general discussion of the principal series from Appendix A can 
be straightforwardly specialized to the case of ${\rm SO_0}(1,N)$. The
$A$ homomorphisms $\nu=:\nu_{\om}$ and $\delta$ are given by 
\ba
\nu_{\om}\!\!&:& \!\!A \ra U(1)\,,\,\sspace  
\nu_{\om}(a(\th)) = e^{i\om \th} \,,\quad 
\om \geq 0\,,
\nonum
\delta\!\! &:& \!\!A \ra \R_+\,, \sspace \quad 
\delta(a(\th)) = e^{-\th \frac{N-1}{2}}\,.
\label{Phomeos}
\end{eqnarray} 

For $N=2 r$ even there are also discrete series representations which 
enter the Plancherel decomposition. One has \cite{Hiraiplanch,Mack}
\be
\pi^{\pm}({\rm disc})\;:\quad \pm s \leq \xi_1 \leq \ldots \leq \xi_{r-1} \,,
\quad \pm s \in \N, \;\xi_i \in \N\,,
\label{Dseries} 
\end{equation}
where $\N$ are the positive integers and the $\pi^{\pm}$ series are labeled
by the sign of $s$. The $(-s,\xi)$ representation is the mirror image of
the $(s, \xi)$ representation in the following sense. Define a spatial reflection 
by $\th (q_0,q_1,\ldots, q_N) =  (q_0,-q_1,\ldots, q_N)$, which induces an
outer automorphism $g \mapsto \th g \th^{-1}$ of ${\rm SO}_0(1,N)$ (in the 
defining matrix representation). Then the mirror image $\pi_{\th(s,\xi)}$ is 
defined by $\pi_{\th(s,\xi)}(g) = \pi_{s,\xi}(\th g \th^{-1})$; the result
mentioned is that $\pi_{\th(s,\xi)}$ is unitary equivalent to $\pi_{-s, \xi}$.    

An explicit formula for the value of the quadratic 
Casimir on any of these UIR (in terms of their parameters) is known 
\cite{Hiraiplanch,Dobrevetal}.

As described in Appendix A.1 the Cartan subgroups 
are relevant for the harmonic analysis on $G$. In the case of 
$G={\rm SO}_0(1,N)$, there is a single conjugacy class of Cartan subgroups 
when $N$ is odd and two when $N$ is even 
\cite{Wallach}, p.188,212. In the notation of the previous section this arises
because $\dim A =1$ in the Iwasawa decomposition $G = KAN$. In this 
case the number of conjugacy classes of $G$ is $2^{{\rm rank}K - {\rm rank}M}$. 
(Recall that ${\rm rank}\,{\rm SO}(N) = [N/2]$, {\rm rank}\,{\rm SO}(1,N)
=[(N+1)/2]). Explicitly, if $H_M$ is the (up to conjugacy unique)
Cartan subalgebra of $M$ one can adjoin either the generator of $A$ 
or the generator of $H_K/H_M$ (where $H_K$ is the up to conjugacy unique 
Cartan subalgebra of $K$) in order to obtain an abelian subgroup of $G$. 
For $N$ even both subgroups obtained in this way have the same dimension 
$N/2 = {\rm rank}{\rm SO}(N) +1$. For $N$ odd $H_M$ and $H_K$ have the 
same dimension $\frac{N}{2}-1$ (a $2 \times 2$ block was needed to 
add a new Cartan generator). The compact abelian subalgebra $H_M = H_K$ 
is thus not maximal and only the noncompact Cartan subalgebra exists. 
According to the general discussion the Plancherel decomposition 
takes the form 
\ba 
\nspace L^2({\rm SO}_0(1,N)) \is L^2({\rm SO}_0(1,N))_{\rm disc} \oplus 
\bigoplus_{\xi \in \widehat{M}} \int_0^{\infty} \!
d\mu(\om,\xi) \cL_{\om,\xi}\,,
\nonum
L^2({\rm SO}_0(1,N))_{\rm disc}  \is 
\left\{ \begin{array}{cl} \{0\} & \;\;\;N\;\;\mbox{odd}\,,
\\[4mm] 
\bigoplus_{\sigma \in {\rm disc}} 
L^2_{\sigma}({\rm SO}_0(1,N)) & \quad N\;\,\;\mbox{even}\,,
\end{array} \right.
\end{eqnarray} 
where the sum in  the discrete part ranges over the set in (\ref{Dseries}). 
An explicit formula for $d\mu(\om,\xi)$ is known \cite{Hiraiplanch}.

{\it 3.~K-content of principal and discrete series:} Restricted to the subgroup 
$K = {\rm SO}(N)$ the irreducible representations $\pi_{\sigma}$ of 
${\rm SO}_0(1,N)$ decompose into a direct sum of irreducible representations
$r_{\kappa}$ of $K$, each of which occurs with multiplicity at most one
\cite{Dixmier2}. The subset of $\widehat{K}_{\sigma} \subset \widehat{K}$ which occurs 
with nonzero (and hence unit) multiplicity is called the $K$ content of the 
representation. Here we describe it explicitly for the principal 
and the discrete series. 

Recall from (\ref{Pseries}) that a principal series representation is 
labeled by a real parameter $\om \geq 0$ and by a highest weight 
$\xi$ of $M = {\rm SO}(N\!-\!1)$. Explicitly the latter means 
\be 
\label{SOprinc1}
\begin{array}{ccc} 
N \;\mbox{even}\;:& \xi = (\xi_1, \ldots ,\xi_{\frac{N-2}{2}})\,,\;\;& 
0 \leq \xi_1 \leq \ldots \leq \xi_{\frac{N-2}{2}}\,,
\\[2mm]
N \;\mbox{odd}\;:& \xi = (\xi_1, \ldots ,\xi_{\frac{N-1}{2}})\,,\;\;& 
|\xi_1|  \leq \xi_2 \leq \ldots \leq \xi_{\frac{N-1}{2}}\,.
\end{array}
\end{equation}
Here we used that UIR of the orthogonal groups are labeled by highest weights
which are ordered sets $(m_1, \ldots ,m_r)$, where $r = 
{\rm rank}\, {\rm SO}(N)$ equals $N/2$ and  $(N\!-\!1)/2$ for $N$ even and odd,
respectively. The $m_i$ are either all integer or all halfinteger and are 
subject to the constraints 
\ba 
{\rm SO}(2 r) &:& |m_1| \leq m_2 \leq \ldots \leq m_r\,,
\nonum
{\rm SO}(2 r+1) &:& \quad 0 \leq m_1 \leq \ldots \leq m_r\,.
\label{SOweights}
\end{eqnarray}
The singlet corresponds to $\kappa = (0, \ldots, 0)$; the symmetric traceless 
tensor representions have $\kappa =(0,\ldots, 0, m_r)$, $m_r \in \N$. The mirror image 
$\th \kappa$ of an UIR $\kappa$ is defined as follows. Let $\th \in 
{\rm O}(N)$ be a reflection, i.e.~$\th^2 = \1,\;\det \th =-1$. Since 
$\th k \th^{-1} \in {\rm SO}(N)$ for all $k$, we can define a 
representation of ${\rm SO}(N)$ by $r_{\th \kappa}(k) := 
r_{\kappa}(\th k \th^{-1})$ called the mirror image of $\kappa$.  
It is again an irreducible highest weight representation and 
unitarily equivalent to $r_{\kappa'}$, where 
\ba 
{\rm SO}(2 r) &:& \kappa' = (-m_1, m_2,\ldots,m_r)\,,
\nonum
{\rm SO}(2 r+1) &:& \kappa' = (m_1, m_2, \ldots, m_r)=\kappa\,.
\label{SOmirrorweights}
\end{eqnarray}

The $K = {\rm SO}(N)$ content $\pi_{\om,\xi}$ can now be described explicitly
\be 
\label{SOprinc3}
\pi_{\om \xi}\Big|_K = \bigoplus_{\kappa \in \widehat{K}_{\xi}} 
r_{\kappa}\,,
\end{equation}
where each UIR of $K$ occurs with multiplicity precisely one and the 
subsets $\widehat{K}_{\xi} \subset \widehat{K}$ are characterized by 
\cite{Ottoson,Schwarz}
\be 
\label{SOprinc4}
\begin{array}{ccc} 
N \;\mbox{even}\;:& \kappa = (m_1, \ldots ,m_{\frac{N}{2}})\,,\;\;& 
|m_1| \leq \xi_1 \leq m_2 \leq \ldots \leq 
\xi_{\frac{N-2}{2}} \leq m_{\frac{N}{2}}\,,
\\[2mm]
N \;\mbox{odd}\;:& \kappa = (m_1, \ldots ,m_{\frac{N-1}{2}})\,,\;\;& 
\;\;|\xi_1|  \leq m_1 \leq \xi_2 \leq \ldots \leq 
\xi_{\frac{N-1}{2}} \leq m_{\frac{N-1}{2}}\,.
\end{array}
\end{equation}
These are precisely the same conditions under which the UIR $\xi$ of 
$M = {\rm SO}(N\!-\!1)$ occurs (with unit multiplicity) in the restriction 
of $r_{\kappa}|_M$ (see e.g.~\cite{BR}). The result (\ref{SOprinc1}) 
thus exemplifies the general reciprocity rule mentioned in part A:
if $\pi_{\om\xi}|_K$ contains $\kappa \in \widehat{K}$ (with unit 
multiplicity) then $r_{\kappa}|_M$ contains $\xi \in \widehat{M}$ 
(with unit multiplicity). 

Since the last label $m_r$, $r = {\rm rank}\,{\rm SO}(N)$, can be made 
arbitrarily large the subsets $K_{\xi}$ are always infinite. However with 
the exception of $\xi =0$ (the $M$-singlet) they never contain $\kappa =0$ 
(the $K$ singlet). The $K$ content of $\pi_{\om,\xi =0}$ (the 
spherical principal series) is given by 
\be 
\widehat{K}_{\xi =0} = \{ (0,\ldots, 0, m_r)\,,\;\;\; m_r \geq 0\}\,.
\label{SOprinc5}
\end{equation}
As can be seen from (\ref{SOdisc1}) below the discrete series representations
never contain a $K$ singlet. Irreducible representations containing a 
vector invariant under a compact subgroup are often called ``class 1''.
One sees that among the representations of the restricted dual of 
${\rm SO}_0(1,N)$ the only class 1 representations are those of the 
spherical principal series $\pi_{\om,0}$, and for them the 
class 1 property with respect to $M$ and with respect to $K$ is 
equivalent.   

The $K$ content of the discrete series (\ref{Dseries}) comes out as follows 
\be
\label{SOdisc1}
\pi^{\pm}({\rm disc})\; :\;\;\; 
\pm s \leq m_1 \leq \xi_1 \leq m_2 \leq \ldots \leq \xi_{r-1} \leq m_r\,,
\end{equation}
either by direct investigation of the Harish-Chandra characters \cite{Mack}, or by 
specialization of Blattner's formula \cite{HechtSchmid}. 
One sees from (\ref{SOmirrorweights}) that a discrete series representation 
never contains some $\kappa \in \widehat{K}$ together with its mirror image 
$\th \kappa$: $m_1$ is strictly positive for 
the $\pi^+$ series and strictly negative for the $\pi^-$ series. 
Equivalently $\widehat{K}_{\pi^+} \cap \widehat{K}_{\pi^-} = \emptyset$. 
In particular the $\pi^{\pm}$ series never contain the $K$ singlet.

{\it 4.~Harmonic analysis on $\H_N$:} 
Here we explicate the reduction of the harmonic analysis on 
$L^2({\rm SO}_0(1,N))$ to that on $L^2(\H_N)$, where  
 $\H_N = {\rm SO}_0(1,N)/{\rm SO}(N)$, $N \geq 2$, is the upper part of 
the two-sheeted hyperboloid in $\R^{N+1}$. The result 
will be the decomposition (\ref{GKharmonic}) of the quasiregular representation 
$\ell_1$ of  ${\rm SO}_0(1,N)$ on $\H_N$. Since 
$K/M \simeq S^{N-1}$ all fiber spaces will be isometric to 
$L^2(S^{N-1})$. The matrix elements (\ref{Epi2}) come out to 
be certain Legendre functions which are also generalized 
eigenfunctions of the Laplace-Beltrami operator on $\H_N$.

Let $-\Delta^{\H_N}$ be minus the Laplace-Beltrami operator on $\H_N$. 
Its spectrum is absolutely continuous and is 
given by the interval $\frac{1}{4}(N\!-\!1)^2 + \om^2$, $\om >0$. There are 
several complete orthogonal systems of improper eigenfunctions. From a 
group theoretical viewpoint the most convenient system 
are the `principal plane waves' ${\Large \eps}_{\om,p}(q)$ (see 
\cite{VilKlim} and 
the references therein) labeled by $\om >0$ and a `momentum'
vector $\vec{p} \in S^{N-1}$. Parameterizing $q = (\xi, \sqrt{\xi^2 -1} \, \vec{s})$, 
they read
\be 
\eps_{\om,p}(q) := [\xi - \sqrt{\xi^2 -1} \, 
\vec{s} \cdot \vec{p} ]^{-\frac{1}{2}(N-1) - i \om}\,.
\label{Edef}
\end{equation}
The completeness and orthogonality relations take the form 
\ba 
\int \!d\gamma_Q(q) \,\eps_{\om,p}(q)^* \eps_{\om',p'}(q) \is 
d(\om)^{-1} \delta(\om - \om') \delta(\vec{p}, \vec{p}')\;,
\nonum
\int_0^{\infty} \!d \om\, d(\om) \int_{S^{N-1}} \!\!dS(p)   
\;\eps_{\om,p}(q)^* \eps_{\om,p}(q') \is \delta(q,q')\,,
\label{Ebasis}
\end{eqnarray}
where $\delta(q,q')$ and $\delta(\vec{p},\vec{p}')$ are the normalized delta 
distributions with respect to the invariant measures $d\gamma_Q(q)$ and
$dS(p)$ on $\H_N$ and  $S^{N-1}$, 
respectively. In terms of the coordinates $(\xi, \vec{s})$ the former reads
\be 
\int \! d\gamma_Q(q)  = \int_1^{\infty} \!d\xi (\xi^2 -1)^{N/2 -1} 
\int_{S^{N-1}} \! dS(\vec{s})\,.
\label{measurefact}
\end{equation}
The spectral weight is determined by the Harish-Chandra c-function for 
${\rm SO}_0(1,N)$ and is given by 
is 
\be 
d(\om) = \frac{1}{(2\pi)^N} 
\left| \frac{\Gamma(\frac{N-1}{2} + i\om)}{\Gamma(i\om)} \right|^2 .
\label{Emu}
\end{equation}
The main virtue of these functions is their simple transformation law 
under ${\rm SO}_0(1,N)$, see e.g.~Appendix A of \cite{O1Nmodel}. 
It characterizes the spherical principal unitary series 
$\pi_{\om,0},\,\om \geq 0$, of ${\rm SO}_0(1,N)$, where 
$\pi_{\om,0}$ and its complex 
conjugate are unitary equivalent (see e.g.~\cite{VilKlim}, Sections 9.2.1 
and 9.2.7). The orthogonality and completeness relations (\ref{Ebasis}) 
amount to the decomposition (\ref{GKharmonic}) of the quasi-regular 
representation $\ell_1$ on $L^2(\H_N)$. 

Spectral projectors $E_I$ commuting with $\ell_1$ are defined in terms of 
their kernels $E_I(q\cdot q')$, $I \subset \R_+$ by 
\ba
&& E_I(q\cdot q') := \int_I d\om\, d(\om) 
\int_{S^{N-1}} \!dS(p) \,\eps_{\om,p}(q)^* \eps_{\om,p}(q')\,,    
\nonum
&& \int \! d\Omega(q') E_I(q\cdot q') E_J(q'\cdot q'') = E_{I \cap J}(q\cdot q'')\;.
\label{PIspec}
\end{eqnarray}
Combined with the completeness relation in (\ref{Ebasis}) this shows that the spectrum 
of $-\Delta^{\H_N}$ is absolutely continuous. 
  
A complete orthogonal set of real eigenfunctions of $-\Delta_{\H_N}$ is
obtained by taking the $dS(p)$ average of the product of $\eps_{\om,p}(q)$
with some spherical harmonics on the 
$p$-sphere. This amounts to a decomposition in terms of ${\rm SO}^{\uparrow}\!(N)$ 
UIR where the `radial' parts of the resulting eigenfunctions are given by 
Legendre functions.  Using the normalization and the integral 
representation from (\cite{Grad} p.~1000) one has in particular 
\be 
\int_{S^{N-1}}\! dS(p) \, \eps_{\om,p}(q) = (2\pi)^{N/2} (\xi^2 -1)^{\frac{1}{4}(2-N)} \,
\cP_{-1/2 + i \om}^{1-N/2}(\xi) \,.
\label{EPrel}
\end{equation}
As a check on the normalizations one can take the $\xi \ra 1^+$ limit in (\ref{EPrel}).
The limit on the rhs is regular and gives $2 \pi^{N/2}/\Gamma(N/2)$, which equals 
the area of $S^{N-1}$ as required by the limit of the lhs. 
Denoting the set of real scalar spherical harmonics by $Y_{\ell m}(k)$, 
$\ell \in \N_0$, 
$m =0,\ldots, d_\ell \!-\!1$, with $d_\ell = 
(2 \ell + N-2)(\kappa + N-3)!/(\ell!(N-2)!)$ we set
\begin{subeqnarray}
\label{Hdef}
E_{\om,\ell m}(q) &:=&  \int \!dS(p) \,Y_{\ell m}(p) \eps_{\om,p}(q)  
\\
&=&  n_\ell(\om)\, Y_{\ell m}(\vec{s}) \,(\xi^2 -1)^{\frac{1}{4}(2-N)} 
\,\cP_{-1/2 + i \om}^{1 -N/2 -\ell}(\xi)\,,\quad \mbox{with} 
\\
&& n_0(\om) = (2\pi)^{N/2}\,, \quad n_\ell(\om) =  (2\pi)^{N/2} 
\left(\prod_{j=0}^{\ell-1} 
\Big[\om^2 + (\mbox{$\frac{N-1}{2}$} + j)^2 
\Big]\right)^{\!1/2},\;\ell\geq 1\,. 
\nonumber
\end{subeqnarray}
The expression (\ref{Hdef}b) is manifestly real, the equivalence to (\ref{Hdef}a) can be 
seen as follows: from (\ref{Ebasis}), (\ref{measurefact}), and the orthogonality 
and completeness of the spherical harmonics one readily verifies that both (\ref{Hdef}a) 
and (\ref{Hdef}b) satisfy
\ba 
\int \!d\Omega(q) \,E_{\om,\ell m}(q)^* E_{\om',\ell' m'}(q) \is 
d(\om)^{-1} \delta(\om - \om') \delta_{\ell,\ell'} \delta_{m,m'} \,,
\nonum
\int_0^{\infty} d\om\, d(\om) \sum_{\ell,m} 
E_{\om,\ell m}(q)^* E_{\om,\ell m}(q') 
\is \delta(q,q')\,.
\label{Hbasis}
\end{eqnarray}
Further both (\ref{Hdef}a) and (\ref{Hdef}b) transform irreducibly with respect to 
the real 
$d_{\ell}$ dimensional matrix representation of ${\rm SO}^{\uparrow}\!(N)$ 
carried
by the spherical harmonics. Hence they must coincide. A drawback of the 
functions (\ref{Hdef}) is that the $\vec{k}$ integration spoils the simple 
transformation law of the $\eps_{\om,p}$ under ${\rm SO}(1,N)$. The 
transformation law 
can now be inferred from the addition theorem
\be
\sum_{\ell,m} E_{\om,\ell m}(q) E_{\om,\ell m}(q') = 
(2\pi)^{N/2}\, [(q\!\cdot \!q')^2 -1]^{\frac{1}{4}(2-N)}\, 
\cP_{-1/2 + i\om}^{1-N/2}(q\!\cdot \!q')\,.
\label{Haddition}
\end{equation}
For example for $q' = g q^{\uparrow}$ this describes the transformation 
of the  ${\rm SO}^{\uparrow}\!(N)$ singlet $E_{\om,0,0}(q)$ under  
$g \in {\rm SO}(1,N)$. 

\newpage
%%%%%%%%%%%%%%%%%%%%%%%%%%%%%%%%%%%%%%%%%%%%%%%%%%%%%%%%%%%%%%%%%%%%%%%%%%%%%
\newappendix{The amenable case ISO(N)}

Here we outline counterparts of our main results for 
the coset space ${\rm ISO}(N)/{\rm SO}(N)\simeq \R^N$. 
This coset can be viewed as the flat space limit of 
the hyperboloid ${\rm SO}_0(1,N)/{\rm SO}(N)$. The underlying 
Euclidean group ${\rm ISO}(N)$ is noncompact but amenable;
in accordance with the general picture \cite{hchain} 
the generalized spin systems turn out to  
have a unique non-normalizable ground state.

The group ${\rm ISO}(N)$ is the semi-direct product 
of the amenable groups ${\rm SO}(N)$ and $\R^N$, and 
hence is itself amenable, \cite{Dixmier1}, Lemma 18.3.7. 
As the defining representation one can take a 
subgroup of $N\!+\!1 \times N\!+\!1$ matrices 
acting on vectors $(x,1)^T$ in $\R^{N+1}$
\be 
g(k,a) = { k \;\; a \choose 0\;\;1} \,,\quad 
g(k,a) { x \choose 1} = { kx + a \choose 1}\,,
\quad k \in {\rm SO}(N)\,,\;\; a, x \in \R^N\,.
\label{C1}
\end{equation} 
The composition law is $g(k_1,a_1) g(k_2, a_2) = 
g(k_1 k_2 ,a_1 + k_1 a_2)$, which gives 
$g(k,a)^{-1} = g(k^{-1}, - k^{-1} a)$ and 
$g(k,a) = g(e,a) g(k,0) = g(k,0) g(e, k^{-1} a)$. 
Haar measure on ${\rm ISO}(N)$ is $dk da$, 
where $dk$ is the normalized Haar measure on ${\rm SO}(N)$ 
and $da$ Lebesgue measure on $\R^N$. 

We shall need a section $g_s: \R^N \ra {\rm ISO}(N)$
such that $x = g_s(x) x^{\uparrow}$, where $x^{\uparrow} 
= (0,1)^T$ is fixed by the rotation subgroup. An obvious choice is 
\be 
g_s(x) = { e\;\; x \choose 0 \;\; 1 } \,,
\quad k_s(g(k,a), x) = k^{-1}\,, 
\label{C2}
\end{equation}
for which the cocycle $k_s$ in (\ref{section}) is independent 
of $x$ and $a$.

The configuration manifold is $\cM= Q^{\nu} = 
\R^{N \nu}$, the state space is $L^2(\cM) = L^2(\R^{N \nu})$,
and ${\rm ISO}(N)$ acts on it via the $\nu$-fold inner 
product of the left quasiregular representation $\ell_1$.
Explicitly 
\be 
[\ell_{\cM}(k,a)\psi](x) = \psi(k^{-1}(x-a))\,,
\quad x = (x_1,\ldots, x_{\nu})\,,
\label{C3}
\end{equation}
where we wrote $\ell_{\cM}(k,a)$ for $\ell_{\cM}(g(k,a))$. 
A transfer operator $\bt$ in the sense of definition 2.1 is 
described by a kernel $T: \R^{N \nu} \times \R^{N \nu} \ra 
\R_+$, which is symmetric, continuous, pointwise strictly positive,
and subject to the condition (\ref{Tassumpt1}). The  
invariance $\ell_{\cM} \circ \bt = \bt \circ \ell_{\cM}$
translates into $T(k(x+a), k(y+a)) = T(x,y)$, for all 
$k \in {\rm SO}(N)$ and $a \in \R^N$. To exploit this symmetry 
we proceed as in Section 2 and switch to a model of $L^2(\cM)$ 
where $\ell_{\cM}$ acts via right multiplication on a single 
group-valued argument.

Recall that $\cM_r = (G\times \cN)/d(K)$ and that the 
isometry to $\cM$ is constructed from the map 
$\tilde{\phi}: \cM \ra {\rm ISO}(N) \times \cN$, 
$\cN \simeq \R^{N(\nu -1)}$ 
\ba
\tilde{\phi}(x_1,\ldots,x_{\nu}) \is 
(g_s(x_1)^{-1}, x_2 - x_1, \ldots, x_{\nu} - x_1)\,,
\nonum
\tilde{\phi}(g^{-1} x_1, \ldots, g^{-1} x_{\nu}) 
\is (g_s(x_1)^{-1} g, x_2 - x_1, \ldots , x_{\nu} - x_1)\,.
\label{C5}
\end{eqnarray}
As expected the group acts on the image points by right 
multiplication on the first argument. Since 
$g(k_0,0) g_s(x_1)^{-1} g(k,a) = g(k_0 k, - k_0(x_1 -a))$
the left ${\rm SO}(N)$ invariant functions on 
${\rm ISO}(N) \times \cN$ are characterized by 
\be 
\psi_r( g_s(e,-k^{-1}(x_1 - a)), k^{-1} n) = 
\psi_r(g(k, -x_1 + a), n)\,.
\label{C6}
\end{equation}
Effectively the functions $\psi_r$ thus project onto functions 
on $Q \times \cN$, however at the price of a more complicated 
group action. In fact $\psi_s(x,n) := 
\psi_r(g(e,-x),n)$ defines an element of $L^2(\cM_s)$ with 
the group action $[\ell_s(k,a) \psi_s](x,n) = 
\psi_s(k^{-1}(x-a), k^{-1}n)$.

The unitary irreducible representations (UIR) entering the 
decomposition of the regular representation of ${\rm ISO}(N)$ 
on $L^2({\rm ISO}(N))$ can be described as follows (Gross and Kunze,  
\cite{GrKunze}). For $0\neq \nu \in \R^N$ let 
$K_{\nu}$ be the isotropy group of $\nu$ in $K = {\rm SO}(N)$. 
Then each $K_{\nu}$ is conjugate to $M:= {\rm SO}^{\uparrow}(N-1)$, 
defined as the subgroup leaving $e_1 = (1,0,\ldots ,0)^T \in \R^N$ 
invariant. Let $m \mapsto r_{\xi}(m)$, $\xi \in \widehat{M}$, 
be an irreducible representation of $M$ on $V_{\xi}$ and consider  
(as in Eq.~(\ref{prep1}))    
\ba
&& L^2_{\xi}(K) = \{ f \in L^2(K, V_{\xi})\,,\;
f(m k) = r_{\xi}(m) f(k) \,,\; m \in M\}\,,
\nonum
&& (f_1,f_2) := \int_K \! dk \bra f_1(k), f_2(k) \ket_{V_{\xi}}\,.
\label{C8}
\end{eqnarray}
On $L^2_{\xi}(K)$ define a unitary representation by 
\be 
[\pi_{\nu,\xi}(k_0,a_0) f](k) = e^{i \nu \cdot k a_0} f(kk_0)\,.
\label{C9}
\end{equation}
This is well-defined because $\pi_{\nu,\xi}$ commutes with 
the left regular representation of $M$. In particular 
$[\pi_{\nu,\xi}(k_0,a_0)f](m k) = r_{\xi}(m) 
[\pi_{\nu,\xi}(k_0,a_0)f](k)$. Moreover \cite{GrKunze}:
\begin{itemize}
\item[{--}] $\pi_{\nu,\xi}$ is irreducible for all $0 \neq \nu \in \R^N$  
and  $\xi \in \widehat{M}$. 
\item[{--}] Every infinite dimensional unitary representation is 
equivalent to some such $\pi_{\nu,\xi}$. 
\item[{--}] Given $0 \neq \nu,\,\nu' \in \R^N$ and 
$\xi,\,\xi' \in \widehat{M}$, the representations $\pi_{\nu,\xi}$ 
and $\pi_{\nu',\xi'}$ are equivalent if and only if first
$\nu$ and $\nu'$ belong to the same ${\rm SO}(N)$ orbit 
and second $\xi$ and $\xi'$ are equivalent under the 
identification of $K_{\nu}$ with $K_{\nu'}$.   
\end{itemize}
These representations constitute the principal series 
of ${\rm ISO}(N)$, the subset with $\xi = 0$ ($M$-singlets)  
is called the spherical principal series. 
We now fix a representative from each 
equivalence class as follows. 
If $\nu \cdot \nu = \nu' \cdot \nu' = \om^2$, $\om \in \R_+$, then 
by abuse of notation we denote the $N$-tuple $(0,\ldots, 0,\om)^T 
\in \R^N$ also by $\om$. In this case we write $K_{\om} = 
{\rm SO}^{\uparrow}(N\!-\!1)$ for $K_{\nu}$ and $\pi_{\om,\xi}$ for 
$\pi_{\nu,\xi}$. The $K$-content of these representations is 
the same as that of the corresponding principal series 
representations of ${\rm SO}_0(1,N)$, as their restrictions to 
$K$ coincide. This gives
\be 
\pi_{\om,\xi}\Big|_K = \bigoplus_{\ell \in \widehat{K}_{\xi}}
r_{\ell} \,,
\label{C9a}
\end{equation}
with $\widehat{K}_{\xi} \subset \widehat{K}$ as in (\ref{SOprinc4}). 

The finite dimensional representations of ISO(N), obtained from the 
irreducible representations of SO(N) by representing the abelian normal 
subgroup of translations trivially are not in the support of the regular 
representation.
 
So the Plancherel decomposition takes the form 
\ba 
L^2({\rm ISO}(N)) \is 
\int_0^{\infty} \! d\om d(\om)\, 
\bigoplus_{\xi \in \widehat{M}} \dim V_{\xi}\; 
\cL_{\om,\xi} \otimes \check{\cL}_{\om,\check\xi}\,,
\nonum
\rho \times \ell \is 
\int_0^{\infty} \! d\om d(\om)\,
\bigoplus_{\xi \in \widehat{M}} \dim V_{\xi}\; 
\pi_{\om,\xi} \otimes \check{\pi}_{\om,\check\xi}\,.
\label{C10}
\end{eqnarray} 
with 
\be 
d(\om) = \frac{\om^{N-1}}{(2\pi)^N} \frac{2 \pi^{N/2}}{\Gamma(N/2)}\,,
\label{C10a}
\end{equation}
(which can be viewed as the squared inverse of the Harish-Chandra $c$ function 
for ${\rm ISO}(N)$; see \cite{Sugiura} for $N\!=\!2$). The formulae for the harmonic 
analysis and synthesis will 
be given below. In relation to (\ref{C10}) the singlet 
representation deserves special consideration. Since ${\rm ISO}(N)$ 
is amenable the singlet must weakly be contained in (equivalently: lie 
in the support of) the regular 
representation (see e.g.~\cite{Dixmier1}, Prop.18.3.6; and Definitions
18.3.1, 18.1.7). By definition this support is the restricted dual 
of the locally compact group under consideration, here 
${\rm ISO}(N)$. As a matter of fact \cite{Dixmier1}, 18.8.4, it 
also coincides with the carrier of the Plancherel measure. 
The upshot is that the limit $\lim_{\om \ra 0}\pi_{\om,\xi =0}$ 
is weakly contained in the decomposition (\ref{C10}) and coincides with 
the singlet representation of ${\rm ISO}(N)$. We shall write
$\pi_{00}$ for it. 

To every $\phi \in L^1({\rm ISO}(N))$ one can asign a compact 
operator as its Fourier transform 
\be 
\widehat{\phi}(\nu,\xi)^{\dagger} = 
\int \!dk da \, \phi(k,a) \, \pi_{\nu,\xi}(k,a) 
= \pi_{\nu,\xi} (\phi)\,,
\label{C11}
\end{equation}
and for $\phi \in (L^1 \cap L^2)({\rm ISO}(N))$ the image 
is a Hilbert-Schmidt operator for almost all $(\om,\xi)$ 
with respect to the Plancherel measure. Since the latter 
is for fixed $\xi$ absolutely continuous with respect 
to Lebesgue measure on $\R_+$, the Hilbert-Schmidt property 
will hold for almost all $\om >0$ with respect to the 
Lebesgue measure. Indeed, $\pi_{\om,\xi}(\phi)$ can be realized 
explicitly as an integral operator on $L^2_{\xi}(K)$. 
Repeating the steps in Appendix A.8 one finds 
\ba
[\pi_{\nu,\xi}(\phi)f](k) \is  \int \! dk'\, 
\pi_{\nu,\xi}(\phi)(k,k') f(k')\,,
\nonum
\pi_{\nu,\xi}(\phi)(k,k') \is 
\int_{M \times \R^N} \! dm da \, \phi( k^{-1} m k', a) 
e^{i \nu \cdot k a} r_{\xi}(m)\,.
\label{C12}
\end{eqnarray}
The formula for the Fourier synthesis reads 
\be 
\phi(k,a) = 
\int_0^{\infty} \! d\om \, d(\om)\,\sum_{\xi \in \widehat{M}} 
\dim V_{\xi}\, \Tr[ \widehat{\phi}(\om,\xi) 
\pi_{\om,\xi}(k,a)] \,.
\label{C13}
\end{equation}
Formally this can be verified by evaluating the trace in terms of a kernel 
of the form (\ref{C12}) and freely exchanging the order of integrations. 
For a proof see \cite{GrKunze}.

With these preparations at hand we can proceed with the group 
theoretical decomposition of the Hilbert space $L^2(\cM_r)$ 
and of standard invariant selfadjoint operators $\bA$ acting on it. 
The constructions of Sections 2 and 3 carry over with minor 
modifications; mainly to fix the notations we run through 
the main steps. Proposition 2.2 remains valid with the 
Plancherel measure from (\ref{C10}) substituted and with the $K$-content 
from (\ref{C9a}). We use pairs $\sigma = (\om,\xi)$ and 
$\check{\sigma} = (\om,\check{\xi})$ to label the representations
and their conjugates. Proposition 3.4 likewise carries over 
and provides the decomposition of the operators. We write 
\ba
L^2(\cM_r) \is \int_0^{\infty} \! d\om\, d(\om)
\, \bigoplus_{\xi \in \widehat{M}}\,\dim V_{\xi}\; 
\cL_{\om \xi}^2(\cM_r)\,,
\nonum
\bA \is   \int_0^{\infty} \! d\om\, d(\om)  
\, \bigoplus_{\xi \in \widehat{M}}\, \dim V_{\xi}\;
(\1 \otimes \check{\bA}_{\om \check{\xi}})\,,
\label{C14}
\end{eqnarray}
for the respective decompositions. The fiber spaces 
$\cL^2_{\om \xi}(\cM_r)$ are isometric to $\cL_{\om \xi} \otimes 
\check{\cL}_{\om \check\xi} \otimes L^2(\cN)$ and $\1 \otimes 
\check{\bA}_{\om \check\xi}$ acts for almost all $\om >0$ as a
bounded linear and selfadjoint operator on these fiber spaces. 

The spectral problems of $\bA$ and $\bA_{\om\xi}$ can now be related as in 
Sections 3.3 and 3.4. We maintain the definitions of the generalized 
eigenspaces $\cE_{\lb,\om\xi}(\bA)$ and $\cE_{\lb}(\bA_{\om\xi})$, 
in Eqs.~(\ref{ECdef}) and (\ref{EMdef}), respectively. Then the map 
\ba 
&& \tau_{v,\om\xi}: \cE_\lb(\bA_{\om\xi}) 
\ra \cE_{\lb,\om\xi}(\bA) \,,\quad 
f \mapsto \tau_{v,\om \xi}(f)\,,
\nonum
&& \tau_{v,\om \xi}(f)(k,a,n) = 
\int_K \!dk'\, f(n,k')^* [\pi_{\om \xi}(k,a)v](k'), \;\; 
v \in L_{\xi}^2(K)\,,
\label{C15}
\end{eqnarray}
again provides an isometry onto its image. In the second line 
we used the realization of $\cL_{\om,\xi}$ as $L^2_{\xi}(K)$
in (\ref{C8}).  
The intertwining properties $\tau_{v,\om \xi}(f)(gg_0,n) 
= \tau_{\pi_{\om \xi}(g_0) v, \om \xi}(f)(g,n)$ 
and $\bA \tau_{v,\om \xi}(f) = 
\tau_{v,\om \xi}(\bA_{\om \xi}f)$ remain valid.  
Then Propositions \ref{relacomp} and \ref{relanoncomp} carry over.

Major modifications however occur in the structure of the 
ground state sector of a transfer operator $\bt$. Its fiber 
operators $\bt_{\om\xi}$ can be realized as integral operators 
on $L^2_{\xi}(K)$ with the following kernel
\be 
\cT_{\om \xi}(n,n';k,k') = 
\int_{M \times \R^N} \! dm da \, \cT(k^{-1} m k', a, n',n) 
\, e^{i \om \,e_1 \cdot ka} r_{\xi}(m) \,.
\label{C16}
\end{equation}
For these integral operators a counterpart of Proposition 
4.1 holds: the operators $\bt_{\om \xi}: L^2_{\xi}(\cN) 
\ra L^2_{\xi}(\cN)$ are bounded for all (not almost all) 
$\om \geq 0$, $\xi \in \widehat{M}$. Their norms are continuous 
functions of $\om$ and obey 
\begin{equation} 
\Vert \bt_{\om \xi} \Vert \leq \Vert \bt_{00} \Vert 
\quad \mbox{for all} \;\; \xi \in \widehat{M},\;
\om \geq 0\,,
\label{C17}
\end{equation}
where the inequality is strict unless $\xi =0$ and $\om =0$. 
Further $\bt_{00}$ is a transfer operator in the sense of 
Definition 2.1. 

This result entails that all the generalized eigenspaces    
$\cE_{\Vert \bt \Vert}(\bt_{\om \xi})$, $\om > 0,\,\xi \in 
\widehat{M}$, must be empty. The remaining 
$\cE_{\Vert \bt \Vert}(\bt_{00})$ coincides with the ground 
state sector  $\cG(\bt_{00})$ of $\bt_{00}$. Provided the 
map (\ref{C15}) is defined also for $\om =0, \xi =0$ it 
asigns to every generalized ground state $f \in \cG(\bt_{00})$ 
a generalized ground state $\tau_{v,00}(f)$ of $\bt$. 
Moreover by forming linear combinations $\sum_i 
\tau_{v_i,00}(f_i)$ one can generate a dense set 
in $\cG(\bt)$. Since the singlet representation $\pi_{00}$ is 
one-dimensional, while all the $\pi_{\om \xi},\,\om >0$, 
are infinite dimensional, the family of maps $\tau_{v,\om 0}$ 
is of course not continuous for $\om \ra 0$. However from 
(\ref{C16}) one can verify directly that every eigenfunction 
of $\bt_{00}$, viewed as a function on $G/K \times 
\cN/d_{\cN}(K)$ constant in the first argument, 
is also a generalized eigenfunction of $\bt$. Indeed 
the counterpart of (\ref{C15}) for the singlet representation
is simply $\tau_{00}(f)(e,n) := \int \! dk\, f(n,k)^*$, 
as the function $v$ is constant and can be omitted. 
Then $[\bt \tau_{00}(f)](e,n) = \tau_{00}(\bt_{00}f)(e,n)$,  
where the relevant kernel $\int \! dk da\, \cT(k,a,n,n')$
is symmetric in $n$ and $n'$. It follows that linear combinations 
$\sum_i c_i \tau_{00}(f_i),\,c_i \in \C$, 
generate a dense subspace of $\cG(\bt)$. In particular 
all generalized ground states of $\bt$ are functions 
on $G/K \times \cN/d_{\cN}(K)$ constant in the first 
argument. This means $\tau_{00}: \cG(\bt_{00}) 
\ra \cG(\bt)$ is an isometry. Whenever $\Vert \bt \Vert$ 
is an eigenvalue of $\bt_{00}$ both $\cG(\bt_{00})$ and 
$\cG(\bt)$ are one-dimensional. Viewed as an element of $\cG(\bt)$ 
however the wave function is not normalizable as the infinite 
volume of ${\rm ISO}(N)$ is overcounted. In summary, we arrive at 

{\bf Theorem C.1} Let $\bt$ be a transfer operator on $L^2(\R^{N\nu})$
commuting with the unitary representation (\ref{C3}) of 
${\rm ISO}(N)$. Let $\cG_{\om\xi}(\bt)$ denote the space of generalized 
ground states whose elements transform equivariantly according
to $\pi_{\om \xi},\,\om \geq 0,\,\xi \in \widehat{{\rm SO}}(N\!-\!1)$,
and let $\1 \otimes \check{\bt}_{\om \check{\xi}}$ be the component of 
$\bt$ in the fiber $\pi_{\om \xi}$. Then:  
\begin{itemize}
\item[(a)] $\cG_{\om \xi}(\bt)$ is empty unless $\om=0,\,\xi=0$, 
where $\pi_{00}$ is the singlet representation. 
\item[(b)] $\cG(\bt)$ can isometrically be identified with $\cG(\bt_{00})$ 
and is generated by rotationally invariant functions of $x_2-x_1,
\ldots, x_{\nu} - x_1$. Whenever $\Vert \bt \Vert$ is an 
eigenvalue of $\bt_{00}$ the transfer operator $\bt$ has a 
unique ground state, which is up to a phase an a.e.~strictly 
positive function of the above type. 
\end{itemize}

%%%%%%%%%%%%%%%%%%%%%%%%%%%%%%%%%%%%%%%%%%%%%%%%%%%%%%%%%%%%%%%%%%%%%%%%%%%%%%%

\newpage 

\end{document}